\documentclass[12pt]{article}

\usepackage[colorlinks=true,linkcolor=blue,citecolor=blue]{hyperref}
\usepackage{bm}
\usepackage{algorithm}
\usepackage{algpseudocode}
\usepackage{amsmath}
\usepackage{bbm}
\usepackage{cite}
\usepackage{makecell}
\usepackage{amsfonts}
\usepackage{amssymb}
\usepackage{amsthm}
\usepackage{fancybox}
\usepackage{multirow}
\usepackage{tikz}
\usepackage{thm-restate}
\usepackage{wrapfig}

\newtheorem{theorem}{Theorem}[section]
\newtheorem{corollary}[theorem]{Corollary}
\newtheorem{lemma}[theorem]{Lemma}
\newtheorem{definition}{Definition}

\newtheorem*{lemma*}{Lemma}

\newcommand{\mat}[1]{\boldsymbol{#1}}
\renewcommand{\vec}[1]{\boldsymbol{#1}}
\providecommand{\eye}{\ensuremath{\mat{I}}}
\providecommand{\mA}{\ensuremath{\mat{A}}}
\providecommand{\mB}{\ensuremath{\mat{B}}}
\providecommand{\mC}{\ensuremath{\mat{C}}}
\providecommand{\mD}{\ensuremath{\mat{D}}}

\providecommand{\mF}{\ensuremath{\mat{F}}}
\providecommand{\mG}{\ensuremath{\mat{G}}}

\providecommand{\mI}{\ensuremath{\mat{I}}}

\providecommand{\mK}{\ensuremath{\mat{K}}}
\providecommand{\mL}{\ensuremath{\mat{L}}}
\providecommand{\mM}{\ensuremath{\mat{M}}}

\providecommand{\mP}{\ensuremath{\mat{P}}}

\providecommand{\mR}{\ensuremath{\mat{R}}}

\providecommand{\mU}{\ensuremath{\mat{U}}}
\providecommand{\mV}{\ensuremath{\mat{V}}}
\providecommand{\mW}{\ensuremath{\mat{W}}}

\providecommand{\mY}{\ensuremath{\mat{Y}}}

\providecommand{\vone}{\ensuremath{\vec{1}}}
\providecommand{\vzero}{\ensuremath{\vec{0}}}
\providecommand{\va}{\ensuremath{\vec{a}}}
\providecommand{\vb}{\ensuremath{\vec{b}}}
\providecommand{\vc}{\ensuremath{\vec{c}}}

\providecommand{\ve}{\ensuremath{\vec{e}}}
\providecommand{\vf}{\ensuremath{\vec{f}}}

\providecommand{\vw}{\ensuremath{\vec{w}}}
\providecommand{\vx}{\ensuremath{\vec{x}}}
\providecommand{\vy}{\ensuremath{\vec{y}}}
\providecommand{\vz}{\ensuremath{\vec{z}}}
\providecommand{\vchi}{\ensuremath{\vec{\chi}}}

\newcommand{\norm}[1]{\left\lVert #1\right\rVert}

\newcommand{\Z}{\ensuremath{\mathbb{Z}}}
\renewcommand{\bar}[1]{\overline{#1}}
\newcommand*{\defeq}{\stackrel{\text{def}}{=}}

\newcommand{\Otil}{\widetilde{O}}

\newcommand{\bound}{O\left(\log^2p/\log\log p\right)}

\newcommand{\lscp}{SLS}
\newcommand{\lsdp}{DPLS}
\newcommand{\ddp}{DDP}
\newcommand{\smp}{SMP}
\newcommand{\GS}{\ensuremath{\mathcal{GS}}}
\newcommand{\LS}{\ensuremath{\mathcal{LS}}}
\newcommand{\WS}{\ensuremath{\mathcal{WS}}}
\newcommand{\UWLS}{\ensuremath{\mathcal{UWLS}}}
\newcommand{\DLS}{\ensuremath{\mathcal{DLS}}}

\renewcommand{\parallel}{\textsc{Para}}
\newcommand{\series}{\textsc{Seri}}
\newcommand{\size}{\mathrm{size}}

\newcommand{\maxdegree}{\mathrm{maxdeg}}

\newcommand{\maxdeg}{\maxdegree}

\newcommand{\pdeg}{\mathrm{pdeg}}
\newcommand{\maxm}{\mathrm{maxm}}
\newcommand{\sgn}{\mathrm{sgn}}

\newcommand{\m}{\mathrm{m}}

\newcommand{\nnz}{\mathrm{nnz}}
\renewcommand{\sc}{\mathbf{SC}}

\title{Hardness of Graph-Structured Algebraic and Symbolic Problems}

\author{
Jingbang Chen\\
Georgia Institute of Technology\\
\texttt{chenjb@gatech.edu}
\and Yu Gao\\
Georgia Institute of Technology\\
\texttt{ygao380@gatech.edu}
\and
Yufan Huang\\
Purdue University\footnote{Part of this work was done while at Georgia Institute of Technology}\\
\texttt{huan1754@purdue.edu}
\\
\and
Richard Peng\\
University of Waterloo\footnotemark[1]\\
\texttt{y5peng@uwaterloo.ca}
\and
Runze Wang\\
Carnegie Mellon University
\thanks{Part of this work was done while at Swarthmore College}\\
\texttt{runzew@andrew.cmu.edu}
}

\begin{document}
\maketitle
\begin{abstract}

In this paper, we study the hardness of solving graph-structured linear systems with coefficients over a finite field $\mathbb{Z}_p$ and over a polynomial ring $\mathbb{F}[x_1,\ldots,x_t]$. 

We reduce solving general linear systems in $\mathbb{Z}_p$ to solving unit-weight low-degree graph Laplacians over $\mathbb{Z}_p$ with a polylogarithmic overhead on the number of non-zeros. Given the hardness of solving general linear systems in $\mathbb{Z}_p$ \cite{CK22}, this result shows that it is unlikely that we can generalize Laplacian solvers over $\mathbb{R}$, or finite-element based methods over $\mathbb{R}$ in general, to a finite-field setting. We also reduce solving general linear systems over $\mathbb{Z}_p$ to solving linear systems whose coefficient matrices are walk matrices (matrices with all ones on the diagonal) and normalized Laplacians (Laplacians that are also walk matrices) over $\mathbb{Z}_p$.

We often need to apply linear system solvers to random linear systems, in which case the worst case analysis above might be less relevant. For example, we often need to substitute variables in a symbolic matrix with random values. Here, a symbolic matrix is simply a matrix whose entries are in a polynomial ring $\mathbb{F}[x_1, \ldots, x_t]$. We formally define the reducibility between symbolic matrix classes, which are classified in terms of the degrees of the entries and the number of occurrences of the variables. We show that the determinant identity testing problem for symbolic matrices with polynomial degree $1$ and variable multiplicity at most $3$ is at least as hard as the same problem for general matrices over $\mathbb{R}$.

\end{abstract}
\newpage
\tableofcontents

\newpage
\section{Introduction}
\label{sec:intro}

Linear system solvers have diverse applications in optimization \cite{LS19}, data science \cite{DGR05}, computer vision \cite{Fus06}, physics simulations \cite{LZL03, JJT07}, and stochastic processes \cite{BMSV06,CKP17}. Hence, faster linear system solvers are appealing to a wide range of researchers.

The current state-of-art sparse linear system solvers over finite fields \cite{CK22} and over the reals \cite{PV21,N22} run in $\omega(n^2)$ time, so people have been studying structured linear systems that admit more efficient solvers. There are two main approaches to achieve efficient solvers for structured linear systems. One is separator-based solvers like nested dissection \cite{Geo73} and its generalizations \cite{LRR79, AY10}. Alon and Yuster \cite{AY10} proved that if $\mA \vx = \vb$ satisfies $\mA$ is well-separable, i.e., the underlying graph of $\mA$ is planar or avoids a fixed minor, then we can solve the linear system over an \emph{arbitrary} field in subquadratic time \cite{AY10}. Another approach is the finite-element based solvers where the structured linear system solvers utilize dependencies between the coefficients to accelerate computation. These dependencies usually lead to nice algebraic and combinatorial properties.
One such example is graph Laplacians and symmetric diagonally-dominant (SDD) matrices.
After Spielman and Teng proposed the first near-linear time Laplacian solver \cite{ST04}, there has been a long line of research on further improving the runtime for solving Laplacian systems \cite{DS08, KMP10,KMP11, KOSZ13}. The current state-of-art runs in $\tilde{O}(m \sqrt{\log m} \log (1/\epsilon))$ \footnote{$\tilde{O}()$ hides a factor of $O(poly(\log \log n))$.} time \cite{JS21}.
Various generalizations to graph Laplacians have also been proved to admit near-linear time solvers \cite{Coh+18,KLPSS16,DS07,KyngPSZ18}. 

While finite-element based solvers make fewer assumptions about the structure of the support of the coefficient matrix, they usually work over $\mathbb{R}$ only. This is in contrast to the scenario with separator-based solvers and even general (sparse) linear system solvers.
For example, the conjugate gradient method \cite{O84} and the block Lanczos algorithm \cite{L50,M95,C93} are classical examples of generalizing techniques over $\mathbb{R}$ to a finite-field setting. More recently, Peng and Vempala \cite{PV21} proved that we can adapt Krylov space based finite-field linear system solvers \cite{EGGSV06,EGGSV07} to a floating-point setting, breaking the $O(n^\omega)$ barrier for solving sparse linear systems. However, all these methods suffer from an $O(n \cdot \nnz)$ barrier. So, it is natural to ask whether we can break the barrier by generalizing finite-element based solvers and achieve subquadratic running time on a wider class of linear systems over finite fields, which has not been studied carefully yet.

In this paper, we show that it is unlikely that we can generalize finite-element based methods over $\mathbb{R}$ to a finite-field setting. Specifically, we provide hardness results showing that solving the following classes of linear systems over $\mathbb{Z}_p$ is as hard as solving general linear systems over $\mathbb{Z}_p$: \begin{itemize} \item Laplacians over $\mathbb{Z}_p$;  \item Unit-weight Laplacians over $\mathbb{Z}_p$; \item Unit-weight Laplacians over $\mathbb{Z}_p$ with degrees at most $O(\log p)$; \item Walk matrices over $\mathbb{Z}_p$.\end{itemize}

Even though we showed that solving graph-structured matrices over $\mathbb{Z}_p$ is hard, we do not always feed the solver with the hardest inputs. For example, the Schwartz-Zippel lemma \cite{DL78,Z79,S80} allows one to decide if a symbolic matrix has full rank by plugging in \textit{random} values to the variables in the symbolic matrix and solving the system against a random vector. Solving such randomly generated matrices may be easier. Applying the Schwartz-Zippel lemma to symbolic matrices has many applications in graph algorithms, including maximum matching \cite{T47}, shortest path \cite{Sankowski05}, transitive closure \cite{S04}, and cycle detection \cite{JNS19}. This motivates us to consider the hardness of problems over symbolic matrices.

We classify graph-structured symbolic matrices in terms of the degrees of the entries and the maximum number of occurrences of each variable in the matrix. We suggest that this is a possible way to systematically and formally study the hardness of symbolic matrix problems by establishing connections between symbolic matrices and other structures.
We focus on the symbolic determinant identity testing problem: determining whether $\det(\mA)= 0$ for symbolic matrices. Specifically, let $\mA$ be a symbolic matrix whose entries are linear combinations of variables. We can state the following hardness results:
\begin{itemize}
    \item If variables appear at most $1$ time, then $\mA$ can be reduced to an Edmonds matrix and thus deciding if $\det(\mA) = 0$ can be solved by bipartite matching algorithms. \cite{motwani1995randomized}
    \item If variables appear at most $2$ times, since Tutte matrices falls into this category, checking if $\det(\mA) = 0 $ is at least as hard as general graph matching \cite{T47}.
    \item If variables appear at most $3$ times, it is as hard as checking whether $\det(\mB) = 0$ for a scalar matrix $\mB$ (Theorem~\ref{thm:symbolic}).
\end{itemize}

\paragraph{Outline} In Section~\ref{sec:related}, we provide an overview of the  literature on solving linear systems. In Section~\ref{sec:prelims}, we formally define notations and concepts relevant to the rest of the paper. We then summarize our results in Section~\ref{sec:ourresults}, including the hardness results for different structured system classes and useful tools that enable the reduction. In Section~\ref{sec:schur}, we introduce Schur complements and resistance circuits over $\mathbb{Z}_p$. We prove properties of Schur complements that enable us to replace an edge with a gadget of the same weight (or resistance). We also provide a construction of unit-weight gadget of arbitrary resistance over $\mathbb{Z}_p$. These gadgets are central to our following reductions. In Section~\ref{sec:hardnessL}, we give the reduction from arbitrary linear systems to $3$ classes of Laplacian systems, and unit-weight Laplacians with maximum degree  $O(\log p)$  is the one with strongest restrictions. We also provide methods to control the diagonal values of Laplacian systems. In Section~\ref{sec:hardnessW}, we give two reductions for proving the hardness of walk matrix systems. In Section~\ref{sec:symbolic}, we analyze the symbolic systems solving by discussing existing results, formalizing problem space, and giving a reduction to special symbolic class.

\section{Related Works}
\label{sec:related}
Before diving into the literature on linear system solvers, we formally define what solving $\mA \vx = \vb$ means. The formal definition of solving a linear system depends on the actual application, e.g.~finding exact rational solutions, finding a solution modulo some prime $p$, or finding an approximate solution up to an relative or absolute error of $\epsilon$. Hadamard's inequality states that if an $n \times n$ matrix $\mA$ over $\mathbb{R}$ has entries bounded by $B$, then $|\det(\mA)| \leq (Bn^{1/2})^n$. By applying Cramer's rule and Hadamard's inequality, if the linear system $\mA \vx = \vb$ has entries bounded by $B$, then we need $\Omega(n(\log B + \log n))$ bits to represent an entry in the solution $\vx$. Thus, computing exact rational solutions can be expensive. For most applications we are more interested in an approximate solution or solution modulo some prime $p$ that fits in the word size $w$. In the rest of the discussion, we focus on these two cases for linear systems with scalar coefficients. In other words, for linear systems with real-valued coefficients, solving the linear system means finding a solution up to an relative error of $\epsilon$. For linear systems over finite fields, we want to find exact solutions over the finite field, and we assume field operations can be done in $O(1)$ unless otherwise specified.

We may view returning a rounded approximate solution as extracting the most significant bits of the entries in an exact solution, while returning the solution modulo some prime (that fits into a word) is extracting the least significant bits of the entries in an exact solution. We can reconstuct exact rational solutions based on approximate floating-point solutions or modular solutions for linear systems with rational coefficients \cite{CS11}.

The situation is a bit more complicated when $\mA$ is a symbolic matrix. One possibility is define ``solving symbolic linear system $\mA \vx = \vb$" to be finding a vector $\vx$ whose entries are rational functions that satisfy the equation. The problem is since $\det(\mA)$ can have exponentially many terms, it would be hard for us to express the solution precisely as rational functions.  In most applications, we only care about the determinant of $\mA$. In the rest of the text, when we refer to ``solving a symbolic matrix", it means checking if $\det(\mA) = 0$ for a symbolic matrix $\mA$.

\paragraph{Structured Linear Systems Over $\mathbb{R}$}
For structured linear systems over $\mathbb{R}$, it is often possible to obtain subquadratic or even near-linear time solvers. Spielman and Teng found the first near-linear time Laplacian system solver that solves an SDD linear system (which includes Laplacian systems) up to $\epsilon$ error in $O(m \log^c n \log(1/\epsilon))$ time \cite{ST04}, even though the exponent $c$ here is large. Daitch and Spielman then showed that a linear system defined by a symmetric M-matrix, including graph Laplacians, can be solved in $\tilde{O}(m \log \frac{\kappa}{\epsilon})$ time where $\kappa$ is the condition number of the matrix \cite{DS08}. Koutis, Miller, and Peng improved the runtime to $\tilde{O}(m \log^2n \log(1/\epsilon))$ \cite{KMP10} and $\tilde{O}(m \log n \log(1/\epsilon))$ \cite{KMP11}, but the algorithms did not take into account the bit complexity of floating point arithmetic. Kelner et al. proposed a simple combinatorial algorithm for solving SDD linear systems \cite{KOSZ13} in $\tilde{O}(m \log^2n \log(1/\epsilon))$ time not relying on arbitrary-precision arithmetic. Currently, the state-of-art Laplacian solver over the reals runs in $\tilde{O}(m \log(1/\epsilon))$ time \cite{JS21}. Efficient Laplacian solvers lead to breakthroughs in many graph-theoretic problems like maximum flow \cite{DS08, CKMST11, KLS20}, semi-supervised learning \cite{ZGL03, KRSS15}, and generating random spanning trees \cite{KM09, DKP+17, Schlid18}. \cite{Teng10} and \cite{Vish12} are two classic surveys on the applications of Laplacian solvers.

There has been a long line of study on the generalization of (undirected) graph Laplacians. One possible way to generalize the notion of Laplacians is generalizing the notion of graph with geometry or topology. A simple, undirected graph can be treated as an one-dimensional simplicial complex in topology. Combinatorial Laplacians generalize graph Laplacians to simplicial complexes.  Cohen et al. proved that the 1-Laplacian of a collapsible simplicial complex with a known collapsing sequence can be solved in near-linear time \cite{Coh+14}. As Kyng et al. noticed, graph Laplacians can also be viewed as a special case of truss stiffness matrices, in which each vertex is embedded in 1 dimension \cite{KyngPSZ18}. Daitch and Spielman showed that for a 2D planar truss, it is possible to solve linear equations in the stiffness matrix in $\tilde{O}(n^{5/4})$ time \cite{DS07}. For a 3-dimensional truss over $n$ vertices which is formed from a union of $k$ convex structures with bounded aspect ratios, solving the linear system associated with the stiffness matrix can be done in $O(k^{1/3} n^{5/3} \log(1/\epsilon))$ time \cite{KyngPSZ18}. Another possible way of generalizing graph Laplacians is generalizing the notion of graph combinatorially or algebraically instead of using geomtric constructions. It has be shown that connection Laplacians \cite{KLPSS16} (where each vertex is associated with a vector and each edge is associated with a unitary matrix) and directed Laplacian systems \cite{Coh+18} have near-linear time solvers. 

\paragraph{Hardness of Structured Linear Systems}
While many generalizations of graph Laplacians have subquadratic or even near-linear time solvers, some generalizations have been shown to be as hard as general linear systems. For example, even though linear systems arising from stiffness matrices of 2D or 3D trusses with certain structures have subquadratic time solver, this is not the case for general 2D trusses. Kyng and Zhang showed that solving linear systems arising from stiffness matrices of general 2D trusses is as hard as solving general linear systems \cite{KZ17}. In the same paper, Kyng and Zhang also showed that 2-commodity Laplacians are complete for general linear systems. While special classes of combinatorial Laplacians admit near-linear time solver \cite{Coh+14}, Ding et al. showed that even solving combinatorial Laplacians of 2-complexes is as hard as solving linear systems in general \cite{DGKZ22}.

\paragraph{Structured Linear Systems Over Finite Field}

Despite the progress on solving structured linear systems over $\mathbb{R}$, less is known about structured linear systems over finite fields. Alon and Yuster showed it is possible to solve a linear system $\mA \vx = \vb$ in subquadratic time if the underlying graph of $\mA$ is separable,
e.g.~being planar or having bounded genus~\cite{AY10}. Here, given $\mA \in \mathbb{F}^{n \times n}$, its underlying graph $G_{\mA} = (\{1,\ldots,n\},E)$ has edges between vertex $i$ and vertex $j$ ($i \neq j$) if and only if $\mA_{ij} \neq 0$ or $\mA_{ji} \neq 0$.
Their algorithm works for arbitrary fields, including $\mathbb{Q}$, $\mathbb{R}$, and finite fields. The algorithm was based on the generalized nested dissection method first developed by Lipton, Rose, and Tarjan \cite{LRR79}. It remains open whether there exists any class of non-separable linear systems that can be solved in subquadratic time over finite fields. 

One application of graph-structured linear system over finite fields is the network coding problem formulated in \cite{CLL11}, which is a central component to computing $s-t$ edge connectivities. 
Let $G = (V,E)$ be a directed graph with $m$ edges and source $s$, and let $d$ denote the outdegree of the source $s$. Let $\mathbb{F}$ be a large finite field. Now for each pair of adjacent edges $(e_i,e_j)$, we will assign a random local encoding coefficient $k_{e_i,e_j} \in \mathbb{F}$. Let $\mK \in \mathbb{F}^{m \times m}$ be the matrix of local encoding coefficients such that $\mK_{i,j} = k_{e_i,e_j}$ if edges $e_i,e_j$ are adjacent and $\mK_{i,j} = 0$ otherwise. Now we are going to assign a global encoding vector $\vf_e \in \mathbb{F}^d$ for each edge $e \in E$, and let $\mF = [\vf_{e_1} \cdots \vf_{e_m}]$ be the matrix of global encoding vectors. The network coding problem in \cite{CLL11} asks for a solution of global encoding vectors such that $\mF (\mI - \mK) = [\mI_d \; \vzero].$ By construction, the underlying graph of $(\mI - \mK)$ is a subset of the line graph of $G$. Combined with the results from \cite{AY10}, the authors of \cite{CLL11} showed that it is possible to compute all edge connectivites from a source vertex in $O(n^{\omega/2})$ time for simple directed planar graphs with constant maximum degree.

\paragraph{Solving General Linear Systems Over Finite Fields}
There are many successful examples about transferring techniques first developed for solving linear systems with real coefficients to solving linear systems with coefficients in a finite field and vise versa.
On the other hand, finite-field algorithms are much more
closely connected with generating exact fractional solutions.

Some classical examples of technique transfers from the real number setting to the finite field setting include using the conjugate gradient method to solve linear systems over finite fields \cite{O84} and generalizing the Lanczos algorithm for finding eigenvalues over $\mathbb{R}$ \cite{L50} to solving linear systems over $\mathbb{Z}_2$ \cite{C93,M95}.
These methods are known to produce exact answers in
$O(\nnz \cdot n)$ arithmetic operations.
While such bounds are problematic under roundoff errors,
they carry over naturally to finite field settings.

More recently, there are studies on how one can adapt linear system solvers over finite fields to linear system solver over the reals. Peng and Vempala adapted the algorithm for inverting matrices or solving linear systems over finite fields \cite{EGGSV06,EGGSV07} to a floating point setting,
resulting in a running time of $O(n^{2.31})$~\cite{PV21} for solving sparse linear systems. This breaks the $O(n^{\omega})$ barrier for solving sparse linear systems.
Nie~\cite{N22} improved the lower bound on the minimum
singular value of the Krylov space matrix and thus improved the running time to $O(n^{2.27})$.
On the other hand, the running time of inverting a matrix
over finite fields was further improved to $O(n^{2.21})$~\cite{CK22},
asymptotically faster than the current runtime bound on solving sparse linear systems with real coefficients.

Exact rational solutions of a linear system can be reconstructed from an approximate floating point solution to the system or from the exact solution over finite fields \cite{CS11}. Dixon's scheme shows it is possible to reconstruct the exact rational solution of $\mA \vx = \vb$ by first computing $\mA^{-1} \pmod p$ for some prime $p$ and apply $
\tilde{O}(n)$ lifting steps \cite{Dix82}. The Dixon's scheme has been the basis for further work on solving sparse rational systems exactly: The linear system solver over rational numbers proposed by Eberly et. al achieves a $\tilde{O}(n^{2.5}(\log(\Vert \mA \Vert + \Vert \vb \Vert))$ bit complexity by improving the cost required for each lifting step \cite{EGGSV06}. Storjohann showed solving a rational system can be done in $\tilde{O}(n^\omega \log \Vert \mA \Vert)$ bit operations \cite{Storjohann05}.


\paragraph{Hardness of Solving General Symbolic Systems} Our discussion above has been largely focusing on fine-grained complexity within $\mathbf{P}$. Symbolic matrices are matrices whose entries are polynomials in variables. The discussion about the hardness of solving symbolic linear systems will go beyond the scope of $\mathbf{P}$ unfortunately. This is because if $\mA \vx = \vb$ is a linear system where coefficients are polynomials in variables $x_1,\ldots,x_t$, then we might need exponentially many terms to express $\det(\mA)$. For actual applications, we are more interested in checking whether $\det(\mA) = 0$, which is a special instance of more general Polynomial Identity Testing (PIT) problem.

Currently, we do not know if we can decide $\det(\mA) = 0$ for a general symbolic matrix $\mA$ in polynomial time deterministically. As mentioned before, in the worst case, $\det(\mA)$ may have exponentially many terms, so it is infeasible to expand $\det(\mA)$ out explicitly. In fact, we should not even count on computing the exact number of terms in the polynomial $\det(\mA)$. This is because if $\mA$ is the Edmonds matrix of a balanced bipartite graph $G = ((V_L, V_R), E)$, the number of terms in $\det(\mA)$ is the number of perfect matchings for $G$. Valiant has shown that counting perfect matchings is $\# P$-complete \cite{V79}, so we should not expect a polynomial-time exact algorithm for computing the number of terms in $\det(\mA)$. Nevertheless, Garg, Gurvits, Oliveira, and Wigderson proved that it is possible to determine if a symbolic matrix over $\mathbb{Q}$ is invertible or not in polynomial time deterministically assuming the variables in the symbolic matrix do \emph{not} commute \cite{GGOW16}.

Contrary to the hardness of testing if $\det(\mA) = 0$ holds deterministically, it is relatively easy to test if $\det(\mA) = 0$ using randomized algorithms with bounded error probability \cite{L79}. If we allow randomized algorithms, then checking if $\det(\mA)$ is zero is easy because we can just plug in random values to $\det(\mA)$ and work with scalars instead of polynomials. Working with scalars mean we can fully utilize numerical methods and take advantage of efficient arithmetic operations. Thanks to the Schwartz-Zippel lemma \cite{S80,Z79, DL78}, it is possible for us to bound the error probability if we simply plug in random values $x_1 = v_1,\ldots, x_t = v_t$ into $\det(\mA)$:

\begin{lemma}[Schwartz-Zippel Lemma, \cite{S80,Z79, DL78}]
Let $p$ be a non-zero degree $d$ polynomial in $\mathbb{F}[x_1,\ldots,x_n]$ and $S \subset \mathbb{F}$ be a subset of $\mathbb{F}$. If we plug in random $x_i$ where $x_i$ is chosen from $S$ uniformly randomly and independently, the probability that $p(x_1,\ldots,x_n)$ evaluates to 0 is bounded above by $\frac{d}{|S|}$.
\end{lemma}
In other words, to certify that $\det(\mA)$ is not identically zero, we may find a set $S \subset \mathbb{F}$ with size $|S| = 2d$ where $d$ is the maximum possible degree of $\det(\mA)$. Notice that if the maximum total degree of an entry of $\mA$ is at most $M$, then $\det(\mA)$ can have degree at most $nM$.  Next, we can repeatedly evaluate $\det(\mA)$ by plugging in $x_1 = v_1,\ldots,x_t = v_t$ where $v_1,\ldots,v_t$ are sampled independently and uniformly randomly from $S$. Each iteration reduces the error probability by half at least. Furthermore, if $\det(\mA)(v_1,\ldots,v_t) \neq 0$ over random samples, then $\det(\mA) \neq 0$ as a polynomial with high probability.

The dichotomy between the hardness of solving determinantal PIT (i.e.~checking if $\det(\mA) = 0$) deterministically versus randomly with bounded error further motivates us to study the following question: are there any special classes of symbolic matrices such that the hardness of determinantal PIT over these classes are somewhere between very easy (i.e.~near-linear time or almost-linear time) and very hard?
For example, one extreme of the hardness spectrum is the class of Edmonds matrix. Checking if $\det(\mA) \neq 0$ for an Edmonds matrix $\mA$ is equivalent to checking if the corresponding bipartite graph admits a perfect matching. We already know bipartite graph matching, or even general network flow problems, can be solved in almost-linear time \cite{CKL22} with high probability. A possible candidate of intermediate class of symbolic matrices is the class of Tutte matrices \cite{T47}. Testing if $\det(\mA) \neq 0$ for a Tutte matrix reduces to general graph matching, which can be solved in polynomial time using Edmonds' blossom algorithm \cite{E65}. These two results motivate us to study graph-structured symbolic matrices and the fine-grained complexity of determinantal PIT for graph-structured symbolic matrices.

\section{Preliminaries}
\label{sec:prelims}
We introduce some algebraic concepts in Section~\ref{sec:prem-alg}. To formalize our reductions and their efficiencies, 
we adopt the concepts of matrix classes and efficient $f$-reducibility similar to \cite{KZ17} and modify them for our goals. We formally introduce the graph-structured matrix classes that we will discuss throughout the paper in Section~\ref{sec:prem-matrix-class}. In Section~\ref{sec:prem-ls-solving}, we formally define what solving a linear system means. In Section~\ref{sec:prem-f-reduce}, we restate $f$-reducibility for clarity.

\subsection{Algebraic Notations}
\label{sec:prem-alg}
We start by defining the various algebraic objects involved.

For any real number $a$, we use $\lfloor a \rfloor$ to denote the
largest integer $x$ such that $x \le a$, and $\lceil a \rceil$ to
denote the smallest integer $x$ such that $x \ge a$.
For any two real numbers $a, b\ (a \le b)$, we use $[a,b]$ to denote the finite set of integers $x$ such that $a \le x \le b$, e.g.~$[1,n] = \{1, 2, \ldots, n\}$. For convenience, for any real number $a \ (a \ge 1)$, let $[a] = [1,a]$. 

We let $\mathbb{Z}_p$ be the finite field of size $p$. First we state the definition of additive and multiplicative inverse over 
$\mathbb{Z}_p$.

\begin{definition}
    For any $x \neq 0 \in \mathbb{Z}_p$, if $y \in \mathbb{Z}_p$ satisfies $xy = 1 \pmod p$, $y$ is $x$'s multiplicative inverse in $\mathbb{Z}_p$, which is denoted as $x^{-1}$.
    For any $x \in \mathbb{Z}_p$, its additive inverse in $\mathbb{Z}_p$
    is $(p - x) \pmod p$, denoted as $-x$.
\end{definition}

    Unless otherwise stated, inverse denotes multiplicative inverse. Moreover, over the finite field $\mathbb{Z}_p$, we can adopt a similar definition of matrix inverse to the real case.
    
    \begin{definition}
        A square matrix $\mA \in \mathbb{Z}_p^{n \times n}$ is non-singular if and only if there exists some matrix $B\in \mathbb{Z}_p^{n \times n}$ such that
        \begin{align*}
            \mA \mB = \mB \mA = \eye \pmod p,
        \end{align*}
    and $\mB$ is the inverse of $\mA$ over $\mathbb{Z}_p$.
    \end{definition}

We use subscripts to index the rows and columns of matrices. For any matrix $\mA \in \mathbb{Z}_p^{m \times n}$, let $\mA_{R, \cdot}$ denote the sub-matrix indexed by the row index set $R \subseteq [m]$ and $\mA _{\cdot, C}$ denote the sub-matrix indexed by the column index set $C \subseteq [n]$.
For simplicity, we use $a:b$ to represent $[a, b]$ in subscripts.

We use $\vone$ to denote the all-ones vector. We use $\ve_i$ to
denote the $i$-th standard basis vector and let $\vchi_{i, j} = \ve_i -
\ve_j$. For any matrix $\mA \in \mathbb{Z}_p^{m \times n}$,  we use
$\nnz(\mA)$ to denote the number of non-zero entries in $\mA$.



\begin{definition}
    An integer $q$ is called a \emph{quadratic residue} modulo $p$ if it is congruent to a perfect square modulo $p$, i.e.~$\exists x$ such that
    $$x^2 = q \pmod p,$$
    otherwise $q$ is not a \emph{quadratic residue} modulo $p$.
\end{definition}

\paragraph{Field Extension.} If fields $\mathbb{F} \subseteq \mathbb{E}$ satisfy operations in $\mathbb{F}$ are equal to operations in $\mathbb{E}$ restricted to $\mathbb{F}$, then we say $\mathbb{E}$ is an extension field of $\mathbb{F}$. We use $\mathbb{F}(a)$ to denote the smallest field containing $\mathbb{F}$ and $a$. For example, $\mathbb{Q}(\sqrt{2}) = \{ a + b \sqrt{2} | a, b \in \mathbb{Q}\}$. Division in $\mathbb{Q}(\sqrt{2})$ can be conducted using the relation $\frac{1}{a + b\sqrt{2}} = \frac{a - b\sqrt{2}}{a^2+2b^2}$ when $a,b$ are not both zero. Similarly, since $3$ is not a quadratic residue modulo 5, $\mathbb{Z}_5(\sqrt{3}) = \{a + b\sqrt{3} | a,b \in \mathbb{Z}_5\}$. Divisions in $\mathbb{Z}_5(\sqrt{3})$ can be conducted using the relation $\frac{1}{a + b\sqrt{3}} = \frac{a - b\sqrt{3}}{a^2 + 3b^2}$ whenever $a,b$ are not both zero. Another class of fields that will appear in the text is field of rational functions: $\mathbb{F}(x_1,\ldots,x_n)$ denotes the field of rational functions over variables $x_1,\ldots,x_n$ with coefficients in field $\mathbb{F}$.

It should be noted that $\mathbb{F}(x_1,\ldots,x_n)$ is not equal to $\mathbb{F}[x_1,\ldots,x_n]$. The later one denotes the ring of polynomial functions over variables $x_1,\ldots,x_n$ with coefficients in $\mathbb{F}$. In general, given a ring $R$ and $a$, $R[a]$ denotes the smallest ring containing both $R$ and $a$. If $a$ is algebraic over a field $\mathbb{F}$, then $\mathbb{F}[a]$ is  a field, and $\mathbb{F}[a] = \mathbb{F}(a)$. 

The degree of a field extension $\mathbb{F} \subseteq \mathbb{E}$, denoted as $[\mathbb{E} : \mathbb{F}]$, is the dimension of $\mathbb{F}$ when viewed as a vector space over $\mathbb{E}$. For example, $[\mathbb{Q}(\sqrt{2}) : \mathbb{Q}] = 2$ since $\mathbb{Q}(\sqrt{2})$ can be viewed as a 2-dimensional vector space over $\mathbb{Q}$ with basis $(1,\sqrt{2})$.

\subsection{Graph-Structured Matrix Classes}
\label{sec:prem-matrix-class}

A matrix class is an infinite set of matrices with a common structure. 
The class of general matrices over $\mathbb{Z}_p$ is $\GS_{\Z_p}$. Similarly, $\GS_{\mathbb{R}}$ denotes the class of general matrices over $\mathbb{R}$.

The first graph-structured matrix class over $\Z_p$ is  Laplacians because Laplacians is arguably the most well-studied graph-structured matrix class over the reals. Normally, Laplacian matrices are defined over $\mathbb{R}$ and they are symmetric matrices with all off-diagonal entries non-positive and $\vone$ in its null space. 
\begin{definition}[Laplacian Matrix in $\mathbb{R}$]
\label{matrix::r}
For every simple, weighted, undirected graph $G = (V, E, \vw: E \to \mathbb{R}_{+})$ with $|V|$ vertices and $|E|$ edges, the Laplacian matrix $\mL$ is defined as:  
    \begin{equation*}
        \begin{split}
            \mL_{a, b} = 
            \begin{cases}
                \sum_{(a, u) \in E} \vw_{a, u}, & \text{if } a = b  \\
                -\vw_{a, b}, & \text{if } (a, b) \in E \\ 
                0, & \text{otherwise} \\
            \end{cases}
        \end{split}
    \end{equation*}
\end{definition}


Over $\mathbb{Z}_p$, we adopt a similar definition of Laplacian 
matrices using additive inverses. Notice that 
unlike Laplacian matrices over the reals, Laplacians
over $\mathbb{Z}_p$ only need to be symmetric and have
$\vone$ in its null space. This is because we cannot distinguish positive and negative numbers in $\mathbb{Z}_p$.

\begin{definition}[Laplacian Matrix in $\mathbb{Z}_p$, $\LS_{\Z_p}$]
\label{matrix::zp}
    \label{def:laplacian_modp}
For every simple, weighted, undirected graph $G = (V, E, \vw: E \to \mathbb{Z}_p\setminus \{0\})$ with $|V|$ vertices and $|E|$ edges, the Laplacian matrix $\mL\in \mathbb{Z}_p^{n\times n}$ is defined as:  
    \begin{equation*}
        \begin{split}
            \mL_{a, b} = 
            \begin{cases}
                \sum_{(a, u) \in E} \vw_{a, u}, & \text{if } a = b  \\
                -\vw_{a, b}, & \text{if } (a, b) \in E \\ 
                0, & \text{otherwise} \\
            \end{cases}
        \end{split}
    \end{equation*}
\end{definition}

When $G$ contains parallel edges, we can replace them by a single edge whose weight equals to the sum of weights of the parallel edges. We repeat this process until $G$ becomes a simple graph and define the Laplacian of $G$ as the Laplacian of the resulting simple graph.

We will be using the same definition for ``underlying graph" as used in \cite{AY10}. In particular, if $\mL$ is the Laplacian of $G = (V,E,  \vw: E \to \mathbb{Z}_p\setminus \{0\})$, its underlying graph $G_{\mL}$ is equal to $G' = (V,E)$ obtained by discarding the weight information in $G$. Notice the underlying graph is an unweighted graph.

\begin{definition}[Underlying Graph, \cite{AY10}]
Given an $n \times n$ matrix $\mA$, the underlying graph $G_{\mA}$ of $\mA$ has $n$ vertices $\{1,\ldots,n\}$. There is an edge between vertex $i$ and $j$ ($i \neq j$) if and only if $\mA_{ij} \neq 0$ or $\mA_{ji} \neq 0$.
\end{definition}

%
%

One special subclass of Laplacians that is worth attention is those whose underlying graphs have all edges
being unit-weight, i.e. all edges have weight 1.

\begin{definition}[Unit-Weight Laplacian Matrix in $\mathbb{Z}_p$, $\UWLS_{\Z_p}$]
\label{def:uwls}
    A Laplacian matrix $\mL \in \mathbb{Z}_p^{n \times n}$ is
    a unit-weight Laplacian matrix if and only if all of its off-diagonal non-zero 
    entries are -1.
\end{definition}

Alternatively, we may also define the Laplacians of unit-weight graphs via incidence matrices:
\begin{definition}[Incidence Matrix]
Given a unit-weight graph $G=(V,E)$ with $|V| = n$ and $|E| = m$, choose one orientation for each edge $e \in V$ (i.e.~pick one vertex as the head of $e$ and the other one as tail), the (oriented) incidence matrix $\mB$ is a $n \times m$ matrix defined as 
\begin{equation*}
        \begin{split}
            \mB_{v, e} = 
            \begin{cases}
                1, & \text{if  $v$ is the head of $e$;} \\
                -1, & \text{if  $v$ is the tail of $e$;}\\ 
                0, & \text{otherwise}. \\
            \end{cases}
        \end{split}
\end{equation*}
\end{definition}
\begin{definition}[Unit-Weight Laplacian Matrix in $\mathbb{R}$ and $\mathbb{Z}_p$ via Incidence Matrix]
Given a unit-weight graph $G = (V,E)$ with $|V| = n$ and $|E| = m$ with oriented incidence matrix $\mB$, the Laplacian of $G$ is defined to be 
\begin{equation*}
\mL \defeq \mB \mB^T.
\end{equation*}
\end{definition}

It is easy to verify that the incidence matrix definition of the Laplacian of a unit-weight graph is equivalent to the definition that treats the graph as a weighted graph with edge weights 1. Moreover, both definitions work over $\mathbb{R}$ and over $\mathbb{Z}_p$. One advantage of expressing the Laplacian $\mL$ as the product $\mB \mB^T$ is this notation emphasizes that unit-weight Laplacians have factor width at most 2. Here, by the definition of factor width in \cite{BCPT05}, the factor width of a symmetric matrix $\mA$ is the smallest integer $k$ such that there exists a matrix $\mV$ such that $\mA = \mV \mV^T$ and each column of $\mV$ contains at most $k$ non-zero entries.

Besides unit-weight Laplacians, another non-trivial subclass of Laplacians is low-degree Laplacians. Intuitively, low-degree Laplacians refer to those Laplacians
whose underylying graphs do not have a node with a very high degree. Formally,
we define two kinds of degrees.
\begin{definition}[Combinatorial Degree of a Laplacian matrix in $\Z_p$]
\label{def:deg}
Given a Laplacian $\mL \in \Z_p^{n \times n}$ with coordinate (vertex) set $V$, we 
define the combinatorial degree of a vertex $v$ as 
    $$\deg_{\mL} (v) \defeq \sum_{u \neq v} \mathbbm{1}_{\{\mL_{u,v} \neq 0\}},$$
and let $\Delta_{\mL}$ denote the maximal combinatorial degree over $V$, i.e.~
    $$\Delta_{\mL} = \max_{v \in V}\ \deg_{\mL}(v).$$
\end{definition}

\begin{definition}[Weighted Degree of a Laplacian matrix in $\Z_p$]
\label{def:weighteddegree}
Given a Laplacian $\mL \in \Z_p^{n \times n}$  with coordinate (vertex) set $V$, we 
define the weighted degree of a vertex $v$ as 
    $$\deg_{\mL}^{\vw} (v) \defeq \sum_{u \neq v } -\mL_{u, v},$$
and let $\Delta_{\mL}^{\vw}$ denote the maximal weighted degree over
$V$, i.e.~
\begin{align*}
    \Delta_{\mL}^{\vw} = \max_{v \in V} \deg_{\mL}^{\vw} (v)
\end{align*}
\end{definition}

Now we can define the degree restricted matrix class.

\begin{definition}[$t$-weighted-degree Laplacian Matrix in $\mathbb{Z}_p$,$\DLS_{t,\Z_p}$]
\label{def:dls}
A Laplacian matrix $\mL \in \mathbb{Z}_p^{n \times n}$ with coordinate (vertex) set $V$ is a $t$-degree Laplacian matrix if and only if $\Delta_{\mL}^{\vw}$ is in $O(t)$.
\end{definition}

Another interesting class of matrices over $\mathbb{Z}_p$ is those having the form of $\mI - \mP$. This class of matrices has application in graph connectivity \cite{CLL11} and we name it walk matrix because it looks like the matrix involved in the famous linear system $(\mI - \mP) \vx = \vzero$ which computes the stationary distribution of 
a random walk.
\begin{definition}[Walk matrix in $\Z_p$, $\WS_{\Z_p}$]
\label{def:walkmatrix}
Over $\Z_p$,  a walk matrix is a matrix of form $\mI - \mP$ where $\mI$ is the identity matrix and $\mP$ is an arbitrary (symmetric) matrix whose diagonal entries are 0s.
\end{definition}

We may wonder how walk matrices intersect with the (weighted) Laplacians defined earlier. Over the reals, the normalized Laplacian for a graph without isolated vertices is both a walk matrix and a Laplacian. An isolated vertex of a graph is simply a vertex with degree 0:
\begin{definition}[Normalized Laplacians over $\mathbb{R}$, \cite{C96}]
\label{def:normLaplacianR}
Let $G = (V,E,\vw: E \rightarrow \mathbb{R}_+)$ be a simple, weighted, undirected graph without isolated vertices. Let $\mL = \mD - \mA$ be its Laplacian as defined in Definition \ref{matrix::r}. Its normalized Laplacian $\mathcal{L}$ is defined to be
\begin{equation*}
\mathcal{L} = \mD^{-1/2} \mL \mD^{-1/2}.
\end{equation*}
\end{definition}

Analogously, over $\mathbb{Z}_p$, we may define a normalized Laplacian matrix to be a Laplacian that is also a walk matrix.
\begin{definition}[Normalized Laplacian Matrix over $\mathbb{Z}_p$]
A normalized Laplacian matrix over $\mathbb{Z}_p$ is a Laplacian matrix that is also a walk matrix. In other words, a normalized Laplacian matrix over $\mathbb{Z}_p$ is a Laplacian matrix with all 1s on the main diagonal.
\end{definition}

It should be noted that if $\mL$ is the Laplacian of a simple, weighted, undirected graph without isolated vertices over $\mathbb{Z}_p$, $\mL$ can still have zeros on the diagonal. We are also not guaranteed that $\mD^{-1/2}$ exists since not all elements in $\mathbb{Z}_p$ are quadratic residues.

\subsection{Linear Systems Solving}
\label{sec:prem-ls-solving}

In this section we give the formal definition of computational and decision problem of solving a linear system over $\mathbb{Z}_p$. 

\begin{definition}[Solving Linear System over $\mathbb{Z}_p$, \lscp]
\label{def:lscp}
    Given a linear system $(\mA, \vb)$ where $\mA \in \mathbb{Z}_p^{m \times n}, \vb \in \mathbb{Z}_p^{m}$, we define the \lscp{} problem as finding a solution $\vx^*\in \mathbb{Z}_p^{n}$ such that
    \begin{align*}
        \mA \vx^* = \vb \pmod p.
    \end{align*}
\end{definition}

\begin{definition}[Decision Problem for Linear System over $\mathbb{Z}_p$, \lsdp]
\label{def:lsdp}
    Given a linear system $(\mA, \vb)$ where $\mA \in \mathbb{Z}_p^{m \times n}, \vb \in \mathbb{Z}_p^{m}$, we define the \lsdp{} problem as deciding whether any solution $\vx^*\in \mathbb{Z}_p^{n}$ exists such that
    \begin{align*}
        \mA \vx^* = \vb \pmod p.
    \end{align*}
\end{definition}

Throughout this paper we consider solving the linear systems exactly, of which the running time mainly depends on the number of non-zero entries in $\mA$. 
Therefore to measure the difficulty of solving a linear system over $\mathbb{Z}_p$, for any linear system $(\mA \in \mathbb{Z}_p^{m \times n}, \vb \in \mathbb{Z}_p^m)$,
we denote the sparse complexity
of it as 
\begin{align*}
    \mathcal{S}(\mA, \vb) = \nnz(\mA).
\end{align*}

\subsection{Reduction Between Matrix Classes}
\label{sec:prem-f-reduce}

\begin{definition}[Efficient $f$-reducibility] 
    Suppose we have two matrix classes $\mathcal{M}^1$ and $\mathcal{M}^2$ and
    two algorithms $\mathcal{A}_{1 \mapsto 2}$, $\mathcal{A}_{2 \mapsto 1}$ such that given an \lscp{} instance $(\mM^1, \vc^1)$ where $\mM^1 \in \mathcal{M}^1$, the algorithm $\mathcal{A}_{1 \mapsto 2}$ returns an \lscp{} instance ($\mM^2, \vc^2$) = $\mathcal{A}_{1 \mapsto 2}(\mM^1, \vc^1)$ such that $\mM^2 
    \in \mathcal{M}^2$, and  if $\vx^2$ is a solution to the \lscp{} instance $(\mM^2, \vc^2)$ then $\vx^1 = \mathcal{A}_{2 \mapsto 1}(\mM^2, \vc^2, \vx^2)$ is a solution the \lscp{} instance $(\mM^1, \vc^1)$.
    
     Consider a function of $f: \mathbb{R}_{+} \mapsto \mathbb{R}_{+}$ such that every output coordinate is a non-decreasing function of
every input coordinate. If we always have
\begin{align*}
    \mathcal{S}(\mM^2, \vc^2) \le f(\mathcal{S}( \mM^1, \vc^1) )
\end{align*}
and the running times of $\mathcal{A}_{1 \mapsto 2}$ and $\mathcal{A}_{2 \mapsto 1}$ are both bounded by $\Otil(\nnz(\mM^1))$,  we say $\mathcal{M}^1$ is efficiently $f$-reducible to $\mathcal{M}^2$, which we also writes 
\begin{align*}
    \mathcal{M}^1 \le_{f} \mathcal{M}^2
\end{align*}

\end{definition}

We will prove that our reduction works for the decision problem as well.
\begin{definition}[Decisional $f$-reducibility] 
    Suppose we have two matrix classes $\mathcal{M}^1$ and $\mathcal{M}^2$ and an algorithm $\mathcal{A}_{1 \mapsto 2}$ such that given a \lsdp{} instance $(\mM^1, \vc^1)$ where $\mM^1 \in \mathcal{M}^1$, the algorithm $\mathcal{A}_{1 \mapsto 2}$ returns an \lsdp{} instance  ($\mM^2, \vc^2$)  = $\mathcal{A}_{1 \mapsto 2}(\mM^1, \vc^1)$ that has the same output as $(\mM^1, \vc^1)$ and satisfies $\mM^2 \in \mathcal{M}^2$.
    
    Consider a function of $f: \mathbb{R}_{+} \mapsto \mathbb{R}_{+}$ such that every output coordinate is a non-decreasing function of
every input coordinate. If we always have
\begin{align*}
    \mathcal{S}(\mM^2, \vc^2) \le f(\mathcal{S}( \mM^1, \vc^1) )
\end{align*}
and the running times of $\mathcal{A}_{1 \mapsto 2}$ is bounded by $\Otil(nnz(\mM^1))$, we say $\mathcal{M}^1$ is decisional $f$-reducible to $\mathcal{M}^2$.
\end{definition}

\section{Our Results}
\label{sec:ourresults}

We now state and discuss our main results.

\paragraph{Hardness of Laplacian Systems Over $\mathbb{Z}_p$.} Since $\mathbb{Z}_p$ does not distinguish positive and negative numbers, a Laplacian matrix over $\mathbb{Z}_p$ is a symmetric matrix  having $\vone$ in its kernel (see Definition~\ref{matrix::zp}). 
This makes Laplacians over $\mathbb{Z}_p$ more expressive than Laplacians over the reals as both $x$ and $-x$ may appear in the non-diagonal entries.
By utilizing this property,
we can reduce a general linear system to a Laplacian system
over $\mathbb{Z}_p$.
The theorem below says that for any prime $p$,
solving an arbitrary linear system $\mA$ over $\mathbb{Z}_p$
is as hard as solving a weighted Laplacian linear system
$\mL$ where $\nnz(\mL) = O(\nnz(\mA))$. Note that we can use the same reduction to show that over
an arbitrary field, general linear systems are as hard as linear systems whose coefficient matrices are symmetric matrices with $\vone$ in the kernels over the same field.

\begin{restatable}[Hardness for Laplacian systems in $\mathbb{Z}_p$]{theorem}{thmGfL}
\label{theorem_GfL}
      Let \[f(\nnz) = O(\nnz),\] then $\GS_{\mathbb{Z}_p} \le_{f} \LS_{\mathbb{Z}_p}$.
\end{restatable} 

Our reductions preserve the entire solution space, so our reduction works both for deciding whether the linear system has a solution (\lsdp, see Definition~\ref{def:lsdp}) and finding an actual solution (\lscp, see Definition~\ref{def:lscp}).

\paragraph{Hardness of Walk Matrix Systems Over $\mathbb{Z}_p$ and Extended Field.}
As an application of our previous results on Laplacian system hardness, we can show that solving  matrices of form $\mI_n - \mA$ are also as hard as solving general linear systems. Here, $\mA$ is a symmetric matrix with zeros on the diagonals, so $\mA$ can be an adjacency matrix. We refer to matrices of form $\mI_n - \mA$ as walk matrices (Definition~\ref{def:walkmatrix}). As van den Brand, Nanongkai, and Saranurak noted, if $\mA$ is the adjacency matrix, $\mI_n + \mA$ will encode all valid cycle covers of the graph \cite{JNS19}. Normalized Laplacians for graphs without isolated vertices \cite{C96} are also matrices of form $\mI_n - \mA$. We can use a similar construction as the reduction from general linear systems to Laplacian systems over $\mathbb{Z}_p$. This reduction involves constructing a $8 \times 8$ block matrix.

\begin{restatable}[Hardness for Walk Matrix systems in $\mathbb{Z}_p$]{theorem}{thmGfW}
\label{theorem_GfW}
      Let \[f(\nnz) = O(\nnz),\]then $\GS_{\mathbb{Z}_p} \le_{f} \WS_{\mathbb{Z}_p}$.
\end{restatable} 

Alternatively, we can strengthen the conclusion by reducing arbitrary linear systems to normalized Laplacian systems. We have shown that we can reduce an arbitrary linear system $\mA \vx = \vb$ over $\mathbb{Z}_p$ to a Laplacian linear system. To further reduce a Laplacian $\mL = \mD - \mA$ to a normalized Laplacian, we can simply normalize the Laplacian by multiplying $\mD^{-1/2}$ on the left and on the right of $\mL$. This normalization does not work directly over finite fields, since not all elements are quadratic residues and having a 0 on the diagonal does not imply the entire row is zero. Nevertheless, with the help of our low-degree Laplacian construction or other described Laplacian diagonal controlling tools, we are able to control the diagonal entries in the reduced Laplacian matrix so that we don't create zeros on the diagonal unless the entire row is zero. We can also work in a field extension so that all elements in $\mathbb{Z}_p$ can have a square root in the extension field. Finally, we can show it is easy to reduce a solution in the extension to a solution in $\mathbb{Z}_p$ simply by keeping only the part in $\mathbb{Z}_p$.

\begin{restatable}[Hardness for Walk Matrix systems in $\mathbb{Z}_p{[}\sqrt{t}{]}$]{theorem}{thmLfW}
\label{theorem_LfW}
      Let \[f(\nnz) = O(\nnz),\] and $t$ is the primitive root of $p$, then $\LS_{\mathbb{Z}_p} \le_{f} \WS_{\mathbb{Z}_p[\sqrt{t}]}$.
\end{restatable} 

\paragraph{Hardness of Unit-Weight Laplacian Systems Over $\mathbb{Z}_p$.}

The intuition that allows us to reduce a general linear system to a Laplacian system in $\mathbb{Z}_p$ is that $\mathbb{Z}_p$ does not distinguish positive and negative numbers.
We are going to show that the Laplacians corresponding to simple, unweighted, undirected graphs are as hard as general linear systems over $\mathbb{Z}_p$. This is in contrast to the situation over $\mathbb{R}$, since we can solve Laplacian systems where the corresponding graph is a simple, unweighted, and undirecteed graph efficiently. 

One important tool of our reductions is the \emph{Schur complement} (SC).
In Lemma~\ref{lem:replace_edge}, we will show that
Schur complement allows us to replace any edge in a graph
by a constructed gadget circuit, or vice versa,
while preserving its effective resistance.
This process will slightly blow up the dimensions of
the matrices.

Therefore, we further showed regardless of the original magnitudes of non-zero entries in the matrix, we can reduce solving a (sparse) general linear system over $\mathbb{Z}_p$ to solving a unit-weight Laplacian system with a blow-up factor of $O(\log^2  p/\log \log p)$ on the number of non-zero entries. This implies that solving unit-weight Laplacians are as hard as solving general linear systems over $\mathbb{Z}_p$, up to poly-logarithmic overhead in runtime. 

Lemma~\ref{lem:replace_edge} can also be applied to eliminate parallel edges. We can replace each edge with a circuit of weight 1 such that after replacement the parallel edges are no longer parallel. This can be achieved by replacing an edge with two length-2 paths connected in parallel. 
In other words, we are replacing each of the parallel edges with a cycle consisting of four edges.


\begin{restatable}[Hardness for unit-weight Laplacian systems in $\mathbb{Z}_p$]{theorem}{thmGfUL}
\label{thm:GfUL}
    Let \[f(\nnz) = O(\nnz \cdot \log^2 p /\log \log p),\]then $\LS_{\mathbb{Z}_p}\le_f \UWLS_{\mathbb{Z}_p}$.
\end{restatable}

\paragraph{Hardness of Low-Degree Laplacian Systems Over $\mathbb{Z}_p$.} 
Our results on the hardness of unit-weight Laplacians imply that given an arbitrary linear system $\mA \vx = \vb$ over $\mathbb{Z}_p$, we can reduce it to a Laplacian system $\mL \vy = \vc$ such that $\mL$ is the Laplacian of an unweighted sparse graph. The reduction does not guarantee that the degrees of individual vertices in the graph corresponding to $\mL$ are small. For example, this does not preclude we reduce $\mA \vx = \vb$ to a unit-weight Laplacian $\mL \vy = \vc$ where $\mL$ corresponds to a star.
Our results on low-degree Laplacians show that Laplacian systems corresponding to unit-weighted graphs with maximum degree $O(\log p)$ are complete for all linear systems over $\mathbb{Z}_p$.

\begin{restatable}[Hardness for Low-Degree Laplacian systems in $\mathbb{Z}_p$]{theorem}{thmGfDL}
\label{thm:GfDL}
    Let \[f(\nnz) = O(\nnz \cdot \log \nnz \cdot \log^2 p /\log \log p),\] then $\LS_{\mathbb{Z}_p}\le_f \DLS_{\log p,\mathbb{Z}_p}$.
\end{restatable} 

These results indicate that if one can develop efficient solvers of these special structured linear systems, all linear systems over $\mathbb{Z}_p$ can be solved equally efficiently. Moreover, if a fast solver exists, it might imply a non-trivially fast algorithm for solving linear systems exactly in $\mathbb{Q}$ since we may apply Dixon's scheme \cite{Dix82} to reconstruct rational solutions. This remains an interesting open problem.


\paragraph{Construction of Arbitrary Resistances Modulo $p$}

The key component to the reduction from Laplacian linear systems to unit-weight Laplacian is to construct circuits of arbitrary resistance modulo $p$. With the help of Lemma~\ref{lem:replace_edge}, we can then replace off-diagonal entries (or in other words, an edge in the graph corresponding to the Laplacian) with the corresponding circuit. Our results prove that we need only $\bound$ unit-weight edges to represent all resistance values $0,\ldots, p-1 \pmod p$. Furthermore, the maximum degree of the circuit is $O(\log p)$. We can construct the desired circuit in $O(\log^3 p)$ time.


\begin{restatable}[Small Unit Circuits of Arbitrary Resistance
over $\mathbb{Z}_{p}$]{lemma}{lemRConstruct}
\label{lem:resistance_construction}
For any prime $p > 2$
and any integer $r$ in $[1, p-1]$,
we can construct in $O(\log^3 p)$ time
a network $\mL$ of unit resistors (aka. an undirected unweighted graph) 
with two vertices $s$ and $t$ such that the effective
resistance between $s$ and $t$ is $r \pmod p$ and that 
\begin{itemize}
    \item the number of non-zero entries of the circuit, $\nnz(\mL)$, is $\bound$,
    \item the maximum degree of the circuit, $\Delta_{\mL}^{\vw}  = \Delta_{\mL}$, is $O(\log p)$.
\end{itemize}
\end{restatable}

\paragraph{Combinatorial Degree Decrease of Laplacians Over $\mathbb{Z}_p$} The construction of arbitrary resistances modulo $p$ proves we can replace undesired off-diagonal entries with -1. The following lemma proves we can also control the diagonal elements of the resulting Laplacian. In particular, we can guarantee that given a Laplacian linear system, we can reduce the Laplacian to another Laplacian where each row or column contains only $O(1)$ non-zero entries. Specifically, we achieve this by transforming it to a sparser graph. Our result shows that this only increases the number of non-zero entries by a factor of $O(\log \nnz)$.
\begin{restatable}[$O(1)$-Combinatorial-Degree Laplacian Construction
over $\mathbb{Z}_{p}$]{lemma}{lemcombdegreedecrease}
\label{lemma:combdegree}
For a Laplacian matrix $\mL \in \mathbb{Z}_p^{n \times n}$ with an underlying graph $G = (V,E)$, we can construct a Laplacian matrix $\hat{\mL} \in \mathbb{Z}_p^{k \times k}$ with an underlying graph $H = (V',E')$ such that $V \subseteq V'$ and $ \sc(\hat{\mL},V)=\mL$. We have $\Delta(H) = O(1)$ and $\nnz(\hat{\mL})=\Theta(|V'|)=\Theta(|E'|)=O(|E|\log|E|)=O(\nnz(\mL)\log \nnz(\mL))$.
\end{restatable}

The above two tools directly help constructing the low-degree Laplacian reduction. One can see that if we first decrease the combinatorial degree of a Laplacian matrix to $O(1)$ by Lemma \ref{lemma:combdegree} and then replace every edge by the unit-weight gadget corresponding to its weight from \ref{lem:resistance_construction}, we will have an unit-weight Laplacian of degree $O(\log p)$.

\paragraph{Hardness of Symbolic Matrices}
We now consider checking whether $\det(\mA)=0$ for symbolic matrices $\mA$ whose entries are linear combinations of variables. Edmonds matrices and Tutte matrices \cite{T47} are two important classes of graph-structured symbolic matrices with connections to graph matchings. If each variable appears only once, then entries of $\mA$ are independent, so we can reduce $\mA$ to an Edmonds matrix. If variables appear at most twice, then checking if $\det(\mA) = 0$ is at least as hard as general graph matching since Tutte matrices satisfy each entry is a linear combination of variables and each variable appears at most twice. What if we allow variables to appear at least three times in $\mA$? In this case, we can show checking if $\det(\mA) = 0$ is at least as hard as checking if $\det(\mB) = 0$ for a matrix $\mB \in \GS_{\mathbb{R}}$:

\begin{restatable}[Hardness of A Special Class of Symbolic Matrices]{theorem}{thmsymb}
\label{thm:symbolic}
   Let $f(\nnz) = O(\nnz)$, then $\mathcal{\GS_{\mathbb{R}}}^{n \times n}$ is decisional $f$-reducible to a symbolic matrix class $\mathcal{SM}[\mathbb{R}[x_1,\ldots,x_{n^2}], 1,3]$.
\end{restatable}

\section{Schur Complements and Resistance Circuits Over $\mathbb{Z}_p$}
\label{sec:schur}
\subsection{Schur Complements}
Schur complement plays an important role in our reductions between subclasses of Laplacians. It enables us to transform a linear system to another one and as long as the Schur complement 
with respect to a specific terminal set does not change, the solution space is
preserved. Therefore we first introduce its definition over $\Z_p$ and some properties. 

\begin{definition}
    For any square matrix $\mA \in \mathbb{Z}_p^{n \times n}$, let $T \subseteq [n]$ be some index set and $S = [n] \setminus T$, then $\mA$ can be written as
    $
    \left[
    \begin{array}{cc}
    \mA_{T,T} & \mA_{T,S} \\
    \mA_{S, T} & \mA_{S, S}
    \end{array}
    \right]
    $ 
    where $\mA_{T, T}, \mA_{T, S}, \mA_{S, T}, \mA_{S, S}$ are block matrices. The Schur complement of $\mA$ with respect to $T$ is defined as    
    \begin{align*}
        \sc(\mA, T) = \mA_{T, T} - \mA_{T, S} \mA_{S, S}^{-1} \mA_{S, T}
    \end{align*}
    which exists when $\mA_{S, S}$ is non-singular.
\end{definition}

The following three lemmas state that the Schur complement always preserves the solution space as it can be regarded as doing a partial Gaussian elimination. 

\begin{lemma}
\label{lem:sc_preserve_sol_1}
    Let $\vx$ be a solution vector for the linear system 
    $\mA \vx = \left[\begin{array}{c} \vb \\ \vzero \end{array}\right]$ and $T$ be the indices of $\vb$. If $\sc(\mA, T)$ exists, then $\sc(\mA, T) \vx_T = \vb$.  
\end{lemma}
\begin{proof}
    Write $\vx = 
    \left[ 
    \begin{array}{c}
    \vx_T \\
    \vx_S
    \end{array}
    \right]$,
    we have
    \begin{align*}
        \left[
        \begin{array}{cc}
           \mA_{T, T}  & \mA_{T, S} \\
           \mA_{S, T}  & \mA_{S, S} \\
        \end{array}
        \right]
        \left[
        \begin{array}{c}
             \vx_T  \\
             \vx_S  \\
        \end{array}
        \right]
        =
        \left[
        \begin{array}{c}
            \vb  \\
            \vzero  
        \end{array}
        \right]
    \intertext{ which implies }
    \mA_{T, T} \vx_T + \mA_{T, S} (-\mA_{S, S}^{-1} \mA_{S,T} \vx_T) =  \sc(\mA, T)\vx_T =  \vb  
    \end{align*}
    where the existence of $\mA_{S, S}^{-1}$ is guaranteed by the assumption $\sc(A, T)$ exists. 
\end{proof}

\begin{lemma}
\label{lem:sc_preserve_sol_2}
Let $\vx_T$ be a solution vector of $\sc(\mA, T) \vx_T=\vb$. Then there exists an extension $\vx$ of $\vx_T$ such that $\mA \vx=\left[\begin{array}{c}\vb\\ \vzero \end{array}\right]$. 
\end{lemma}

\begin{proof}
   By taking $\vx_S = -\mA_{S, S}^{-1} \mA_{S, T} \vx_T$ and extending $\vx_T$ to 
   $\vx = 
   \left[ 
   \begin{array}{c}
      \vx_T    \\
      \vx_S   
   \end{array}
    \right]$, we have 
    \begin{align*}
        \left[
        \begin{array}{cc}
           \mA_{T, T}  & \mA_{T, S} \\
           \mA_{S, T}  & \mA_{S, S} \\
        \end{array}
        \right]
        \left[
        \begin{array}{c}
             \vx_T  \\
             \vx_S  \\
        \end{array}
        \right]
        = 
        \left[
        \begin{array}{c}
             \mA_{T,T} \vx_T + \mA_{T,S} (-\mA_{S, S}^{-1} \mA_{S, T} \vx_T)  \\
             \mA_{S,T} \vx_T - \mA_{S,S} \mA_{S,S}^{-1} \mA_{S,T} \vx_T  \\
        \end{array}
        \right]
        = 
        \left[
        \begin{array}{c}
            \sc(\mA, T) \vx_T \\
            \vzero
        \end{array}
        \right]
        =
        \left[
        \begin{array}{c}
            \vb \\
            \vzero
        \end{array}
        \right]
    \end{align*}
\end{proof}

A proof for the following Lemma can be found in \cite{OSTROWSKI1971}. 
\begin{lemma}
\label{lem:commutative}
Let $\mL\in \mathbb{Z}^{n\times n}_p$ be a Laplacian matrix. Let $T_1\subseteq T_2$ be two subsets of $[n]$. Then $\sc(\mL, T_1)=\sc(\sc(\mL, T_2), T_1)$.
\end{lemma}

The following lemma enables us to replace a subgraph structure in the Laplacian matrix while keeping the Schur complement unchanged, which preserves the solution space. For simplicity of notation, we may add two matrices whose dimensions do not match. In such additions, we assume both matrices are in $\mathbb{Z}_p^{[n]\times[n]}$ by padding zeros.

\begin{lemma}
\label{lem:subgraphreplace}
Let $C,X,F \in [n]$ be three disjoint subset of $[n]$ such that $C\cup X\cup F=[n]$. Let $\mL_1 \in \mathbb{Z}_p^{(C\cup X) \times (C\cup X)}$ and $\mL_2 \in \mathbb{Z}_p^{(C\cup F) \times (C\cup F)}$ be two Laplacian matrices. Let $\mL \in \mathbb{Z}_p^{(C\cup F\cup X) \times (C\cup F\cup X)}$ be the sum of $\mL_1$ and $\mL_2$. If $\sc(\mL_2,C)$ exits, we have $\sc(\mL,C\cup X)=\mL_1+\sc(\mL_2,C)$, where $\sc(\mL_2,C)$ is considered as a matrix in $\mathbb{Z}_p^{(C\cup X) \times (C\cup X)}$ by padding zeroes.
\end{lemma}
\begin{proof}
We know $\mL$ is a Laplacian matrix so $\sc(\mL, C \cup X)$ is well-defined. We can verify the equation by the definition of Schur complement. The RHS can be written as 
\[
\mL_1+(\mL_2)_{C, C}-(\mL_2)_{C, F}(\mL_2)_{F,F}^{-1}(\mL_2)_{F, C}.
\]
By the construction of $\mL$, we have
\[
\mL_{C\cup X, F}=\left[\begin{array}{c}\mL_{C, F}\\\mL_{X, F} \end{array}\right]=\left[\begin{array}{c}(\mL_2)_{C, F}\\ \vzero \end{array}\right],
\]
\[
\mL_{F,F}=(\mL_2)_{F, F}, 
\] and 
\[
\mL_{C\cup X, C\cup X}=\mL_1+(\mL_2)_{C, C}.
\]
Thus,
\begin{align*}
    &\sc(\mL,C\cup X)\\
    =&\mL_{C\cup X, C\cup X}-\mL_{C\cup X, F}\mL_{F,F}^{-1}\mL_{F, C\cup X}\\
    =&\mL_1+(\mL_2)_{C, C}+\left[\begin{array}{c}(\mL_2)_{C, F}\\ \vzero \end{array}\right](\mL_2)_{F,F}^{-1}\left[\begin{array}{cc}(\mL_2)_{F, C} & \vzero \end{array}\right]\\
    =&\mL_1+\sc(\mL_2,C).
\end{align*}
\end{proof}



\subsection{Resistance Circuits Over $\mathbb{Z}_p$}
\label{subsec:circuits}
\begin{definition}
\label{def:circuit}
    A Laplacian matrix $\mL\in \mathbb{Z}^{n\times n}_p$ is called a circuit with weight $r$ (or equivalently resistance $r^{-1}$) if  $\sc(\mL, \{1, 2\})=\vchi_{1, 2}^\top r \vchi_{1, 2}$. 
    We call $\left[\begin{array}{cc} 1 & -1 \\ -1 & 1\end{array}\right]$ the unit circuit.
\end{definition}

We will say the resistance $r$ of a circuit $\mL$ is equivalent to $w$ for any $w\in \mathbb{Z}$ such that $r\equiv w\pmod p$. 

Our construction depends on the following physical rule of connecting circuits in parallel or in series.
It is forklore for $\mathbb{R}$. Here we prove it again for $\mathbb{Z}_p$ for completeness.
\begin{lemma}
\label{lem:pa_and_se}
Let $\mL \in \mathbb{Z}^{n\times n}_p$ and $\mU \in \mathbb{Z}^{m\times m}_p$ be two circuits with resistance $a$ and $b$, respectively. Then 
\begin{enumerate}
    \item There exists a circuit $\parallel(\mL, \mU)\in \mathbb{Z}^{(n+m-1)\times (n+m-1)}_p$ such that 
    \begin{enumerate}
        \item The resistance of $\parallel(\mL, \mU)$ is $\frac{ab}{a+b}$,
        \item $\nnz(\parallel(\mL, \mU))\le \nnz(\mL)+\nnz(\mU)$.
    \end{enumerate}\label{itm:parallel}
    \item There exists a circuit $\series(\mL, \mU)\in \mathbb{Z}^{(n+m-2)\times (n+m-2)}_p$ such that 
    \begin{enumerate}
        \item The resistance of $\series(\mL, \mU)$ is $a+b$,
        \item $\nnz(\series(\mL, \mU))\le \nnz(\mL)+\nnz(\mU)$,
        \item $\Delta_{\series(\mL, \mU)} \le \Delta_{\mL}+\Delta_{\mU}$,
        \item $\max(\deg_{\series(\mL, \mU)}^{\vw}(1), \deg_{\series(\mL, \mU)}^{\vw}(2)) \le \max(\Delta_{\mL},\Delta_{\mU})$.
    \end{enumerate}\label{itm:series}
\end{enumerate}
\end{lemma}
\begin{proof}

Proof for Item~\ref{itm:series}: Define 
\begin{align*}
\series(\mL, \mU)\defeq \left[
\begin{array}{ccccc}
\mL_{1, 1} & 0 &  \mL_{1, 2}&\mL_{1, 3:n}& \vzero \\
0&\mU_{1, 1}&\mU_{1, 2}&\vzero & \mU_{1, 3:m}\\
\mL_{2, 1}&\mU_{2, 1} & \mL_{2, 2}+\mU_{2, 2} & \mL_{2, 3:n} &\mU_{2, 3:m} \\
\mL_{3:n, 1}& \vzero &  \mL_{3:n, 2} & \mL_{3:n, 3:n} & \vzero\\
\vzero& \mU_{3:m, 1} & \mU_{3:m, 2} & \vzero & \mU_{3:m, 3:m} \\
\end{array}\right].
\end{align*}
$\nnz(\series(\mL, \mU))\le \nnz(\mL)+\nnz(\mU)$ is satisfied
because $\nnz(\mA)+\nnz(\mB)\ge \nnz(\mA+\mB)$ for any matrices $\mA, \mB$ over $\mathbb{Z}_p$. 

By the definition of $\deg$ (Definition~\ref{def:deg}) and by construction, the degree of a vertex in $\series(\mL, \mU)$ is either equal to the degree of some vertex in $\mL$ or $\mU$, or equal to the sum of two vertices from $\mL$ and $\mU$ respectively. Thus,  $\Delta_{\series(\mL, \mU)} \le \Delta_{\mL}+\Delta_{\mU}$. For the last property,
$$\max(\deg_{\series(\mL, \mU)}^{\vw}(1), \deg_{\series(\mL, \mU)}^{\vw}(2)) \le \max(\Delta_{\mL},\Delta_{\mU}),$$ we note that $\deg_{\series(\mL, \mU)}^{\vw}(1)=\deg_{\mL}(1)$ and $\deg_{\series(\mL, \mU)}^{\vw}(2)=\deg_{\mU}(1)$. Both $\deg_{\mL}(1)$ and $\deg_{\mU}(1)$ are bounded by $\max(\Delta_{\mL},\Delta_{\mU})$.

Let $T=\{1, 2, 3\}, F=\{4, \ldots, n+m-2\}$.
By definition, we have  
\begin{align*}
    & \quad \sc(\series(\mL, \mU), \{1, 2, 3\})\\
    &= \series(\mL, \mU)_{T, T}-\series(\mL, \mU)_{T, F}
    \series(\mL, \mU)_{F, F}^{-1}
    \series(\mL, \mU)_{F, T}\\
    &=\left[
    \begin{array}{ccc}
    \mL_{1, 1} & 0  & \mL_{1, 2}\\ 
    0 & \mU_{1, 1} & \mU_{1, 2} \\
    \mL_{2, 1} & \mU_{2, 1} & \mL_{2, 2} + \mU_{2, 2}
    \end{array}
    \right] \\
    &-\left[
    \begin{array}{cc}
    \mL_{1, 3:n}& \vzero \\
    \vzero & \mU_{1, 3:m} \\
    \mL_{2, 3:n} &\mU_{2, 3:m}
    \end{array}
    \right]
    \left[
    \begin{array}{cc}
    \mL_{3:n, 3:n} & \vzero\\
    \vzero & \mU_{3:m, 3:m}
    \end{array}
    \right]^{-1}
    \left[
    \begin{array}{ccc}
    \mL_{3:n, 1}& 0 & \mL_{3:n, 2}\\
    0& \mU_{3:m, 1}&\mU_{3:m, 2}
    \end{array}\right]\\
    &=\left[
    \begin{array}{ccc}
    \mL_{1, 1} & 0  & \mL_{1, 2}\\
    0 & \mU_{1, 1} & \mU_{1, 2} \\
    \mL_{2, 1} & \mU_{2, 1} & \mL_{2, 2} + \mU_{2, 2}
    \end{array}
    \right] \\
    &-\left[
    \begin{array}{ccc}
    \mL_{1, 3:n} \mL_{3:n, 3:n}^{-1} \mL_{3:n, 1} & \vzero & \mL_{1, 3:n}\mL_{3:n, 3: n}^{-1} \mL_{3:n, 2}\\
    \vzero & \mU_{1, 3:m}\mU_{3:m, 3:m}^{-1} \mU_{3:m, 1} & \mU_{1, 3:m}\mU_{3:m, 3:m}^{-1} \mU_{3:m, 2}\\
    \mL_{2, 3:n}\mL_{3:n, 3:n}^{-1} \mL_{3:n, 1} & \mU_{2, 3:m}\mU_{3:m, 3:m}^{-1} \mU_{3:m, 1} & \left(\makecell{\mL_{2, 3:n}\mL_{3:n, 3:n}^{-1} \mL_{3:n, 2}\\+\\\mU_{2, 3:m}\mU_{3: m, 3:m}^{-1} \mU_{3:m, 2}}\right)
    \end{array}\right]\\
    &=\left[\begin{array}{ccc}\sc(\mL, \{1, 2\})_{1, 1} & 0 & \sc(\mL, \{1, 2\})_{1, 2}\\ 0 &\sc(\mU, \{1, 2\})_{1, 1} & \sc(\mU, \{1, 2\})_{1, 2}\\ \sc(\mL, \{1, 2\})_{2, 1}  & \sc(\mU, \{1, 2\})_{2, 1} & \sc(\mL, \{1, 2\})_{2, 2}+\sc(\mU, \{1, 2\})_{2, 2}\end{array}\right]\\
    &=\left[\begin{array}{ccc}a^{-1} & 0 & -a^{-1}\\ 0 &b^{-1} & -b^{-1}\\ -a^{-1}  & -b^{-1} & a^{-1}+b^{-1}\end{array}\right].
\end{align*}
By Lemma~\ref{lem:commutative}, 
\begin{align*}
    & \quad \sc(\series(\mL, \mU), \{1, 2\})\\
    &=\sc(\sc(\series(\mL, \mU), \{1, 2, 3\}), \{1, 2\})\\
    &=\sc\left(\left[
    \begin{array}{ccc}
    a^{-1} & 0 & -a^{-1}\\
    0 &b^{-1} & -b^{-1}\\
    -a^{-1}  & -b^{-1} & a^{-1}+b^{-1}\end{array}\right]
    , \{1, 2\}\right)\\
    &=\left[\begin{array}{cc}\frac{1}{a+b} & -\frac{1}{a+b}\\ -\frac{1}{a+b} &\frac{1}{a+b}\end{array}\right].
\end{align*}
Proof for Item~\ref{itm:parallel}:
Define 
\begin{align*}
\parallel(\mL, \mU)\defeq \left[
\begin{array}{cccc}
\multicolumn{2}{c}{\multirow{2}{*}{$\mL_{1:2, 1:2}+\mU_{1:2, 1:2}$}}& \mL_{1, 3:n} & \mU_{1, 3:m} \\
\multicolumn{2}{c}{}& \mL_{2, 3:n}& \mU_{2, 3:m} \\
\mL_{3:n, 1} & \mL_{3:n, 2}  & \mL_{3:n, 3:n} & \vzero\\
\mU_{3:m, 1} & \mU_{3:m, 2}  & \vzero & \mU_{3:m, 3:m}
\end{array}\right].
\end{align*}
 $\nnz(\parallel(\mL, \mU))\le \nnz(\mL)+\nnz(\mU)$ is satisfied for the same reason as in Item~\ref{itm:series}. Let $T=\{1, 2\}, F=\{3, \ldots, n+m-2\}$. By definition, we have 
\begin{align*}
    & \quad \sc(\parallel(\mL, \mU), \{1, 2\})\\
    &= (\mL+\mU)_{T, T}-\parallel(\mL, \mU)_{T, F}
    \parallel(\mL, \mU)_{F, F}^{-1}
    \parallel(\mL, \mU)_{F, T}\\
    &=(\mL+\mU)_{T, T}-\left[\begin{array}{cc}\mL_{1, 3:n} & \mU_{1, 3:m}\\ \mL_{2, 3: n}&\mU_{2, 3:m}\end{array}\right]
    \left[\begin{array}{cc}\mL_{3:n, 3:n} & \vzero \\ \vzero & \mU_{3:m, 3: m}\end{array}\right]^{-1}
    \left[\begin{array}{cc}\mL_{3:n, 1} & \mL_{3:n, 2}\\\mU_{3:m, 1} & \mU_{3:m, 2}\end{array}\right]\\
    &=(\mL+\mU)_{T, T}-\left(\mL_{1:2, 3:n} \mL_{3:n, 3:n}^{-1} \mL_{3:n, 1:2} + \mU_{1:2, 3:m} \mU_{3:m, 3:m}^{-1} \mU_{3:m, 1:2}\right)\\
    &=\sc(\mL, \{1, 2\})+\sc(\mU, \{1, 2\})\\
    &=\vchi_{1, 2}^\top (a^{-1}+b^{-1}) \vchi_{1, 2} \qedhere
\end{align*}
\end{proof}

Now we prove the lemma that enables us to do some transformations of a linear system without changing the solution space. On a high level, the lemma states that for each Laplacian $\mL$, we can replace one edge $e$ whose weight is $\vw_e$ by one circuit
with the same weight. This lemma empowers us to reduce the edge weight by replacing edges with high weights by circuits with low edge weights. 
\begin{lemma}
\label{lem:replace_edge}
Given any Laplacian system $\mL \in \mathbb{Z}_p^{n\times n}$, a non-zero off-diagonal entry $\mL_{i_0, j_0}$, and a circuit $\mR$ with weight $-\mL_{i_0, j_0}$ , we can construct a Laplacian system $\mU \in \mathbb{Z}_p^{(n+m)\times (n+m)}$ for some $m \ge 0$ such that 

\begin{enumerate}
    \item $\nnz(\mU) \le \nnz(\mL) + \nnz(\mR)$. 
    \label{itm:replaceitem1}
    \item For any non-diagonal entry $\mL_{i, j}$ where $(i, j)\in [n]\times[n]$ except  $(i_0, j_0)$ and $(j_0, i_0)$, $\mU_{i, j}=\mL_{i, j}$.
    \label{itm:replaceitem2}
    \item For any linear system $\mL \vx = \vb$, $\left\{\vx\mid \mL \vx = \vb\right\}=\left\{\vy_{[n]}\mid\mU \vy = \left[\begin{array}{c}\vb \\ \vzero \end{array}\right]\right\}$.
    \label{itm:replaceitem3}
\end{enumerate}
\end{lemma}
\begin{proof}
    Without loss of generality, we may assume $i_0=1$ and $j_0=2$. Suppose $\mR$ is a $k$ by $k$ matrix. Define 
    \[
    \mU=\left[\begin{array}{cccc}
    \mL_{1, 1}+\mL_{1, 2} +\mR_{1, 1} & \mR_{1, 2} & \mL_{1, 3:n} & \mR_{1, 3:k} \\
    \mR_{2, 1} & \mL_{2, 2}+\mL_{2, 1}+\mR_{2, 2} & \mL_{2, 3:n} & \mR_{2, 3:k} \\
    \mL_{3:n, 1} & \mL_{3:n, 2} & \mL_{3:n, 3:n} & \vzero\\
    \mR_{3:k, 1} & \mR_{3:k, 2}  & \vzero & \mR_{3:k, 3:k}\\
    \end{array}\right].
    \] 
    To prove $\nnz(\mU) \le \nnz(\mL) + \nnz(\mR)$, we note that by construction, \begin{align*}
        &\nnz(\mU)=\nnz(\mL)+\nnz(\mR)\\-&[\mL_{1,1}\neq 0]-[\mR_{1,1}\neq 0]-[\mL_{1,2}\neq 0]+[\mU_{1,1}\neq 0]\\-&[\mL_{2,2}\neq 0]-[\mR_{2,2}\neq 0]-[\mL_{2,1}\neq 0]+[\mU_{2,2}\neq 0].
    \end{align*}  If $\mU_{1,1}$ is non-zero, at least one of $\mL_{1,1}, \mR_{1,1}$ and $\mL_{1,2}$ must be non-zero. Similar argument applies to $\mU_{2,2}$. Thus, $\nnz(\mU) \le \nnz(\mL) + \nnz(\mR)$. Item~\ref{itm:replaceitem2} follows from the definition directly.
    Let $m=k-2, S = \{n+1, \ldots, n+m\}$. We have 
    \begin{align*}
        \quad \sc(\mU, [n])
        &=\mU_{[n], [n]} - \mU_{1:2, S} \mU_{S, S}^{-1} \mU_{S, 1:2}\\
        &=\left(\mU_{[n], [n]}-\mR_{1:2, 1:2}\right) 
        + \mR_{1:2, 1:2} -\mU_{1:2, S} \mU_{S, S}^{-1} \mU_{S, 1:2}\\
        &=\left(\mU_{[n], [n]}-\mR_{1:2, 1:2}\right) +\mR_{1:2, 1:2} 
        -\mR_{1:2, 3:k} \mR_{3:k, 3:k}^{-1} \mR_{3:k, 1:2}\\
        &= \left(\mU_{[n], [n]}-\mR_{1:2, 1:2}\right)+\sc(\mR, \{1, 2\})\\
        &= \left(\mL_{[n], [n]}+\vchi_{1, 2}^\top \mL_{1, 2} \vchi_{1, 2}\right)+\sc(\mR, \{1, 2\})\\
        &=\mL.
    \end{align*}
    Since $\sc(\mU, [n])=\mL$, Item~\ref{itm:replaceitem3} follows from Lemma~\ref{lem:sc_preserve_sol_1} and Lemma~\ref{lem:sc_preserve_sol_2}.
\end{proof}

The following lemma is a simplified version of the known ``Star-mesh Transform", which was proven in \cite{V73}. 

\begin{lemma}[Star-mesh Transform]
\label{lemma:starTclique}
A Laplacian $\mL \in \mathbb{Z}_p^{n+1 \times n+1}$ whose corresponding graph has vertex set $V=\{v_0,v_1,\dots,v_n\}$ and edge set $E=\{(v_0,v_i,w_i)\} (1 \leq i \leq n)$ is equivalent to a Laplacian $\hat{\mL} \in \mathbb{Z}_p^{n \times n}$ whose corresponding graph is a clique formed by vertices $V=\{v_1,\dots,v_n\}$ with edge set $E=\{(v_i,v_j,\frac{w_iw_j}{\sum_{i=1}^{k} w_i})\}$ ($i\neq j, 1 \leq i,j \leq n)$. The equivalence implies $\sc(\mL,V-v_0)=\hat{\mL}$. We also call $v_0$ as the center of the star.
\end{lemma}

\subsection{Unit-Weight Circuit Gadget}
\label{sec:gadget}



In this section, we provide a construction such that for any $p$ and any resistance,
we can efficiently construct a unit-weight circuit
of $\bound$ non-zero entries whose $s$-$t$ effective resistance matches
that value, thus proving Lemma~\ref{lem:resistance_construction}.

\lemRConstruct*

\subsubsection{Compositions of Series/Parallel Circuits}
\label{subsec:sp}

To construct any resistance value, we first need the following three lemmas which construct any resistance of the form $ab^{-1}$ with $O(a+b)$ non-zero entries. 
\begin{lemma}
\label{lem:z}
Given two circuits with resistances $1$ and $z$ and $\nnz$ $n$ and $m$ respectively, we can construct the following circuits:
\begin{itemize}
    \item A circuit with resistance $1+z$ and $\nnz$ at most $n+ m$.
    \item A circuit with resistance $z(z+1)^{-1}$ and $\nnz$ at most $n+m+12$.
\end{itemize}
\end{lemma}
\begin{figure}
    \centering
	\begin{tikzpicture}[auto, node distance=3cm, every loop/.style={},
                    main node/.style={circle,draw}]

  \node[main node] (1) at (0, 0) {1};
  \node[main node] (2) at (2, 0) {2};
  \node[main node] (11) at (6, 0) {1};
  \node[main node] (22) at (9, 0) {2};

  \node[main node] (3) at (7.5, 1) {$u$};
  \node[main node] (4) at (7.5, -1) {$v$};
  
  \draw[thick, ->] (3, 0) -- (5, 0);

  \draw  (1) edge node [above] {$1$} (2);
  \draw (11) edge node[above] {$1$} (3);
  \draw (22) edge node[above] {$1$} (3);
  \draw (11) edge node[above] {$1$} (4);
  \draw (22) edge node[above] {$1$} (4);
\end{tikzpicture}
    \caption{Edge Replacement to avoid parallel edges.}
    \label{fig:edge-replace}
\end{figure}
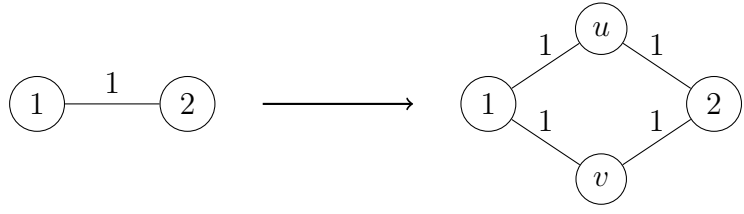

\begin{proof}
Let these two circuits be $\mL$ and $\mU$ such that the effective resistance (between vertex 1 and 2) of $\mL$ is 1 and the effective resistance (between vertex 1 and 2) of $\mU$ is $z$. To avoid parallel edges, for each of $\mL$ and $\mU$, we first replace the edge between $1$ and $2$ (if exists) by $4$ edges $\{1, u\}, \{1, v\}, \{2, u\}, \{2, v\}$ where $u$ and $v$ are two new vertices. The new circuit has the same resistance as the original one by calculation. 
Thus, we may assume there is no edge between $1$ and $2$ in $\mL$ or $\mU$.

By Lemma~\ref{lem:pa_and_se}, $\series(\mL, \mU)$ has resistance $1+z$ and $\nnz$ $n+m$. $\parallel(\mL, \mU)$ has resistance $z(z+1)^{-1}$ and $\nnz$ $n+m$. $\parallel(\mL, \mU)$ has no parallel edges by our assumption.

Notice the circuit of the rhombus gadget (shown in Figure~\ref{fig:edge-replace}) is just given by 
\begin{equation*}
\mR = \begin{bmatrix}2 & 0 & -1 & -1 \\ 0 & 2 & -1 & -1 \\ -1 & -1 & 2 & 0 \\ -1 & -1 & 0 & 2 \end{bmatrix}
\end{equation*}

The rows represent vertex $1,2,u,v$ respectively. By Lemma~\ref{lem:replace_edge}, replacing an edge between vertex 1 and 2 with circuit $\mR$ would increase the number of non-zero entries by 6. Therefore, in the worst case, if we need to replece the edge between vertex 1 and 2 in $\mL$ and in $\mU$, we will increase the $\nnz$ by 12.


\end{proof}
\begin{lemma}
\label{lem:onestep}
Given two circuits with resistances $1$ and $ab^{-1}$ and $\nnz$ $n$ and $m$, we can construct the following circuits:
\begin{itemize}
    \item A circuit of resistance $(a+b)b^{-1}$  and $\nnz$ at most $n+m$.
    \item A circuit of resistance $a(a+b)^{-1}$ and $\nnz$ at most $n+m + 12$.
\end{itemize}
\end{lemma}
\begin{proof}
By Lemma~\ref{lem:z} with $z=ab^{-1}$.
\end{proof}
\begin{lemma}
\label{lemma:sqrtconstruction}
For any $a, b\in \mathbb{Z}$ such that $1\le a, b\le M$, we can construct a circuit of $\nnz\ O(M)$ which only consists of \emph{unit-weight} edges, and its resistance is equivalent to $ab^{-1} \pmod p$ .
\end{lemma}
\begin{proof}
We construct the value $ab^{-1} \pmod p$ recursively. If $a>b$, we can construct $ab^{-1} \pmod p$ from $(a-b)b^{-1} \pmod p$ and another unit circuit by Lemma~\ref{lem:onestep}. If $a<b$, we can construct $ab^{-1} \pmod p$ from $a(b-a)^{-1} \pmod p$ and another unit circuit by Lemma~\ref{lem:onestep}. If $a=b$, we can construct $ab^{-1}\equiv 1 \pmod p$ by just one unit circuit. Since $a, b$ are always positive integers and their initial values are no more than $M$ in $\mathbb{Z}$, this process ends in $O(M)$ steps. Because unit circuit has $\nnz = O(1)$ and by Lemma~\ref{lem:onestep} the $\nnz$ of constructed circuit is $O(M)$.
\end{proof}
Then we are able to construct resistance values near $\frac{pi}{j}$ (one of the points that exactly divide $[1, p-1]$ into $j$ equal parts).
\begin{lemma}
\label{lem:fracK}
Let $k\in \mathbb{Z}$ be an integer in $[1, p-1]$. For any $1\le i\le j\le k$, there exists a circuit $\mL$ with $\nnz(\mL) = O(k)$ which only consists of \emph{unit-weight} edges whose resistance $s$ is equivalent to some integer $w\in [\frac{pi}{j}, \frac{pi}{j}+1]$.
\end{lemma}
\begin{proof}
Let $w = \lceil pi/j \rceil \in \mathbb{Z}$. It follows that $w \in [\frac{pi}{j}, \frac{pi}{j}+1]$. Now $w = \frac{\lceil pi/j \rceil j}{j}$ as integers, and thus $w \equiv (\lceil pi/j \rceil j)j^{-1} \pmod p$. Now $pi \leq \lceil pi/j \rceil j \leq pi+j$, so $(\lceil pi/j \rceil j) \pmod p \leq j$. By Lemma~\ref{lemma:sqrtconstruction}, we can construct a circuit with resistance $w \pmod p$ and $\nnz$ $O(j) = O(k)$.
\end{proof}

Since $-x\equiv p-x \pmod p$, we can also construct resistance values near $-\frac{pi}{j}$ for some $0\le i\le j$.
\begin{corollary}
\label{cor:fracK}
Let $k\in \mathbb{Z}$ be an integer in $[1, p-1]$. For any $1\le j\le k$ and $-j\le i\le j$, there exists a circuit $\mL$ with $\nnz(\mL) = O(k)$ which only consists of \emph{unit-weight} edges whose resistance is equivalent to some integer $w\in [\frac{pi}{j}, \frac{pi}{j}+1]$.
\end{corollary}
\begin{proof}
When $i$ is $0$, we let $\mL$ be the unit circuit. When $i$ is positive, this is equivalent to Lemma~\ref{lem:fracK}. When $i$ is negative, we first construct a circuit $\mL$ with $\nnz(\mL) = O(k)$ and resistance equivalent to $\bar{w}$ for some $\bar{w}\in [p(j+i)/j, p(j+i)/j+1]$ by Lemma~\ref{lem:fracK}. Then the resistance of $\mL$ is also equivalent to $w$ for $w=\bar{w}-p\in [pi/j, pi/j+1]$ because $w\equiv \bar{w}-p \pmod p$. In fact, by the construction in Lemma~\ref{lem:fracK}, $\bar{w} = \lceil \frac{p(j+i)}{j} \rceil = p + \lceil \frac{pi}{j} \rceil $, so in this case we also have $w = \lceil \frac{pi}{j} \rceil$.
\end{proof}

\subsubsection{Composing Multiple SP Circuits}
\label{subsec:composition}

By adding the resistances given by Corollary~\ref{cor:fracK} for several different prime numbers $j_1, \ldots, j_t$, we can construct resistances near $\frac{pi}{j}$ for $j=\prod_{s=1}^t j_s$ and any $-j\le i\le j$.
\begin{lemma}
\label{lem:fracK2}
Let $k\in \mathbb{Z}$ be an integer in $[1, p-1]$. Let $j_1< \ldots< j_t$ be $t$ different primes in $[1, k]$. For any $-j\le i\le j$ where $j\defeq \prod_{s=1}^t j_s$, there exists an integer $w$ and an electric circuit $\mL$ with unit-weight edges, $\nnz(\mL) = O(kt)$, $\Delta_{\mL} = O(k)$ and resistance equivalent to some integer $w\in [\frac{pi}{j}, \frac{pi}{j}+t]$.
\end{lemma}
For proving Lemma~\ref{lem:fracK2}, we need the following lemma that is similar to the reverse of the Chinese Remainder Theorem.
\begin{lemma}
\label{lem:revCRT}
Let $j_1, \ldots, j_t$ be $t$ pairwise coprime positive integers. Let $j$ be the product of $j_1,\ldots, j_t$. For any integer $i$ in $[-j, j]$, we can write $\frac{j}{i}$ as the sum 
$$\sum_{s=1}^t \frac{a_s}{j_s}$$
for some $a_1,\ldots, a_t$ satisfying $\forall 1\le s\le t, a_s\in [-j_s, j_s]$.
\end{lemma}
\begin{proof}
We first write $\frac{i}{j}$ as $\sum_{s=1}^t \frac{a_s}{j_s}$ for some $a_1,\ldots, a_t$ satisfying $\forall 1\le s\le t, a_s\in [-j_s, j_s]$. 

$a_1,\ldots, a_t$ can be found recursively. We first find an $a_t$ such that 
\begin{enumerate}
    \item \label{itm:cdt1} $\frac{i}{j}-\frac{a_t}{j_t}=\frac{\bar{i}}{j/j_t}$ for some integer $\bar{i}$, and 
    \item \label{itm:cdt2} $-j/j_t\le \bar{i} \le j/j_t$.
\end{enumerate} We let 
\[
a_t=\left(i(j/j_t)^{-1}\right)\pmod {j_t}
\] where $(j/j_t)^{-1}$ is the inverse of $j/j_t$ modulo ${j_t}$. The inverse exists since $j/j_t$ is the product of primes less than $j_t$. Then Condition~\ref{itm:cdt1} is satisfied. Because 
\[-1\le\frac{i}{j}\le 1\text{ and }-1\le -\frac{a_t}{j_t}\le 0,\]
we have \[\frac{\bar{i}}{j/j_t}\le 1.\] 
If 
\[\frac{i}{j}-\frac{a_t}{j_t}\ge -1,\]
Condition~\ref{itm:cdt2} is also satisfied. 
Otherwise, 
\[-2\le \frac{i}{j}-\frac{a_t}{j_t}< -1.\] 
In this case, we decrease $a_t$ by $j_t$ and add $\bar{i}$ by $j/j_t$ to fulfill Condition~\ref{itm:cdt2}. 
Condition~\ref{itm:cdt1} still holds because both sides of the equation increase by $1$.

$a_{t-1},\ldots, a_1$ can be constructed recursively by \[\frac{\bar{i}}{j/j_t}=\sum_{s=1}^{t-1} \frac{a_s}{j_s}.\]
\end{proof}
\begin{proof}[Proof for Lemma~\ref{lem:fracK2}]

After finding $a_1,\ldots, a_t$ by Lemma~\ref{lem:revCRT}, we construct $t$ $\nnz$-$O(k)$ circuits $\mL_1,\ldots, \mL_t$ such that $\mL_s$ has resistance equivalent to some integer in 
\[[pa_s/j_s, pa_s/j_s+1]\] by Corollary~\ref{cor:fracK}. Then 
\[\series(\mL_1, \series(\mL_2, \ldots, \series(\mL_{t-1}, \mL_t)\ldots))\] has $\nnz$ $O(tk)$ and resistance $r$ equivalent to some integer $w$ between $[\frac{pi}{j}, \frac{pi}{j}+t]$. By Corollary~\ref{cor:fracK}, each $\mL_i$ has $\nnz$ $O(k)$. Thus, their maximum degrees are bounded by $O(k)$. By Lemma~\ref{lem:pa_and_se}, we prove $\Delta_{(\series(\mL_1, \series(\mL_2, \ldots, \series(\mL_{t-1}, \mL_t)\ldots)))}=O(k)$ by induction. The induction hypothesis is
\begin{itemize}
    \item $\Delta_{(\series(\mL_i, \series(\mL_{i+1}, \ldots, \series(\mL_{t-1}, \mL_t)\ldots)))}=2ck$,
    \item $\deg_{\series(\mL_i, \series(\mL_{i+1}, \ldots, \series(\mL_{t-1}, \mL_t)\ldots))}(1)=ck$, and 
    \item $\deg_{\series(\mL_i, \series(\mL_{i+1}, \ldots, \series(\mL_{t-1}, \mL_t)\ldots))}(2)=ck$.
\end{itemize} We first choose the constant $c$ so that $\Delta_{\mL_i}$ is bounded by $ck$ for each $i$. Then the induction hypothesis holds for $i=t-1$ which we take as the base case. The induction step follows from the last two properties of $\series$ in Lemma~\ref{lem:pa_and_se}.
\end{proof}
\subsubsection{Composing Circuits of the Desired Resistance}
Lemma~\ref{lem:fracK2} can construct any resistance between $[1, p-1]$ up to a small additive error. We prove Lemma~\ref{lem:resistance_construction} by fixing the error.
\begin{proof}[Proof for Lemma~\ref{lem:resistance_construction}]
    Let $t$ be the smallest integer such that $p\le \prod_{s=1}^t j_s$ where $j_s$ is the $s$-th smallest prime. Let $j=\prod_{s=1}^t j_s$ and $k=j_t$. Using Lemma~\ref{lem:fracK2}, for any $i \in [1, j]$, we can construct an nnz-$O(tk)$ circuit $\mL_i$ with resistance equivalent to some integer $x_i \in [\frac{pi}{j}, \frac{pi}{j}+t]$. Since $j\ge p$, we know $|x_{i+1}-x_i|\le 1+2t$. Let 
    \[X=\{x_0\defeq 0, x_1, \ldots, x_{j}, x_{j+1}\defeq p\}.\]
    Since $|x_{i+1}-x_i|=O(t)$ for any $1\le i<j$, $x_1=O(t)$ and $|p-x_{j}|=O(t)$, any number $z$ in $[1, p-1]$ must be between $x_i$ and $x_{i+1}$ for some $0\le i\le j$. Thus, $r$ can be written as the sum of one $x_i\in X$ and another non-negative integer $y=O(t)$, i.e.~$z=\min(x_i, x_{i+1})+y$. Let $\mU_y$ be a circuit with resistance equivalent to $y$ and $\nnz$ $O(y)$, e.g.~$\mU_y$ can be a path of length $y$. If $i=0$, $\mU_y$ is the desired circuit. If $i>0$, by Lemma~\ref{lem:pa_and_se}, $\series(\mL_i, \mU_y)$ is a circuit with resistance equivalent to $r$ and $\nnz$ $O(kt)$. By the Prime Number Theorem \cite{P1896, H1896}, we can show that $t \le \log p/\log\log p$ and $k=j_t\le \log p$. Thus, $O(kt)=\bound$.
    Finally, we can show our construction of circuits with given resistance modulo $p$ can be implemented efficiently by the following lemma. This completes the proof.
\end{proof}

\begin{lemma}
\label{lem:complexity}
For any resistance $r \in [p - 1]$, the construction of the circuit in Lemma~\ref{lem:resistance_construction} with 
resistance $r$ can be done in $O(\log^3 p)$ time.
\end{lemma}

\begin{proof}
Given $\mL \in \mathbb{Z}^{n \times n}_p$ and $\mU \in \mathbb{Z}^{m \times m}_p$, the construction of $\parallel(\mL,\mU)$ and $\series(\mL,\mU)$ in Lemma~\ref{lem:pa_and_se} can be done in $O(\nnz(\mL) + \nnz(\mU))$ time because $\parallel(\mL,\mU)$ and $\series(\mL,\mU)$ can be obtained by re-arranging the non-zero entries of $\mL$ and $\mU$ plus performing a constant number of additions. As a result, the construction in Lemma~\ref{lem:z} can be done in $O(n+m)$ time given two circuits with resistances 1 and $z$ and $\nnz$ $n$ and $m$. The construction in Lemma~\ref{lem:onestep} can be done in $O(n+m)$ time given two circuits with resistances 1 and $ab^{-1}$ and $\nnz$ $n$ and $m$. 

In Lemma~\ref{lemma:sqrtconstruction}, suppose we are given $a,b \in [M]$, we can construct a circuit with resistance $ab^{-1} \mod p$ in at most $M$ steps and each step involves connecting a unit circuit with the circuit constructed by recursing on $(a-b)b^{-1} \pmod p$ or $a(b-a)^{-1} \pmod p$. Therefore, the construction in Lemma~\ref{lemma:sqrtconstruction} can be completed in $O(M^2)$ time. As a result, given $k$ and $i, j$ such that $1 \le i \le j \le  k$, we can construct a circuit $\mL$ with resistance equivalent to some integer $w \in [\frac{pi}{j}, \frac{pi}{j}+1]$ in $O(k^2)$ time. By Corollary~\ref{cor:fracK}, given $k \in [p-1]$, for any $i,j \in [k]$ and $-j \leq i \leq j$, we can construct a circuit $\mL$ with resistance equivalent to some integer $w \in [\frac{pi}{j}, \frac{pi}{j}+1]$ in $O(k^2)$ time.

Now we may analyze the runtime for the construction done by Lemma~\ref{lem:fracK2}. We may find the $a_t$s in $O(t)$ time and construct circuits $\mL_1,\ldots,\mL_t$ where $s$-th circuit $\mL_s$ has resistance equivalent to some integer in $[pa_s/j_s,pa_s/j_s+1]$ in $O(tk^2)$ time and connect them in serial in $O(t^2k)$ time. This results in a complexity of $O(tk^2) + O(t^2k) = O(tk(k+t))$.

Finally, we examine the construction in Lemma~\ref{lem:resistance_construction}. Finding the smallest integer $t$ such that $p \leq \prod_{s=1}^t j_s$ can be done in $O(\log^2 p)$ time since $j_t \leq \log p$. $x_i = \sum_{s=1}^t \lceil \frac{pa_s}{j_s} \rceil$  can be computed exactly in $O(t)$ time for a given $i$. Since $x_0 = 0 < z$, $x_{j+1} := p > z$, we can use divide-and-conquer to find an $i$ such that $z$ is between $x_i$ and $x_{i+1}$ in $O(\log j) = O(\log (kp))$ steps, so we can find the $i$ and write $z$ as $z = \min(x_i,x_{i+1})+y$ in $O(\log(kp)t) = O(\log^2 p)$ time. Since $\mU_y$ is just a path of length $y$, we can construct $\mU_y$ in $O(y^2) = O(t^2)$ time, and by our analysis for Lemma~\ref{lem:fracK2}, $\mL_i$ can be constructed in $O(tk(t+k))$ time. Next, connect $\mL_i$ and $\mU_y$ in serial should be done in $O(tk)$ time since $\nnz(\mL_i) = O(tk)$ and $\nnz(\mU_y) = O(t)$. As a result, given the bounds $t \leq \log p/\log \log p$ and $k \leq \log p$, we can construct the final circuit $\mU$ in $O(\log^3p)$ time.

\end{proof}

\section{Hardness of Solving Laplacian System over $\mathbb{Z}_p$}
\label{sec:hardnessL}
\subsection{Reduction to General Laplacian System}
\label{subsec:reductiongeneralweight}

In this section we show that the class of general matrices $\GS_{\mathbb{Z}_p}$ is efficiently and decisional $f$-reducible to the class of Laplacian matrices $\LS_{\mathbb{Z}_p}$ for specifically chosen $f$.

\thmGfL*

\begin{proof}
\ We first prove that given any linear system $(\mA \in \mathbb{Z}_p^{m \times n}, 
\vb \in \mathbb{Z}_p^{m}) \in \GS_{\mathbb{Z}_p}$, we can compute a Laplacian system 
$(\mL \in \mathbb{Z}_p^{2(m+n) \times 2(m+n)}, \vc \in \mathbb{Z}_p^{2(m+n)})$ such 
that 

\begin{enumerate}
    \item $\nnz(\mL) = O(\nnz(\mA))$ 
    \item $\{\vy_{m+1:m+n} - \vy_{2m + n + 1:2m + 2n} \mid \mL \vy = \vc \} = 
    \{ \vx \mid \mA \vx = \vb \}$
\end{enumerate}

Then we will let $\mathcal{A}_{1 \mapsto 2}(\mA, \vb) = (\mL, \vc)$, $\mathcal{A}_{2 \mapsto 1}(\mL, \vc, \vy) = \vy_{m+1:m+n} - \vy_{2m + n + 1:2m + 2n}$ and show their running times.

\paragraph{Construction of $(\mL, \vc)$}

Inspired by \cite{KOSZ13}, we construct $(\mL, \vc)$ as 

\begin{align*}
     \mL \defeq 
     \left[ 
            \begin{array}{cccc}
                \vzero & \mA & \vzero &  -\mA \\
                \mA^T & \vzero & -\mA^T & \vzero \\
                \vzero & -\mA & \vzero & \mA \\
                -\mA^T & \vzero & \mA^T & \vzero
            \end{array}
            \right],\quad
            \vc \defeq  
            \left[
            \begin{array}{c}
            \vb \\
            \vzero \\
            -\vb \\
            \vzero
            \end{array}
            \right]
\end{align*}
It's easy to verify $\mL$ is symmetric and
    \begin{align*}
        \mL \vone = 
        \left[
        \begin{array}{cccc}
        \vzero & \mA & \vzero & -\mA \\
        \mA^T & \vzero & -\mA^T & \vzero \\
        \vzero & -\mA & \vzero & \mA \\
        -\mA^T & \vzero & \mA^T & \vzero \\ 
        \end{array}
        \right]
        \left[
        \begin{array}{c}
            \vone \\
            \vone \\
            \vone \\
            \vone \\
        \end{array}
        \right]
        = 
        \left[
        \begin{array}{c}
            \mA - \mA \\
            \mA^T - \mA^T \\
            \mA - \mA \\
            \mA^T - \mA^T \\
        \end{array}
        \right]
        = \bm{0}
    \end{align*}
    which implies $\vone$ is in the nullspace of $\mL$. 
    According to Definition~\ref{matrix::zp}, $\mL$ is a Laplacian matrix over the finite field $\mathbb{Z}_p$. 
\paragraph{Proof for $\mathcal{S}(\mL, \vc) \le f(\mathcal{S}(\mA, \vb))$}
    By our construction of $\mL$, we have $\nnz(L) = O(\nnz(\mA))$. Besides we have $\mA \in \mathbb{Z}_p^{m \times n}$, thus $\size(L) = O(p \times \nnz(\mA))$. Therefore
    \begin{align*}
        \mathcal{S}(\mL, \vc) \le (O(\nnz(\mA), O(p \times \nnz(\mA))) = f(\mathcal{S}(\mA, \vb)) 
    \end{align*}
Next we show $\{\vy_{m+1:m+n} - \vy_{2m + n + 1:2m + 2n} \mid \mL \vy = \vc \} = 
    \{ \vx \mid \mA \vx = \vb \}$. 
Considering a solution $\vy = (\vy_1^T, \vy_2^T, \vy_3^T, \vy_4^T)^T$ to the linear system $\mL \vy = \vc$ 
, we have
    \begin{align*}
    \left. 
    \begin{array}{r @{{}={}} l}
        \mA \vy_2 - \mA \vy_4 & \vb   \\
        \mA^T \vy_1 - \mA^T \vy_3 & \vzero   \\
        -\mA \vy_2 + \mA \vy_4 & -\vb   \\
        -\mA^T\vy_1 + \mA^T \vy_3 & \vzero   \\
    \end{array}
    \right\}
    \text{ implies } \mA(\vy_2 - \vy_4) = \vb  
    \end{align*}
     Thus 
     \begin{align*}
        \vy_{m+1:m+n} - \vy_{2m + n + 1:2m + 2n} = \vy_2 - \vy_4 \in \{ \vx \mid \mA \vx = \vb   \} 
     \end{align*}
     For every solution $\vx$ to $\mA \vx = \vb  $, we have
     \begin{align*}
        \left[
        \begin{array}{cccc}
        \vzero & \mA & \vzero & -\mA \\
        \mA^T & \vzero & -\mA^T & \vzero \\
        \vzero & -\mA & \vzero & \mA \\
        -\mA^T & \vzero & \mA^T & \vzero \\ 
        \end{array}
        \right]
        \left[
        \begin{array}{c}
            \vzero \\
            \vx \\
            \vzero \\
            \vzero \\
        \end{array}
        \right]
        = 
        \left[
        \begin{array}{c}
            \mA \vx \\
            \vzero \\
            -\mA \vx \\
            \vzero \\
        \end{array}
        \right]
        = 
        \left[
        \begin{array}{c}
            \vb \\
            \vzero \\
            -\vb \\
            \vzero \\
        \end{array}
        \right]
        = \vc  
 \end{align*}
which implies
\begin{align*}
    \vy = \left[
        \begin{array}{c}
            \vzero \\
            \vx \\
            \vzero \\
            \vzero \\
        \end{array}
        \right]
\end{align*}
is a solution to the linear system $\mL \vy = \vc  $. Since $\vy_{m+1:m+n} - \vy_{2m + n + 1:2m + 2n} = \vx - \vzero = \vx$,
\begin{align*}
    \vx \in\{\vy_{m+1:m+n} - \vy_{2m + n + 1:2m + 2n} \mid \mL \vy = \vc \}
\end{align*}
\paragraph{Running Time of $\mathcal{A}_{2 \mapsto 1}$, $\mathcal{A}_{1 \mapsto 2}$}

By our construction of $(\mL, \vc)$, we have $\mathcal{A}_{1 \mapsto 2}$ run in $O(\nnz(\mA))$ time. Since $\vx = \vy_{m+1: m+n} - \vy_{2m + n + 1:2m + 2n}$,
$\mathcal{A}_{2 \mapsto 1}$ runs in $O(n) = O(\nnz(\mA))$ time.

\end{proof}

\begin{corollary}
\label{cor:GFL}
      Let $f(\nnz) = O(\nnz)$, then $\GS_{\mathbb{Z}_p}$ is decisional $f$-reducible to $\LS_{\mathbb{Z}_p}$.
\end{corollary} 
\begin{proof}
We may use the same $\mathcal{A}_{1 \mapsto 2}$ as in Theorem~\ref{theorem_GfL}. The algorithm runs in the same time and satisfies
\begin{align*}
    \mathcal{S}(\mathcal{A}_{1 \mapsto 2}(\mA, \vb)) \le f(\mathcal{S}(\mA, \vb))
\end{align*}
If $\mathcal{A}_{1 \mapsto 2}(\mA, \vb)$ has a solution $\vy$, $(\mA, \vb)$ has a solution $\vy_{m+1:m+n} - \vy_{2m+n+1:2m+2n}$. If $(\mA, \vb)$ has a solution $\vx$, $\mathcal{A}_{1 \mapsto 2}(\mA, \vb)$ must have a solution $\vy$ because otherwise
\begin{align*}
    \{ \vx \mid \mA \vx = \vb \} \neq \emptyset = \{ \vy_{m+1:m+n} - \vy_{2m+n+1:2m+2n} \mid \mL \vy = \vc \}
\end{align*}
Thus, the outputs to $\mathcal{A}_{1 \mapsto 2}(\mA, \vb)$ and to $(\mA, \vb)$ are always the same.
\end{proof}

\subsection{Controlling Laplacian Diagonals over $\mathbb{Z}_p$}
In this section, we try controlling the diagonals of any Laplacian matrix $\mL \in \mathbb{Z}_p^{n \times n}$ without changing the solution space of the linear system. We can control the range of off-diagonal entries by applying Lemma~\ref{lem:resistance_construction}. Since a diagonal entry of $\mL$ is the sum of off-diagonal entries in the row, we can apply Lemma~\ref{lem:resistance_construction} to indirectly control diagonal entries. In this section, we show that it is also possible to control diagonal entries directly. Specifically, we introduce the following three tools:
\begin{enumerate}
    \item Adjusting Diagonals to be Non-Zero: In Section~\ref{subsec:walkreducefroml}, we need to ensure all diagonal entries $\mL_{i,i}$ are non-zero so that $\frac{1}{\mL_{i,i}}$ is defined. The tool shows it is possible to achieve this guarantee.
    \item Graph Stretching: By spliting an edge to two edges with an intermediate vertex, we are able to isolate the influence from both endpoints. This tool is used in Section~\ref{subsubsub:combdecrease} for decreasing combinatorial degree.
    \item Combinatorial Degree Decrease: The diagonal value is influenced by both the density of the represented graph and the value of entries. This tool is able to decrease the graph density to $\Delta_{\mL}=O(1)$ level. In Section~\ref{section:lowdegree}, this serves as the first step of the reudction.
\end{enumerate}

By Lemma~\ref{lem:sc_preserve_sol_1}~and~\ref{lem:sc_preserve_sol_2}, we know that we can modify the graph structure from $G(V,E)$ of $\mL$ to $G'(V',E')$ of $\hat{\mL}$ as long as  $\sc(\hat{\mL},V)=\mL$. Furthermore, by using Lemma~\ref{lem:subgraphreplace}, we can replace a subgraph with a new one if the Schur complement over the original vertices is equal to $\mL$. We will apply Lemma~\ref{lemma:starTclique} frequently to achieve our goal in this section.

\subsubsection{Adjusting Diagonals to be Non-Zero}
\label{subsub:nonzerodiag}
To simplify our algorithm, we assume $p>3$. Some of our reductions require the diagonal entries of the Laplacian must be non-zero. We show that it is always achievable by replacing some subgraphs. By definition, the diagonal entry $\mL_{i,i}$ is equal to $\sum_{i\neq j} \mL_{i,j}$. We want to deal the case when $\mL_{i,i}=0$. If every $\mL_{i,j}$ entry is $0$, in this case we can remove this row since it is empty.  Otherwise, we select an arbitrary entry $\mL_{i,j}$ with value $w \neq 0$ and try to split it. We can add one intermediate vertex $k$ so that we can replace $(i,j,w)$ with edges $(i,k,w_1)$ and $(k,j,w_2)$. Here, $(i,j,w)$ denotes the weighted edge between $i$ and $j$ whose weight is $w$.

\begin{lemma}
\label{lem:splitedgesc}
    Let $\mL \in \mathbb{Z}_p^{n \times n}$ be the Laplacian of the graph $G = (V,E)$. Let $\mL_0$ be the Laplacian constructed by removing $(i,j,w)$ and adding $(i,k,w_1),(k,j,w_2)$ where $w_1\neq 0, w_2\neq 0$. If $\frac{1}{w_1}+\frac{1}{w_2}=\frac{1}{w}$, we have $\sc(\mL_0,V)=\mL$.
\end{lemma}
\begin{proof}
Let $\mL_1$ be the graph resulting from removing edge $(i,j,w)$ from $G$, $\mL_2$ be Laplacian of the graph formed by the edges $(i,k,w_1),(k,j,w_2)$,  and $\mL_3$ be the Laplacian of the graph formed by the edge $(i,j,w)$. We have $\mL=\mL_1+\mL_3$ where the two operands of the sum are viewed as matrices in $\mathbb{Z}^{n \times n}$ after padding $\mL_3$. If $\frac{1}{w_1}+\frac{1}{w_2}=\frac{1}{w}$, by Lemma~\ref{lemma:starTclique}, we have $\sc(\mL_2,\{i,j\})=\mL_3$. Let $X=V-\{i,j\}$, $C=\{i,j\}$, $F=\{k\}$, by Lemma~\ref{lem:subgraphreplace}, $\sc(\mL_0,V)=\sc(\mL_0,C+X)=\mL_1+\sc(\mL_2,C)=\mL_1+\mL_3=\mL$.
\end{proof}
 
Now we want to prove that we can find $w_1$ and $w_2$ such that $\frac{1}{w_1} + \frac{1}{w_2} = \frac{1}{w}$ and after applying Lemma~\ref{lem:splitedgesc}, $(\mL_0)_{i,i},(\mL_0)_{j,j},(\mL_0)_{k,k}$ are all non-zero. By construction, $\mL_{0_{k,k}} = w_1 + w_2$. Suppose before splitting, $\mL_{j,j} = t$. We hope that $w_1,w_2,(\mL_0)_{k,k}=w_1+w_2,(\mL_0)_{i,i}=0-w+w_1,(\mL_0)_{j,j}=t-w+w_2$ are all non-zero. Because $\frac{w_1+w_2}{w_1 w_2}=\frac{1}{w_1}+\frac{1}{w_2}\not\equiv 0$, $w_1+w_2$ is always non-zero.

Now when $p > 3$, at least two out of $\frac{1}{t-1},\frac{1}{t+1},\frac{1}{t+2}$ are well-defined (i.e. the denominator is non-zero). Furthermore, since $p > 3$, the candidates are distinct if they are well-defined. At most one candidate can be equal to $\frac{1}{w}$ since they are distinct, and none of them can be equal to $\frac{1}{t}$. Thus, there always exist one from $\frac{1}{t-1},\frac{1}{t+1},\frac{1}{t+2}$ that is well-defined and is not equal to $\frac{1}{w}$. We can let it to be $w_1$. We then let $w_2=\frac{1}{\frac{1}{w}-\frac{1}{w_1}}$ to satisfy all other conditions.

We can apply the process repeatedly until all diagonal entries of the Laplacian are non-zero. Let the resulting Laplacian be $\hat{\mL}$. By Lemma~\ref{lem:splitedgesc}, $\sc(\hat \mL,V) = \mL$, so the reduction preserves the solution space. There are at most $n$ diagonal entries in $\mL$, thus the reduction maintains $\nnz(\mL)=O(\nnz(\mL))$.

\subsubsection{Graph Stretching}
\label{subsubsec:stretch}
The main idea of decreasing combinatorial degree is by splitting high degree vertices to a few low degree vertices. Since this operation affects the degrees of the neighbors of the vertex being split, we need to isolate the influence by stretching each edge by an additional vertex. Specifically, for every edge $(x,y,w) (w \in [1,p-1])$, we remove it and add $(x,t,2w)$ and $(t,y,2w)$ where $t$ is a new vertex for each edge. Since $w \in [1,p-1]$, we have $2w \not\equiv 0$. Since $\frac{1}{w}=\frac{1}{2w}+\frac{1}{2w}$, by Lemma~\ref{lem:splitedgesc}, such stretching does not change the Schur complement of $V$. Note that the combinatorial degree of each $t$ is $O(1)$. Since we do such split for every edge with non-zero weight, the number of vertices and edges are increased by $O(\nnz(\mL))$, thus stays in $O(\nnz(\mL))$.
\subsubsection{Combinatorial Degree Decrease}
\label{subsubsub:combdecrease}
With the following lemma, we can decrease the max combinatorial degree of any Laplacian matrix $\mL$ but the weights of edges of the resulting graph are unbounded.

\lemcombdegreedecrease*

To prove this lemma, the following lemma gives a construction to reduce any vertex with combinatorial degree $>2$ by $2$ with at most one extra edge and one extra vertex, while the Schur complement of $V$ stays unchanged.
\begin{lemma}
\label{lemma:StarTsmallStars}
For the Laplacian $\mL$ of any graph $G$ with vertex set $V=\{v_0,v_1,v_2,\dots,v_k\}$ and edge set $E=\{(v_0,v_i,w_i)\}$ ($1 \leq i \leq k$), we construct the Laplacian $\hat{\mL}$ of graph $G'$ where $G'$ is the union of the following parts: 
\begin{itemize}
    \item $H$: $V_{H}=\{v_0,v_3,\dots,v_k\},E_{H}=\{(v_0,v_i,w)\} (3\leq i \leq k)$;
    \item an extra node $t$ and three edges $(t,v_0,a)$, $(t,v_1,b)$, $(t,v_2,c)$;
    \item an extra edge $(v_1,v_2,d)$.
\end{itemize}
If $w_1+w_2\not\equiv 0$, we let $a=2(w_1+w_2),b=2w_1,c=2w_2,d=-\frac{w_1w_2}{w_1+w_2}$. Otherwise, we let $a=1,b=w_1,c=w_2,d=-w_1w_2$. With such a construction, we have $\sc(\hat{\mL},V)=\mL$.
\end{lemma}
\begin{proof}
When $w_1+w_2\not\equiv 0$, since $p>2$, $4(w_1+w_2)\not\equiv 0$. View the part consisting of vertices $\{t,v_0,v_1,v_2\}$ and edges $(t,v_0,a), (t,v_1,b), (t,v_2,c)$ as a star. Let its Laplacian be $\mL_1$. By Lemma~\ref{lemma:starTclique}, we can compute $\sc(\mL_1,\{v_0,v_1,v_2\})$. The edge weight is as following:
\[
w(v_0,v_1)=\frac{ab}{a+b+c}=\frac{2(w_1+w_2)\cdot 2w_1}{2(w_1+w_2)+2w_1+2w_2}=w_1
\]
\[
w(v_0,v_2)=\frac{ac}{a+b+c}=\frac{2(w_1+w_2)\cdot 2w_2}{2(w_1+w_2)+2w_1+2w_2}=w_2
\]
\[
w(v_1,v_2)=\frac{bc}{a+b+c}=\frac{4w_1w_2}{4(w_1+w_2)}=-d
\]

Similarly, when $w_1+w_2\equiv 0$, $w_1+w_2+1=1 \not\equiv 0$, then we have 
\[
w(v_0,v_1)=\frac{ab}{a+b+c}=\frac{1 \cdot w_1}{1+w_1+w_2}=w_1
\]
\[
w(v_0,v_2)=\frac{ac}{a+b+c}=\frac{1 \cdot w_2}{1+w_1+w_2}=w_2
\]
\[
w(v_1,v_2)=\frac{bc}{a+b+c}=\frac{w_1w_2}{1+w_1+w_2}=w_1w_2=-d
\]

Let $\mL_2$ be the Laplacian of the graph formed by $(t,v_0,a),(t,v_1,b),(t,v_2,c),(v_1,v_2,d)$, $\mL_3$ be the Laplacian of the graph formed by $(v_1,v_2,d)$. One can see that $\mL_2=\mL_1+\mL_3$ after padding. By Lemma~\ref{lem:subgraphreplace}, we have $\sc(\mL_2,\{v_0,v_1,v_2\})=\mL_3 + \sc(\mL_1,\{v_0,v_1,v_2\})$. With the above calculation, $\sc(\mL_2,\{v_0,v_1,v_2\})$ is the Laplacian of the graph formed by $(v_0,v_1,w_1),(v_0,v_2,w_2),(v_1,v_2,0)$.

Let $\mL_4$ be the Laplacian of $H$. One can see that $\hat{\mL}=\mL_4+\mL_2$. Let $C=\{v_0,v_1,v_2\}$, $X=\{v_3,\dots,v_k\}$, $F=\{t\}$. By Lemma~\ref{lem:subgraphreplace}, we have $\sc(\hat{\mL},V)=\sc(\hat{\mL},C+X)=\mL_4 + \sc(\mL_2,C)$. Since $\sc(\mL_2,\{v_0,v_1,v_2\})$ is the Laplacian of the graph formed by $(v_0,v_1,w_1),(v_0,v_2,w_2)$, $\sc(\hat{\mL},V)=\mL$.

\end{proof}

\begin{proof}[Proof of Lemma~\ref{lemma:combdegree}]
The construction is processed by applying Lemma~\ref{lemma:StarTsmallStars} repeatedly. We first apply the graph stretching described in Section~\ref{subsubsec:stretch} to $G$, denoted as $G'$. We also mark all original vertices before stretching as the internal vertices. We let $G^{(0)}$ be $G'$, then $G^{(k)}$ is defined iteratively:
\begin{enumerate}
    \item If $G^{(k-1)}$ has some internal vertex $x$ with max combinatorial degree $d$ larger than $2$, we group the $d$ edges out of $x$ into $\lfloor\frac{d}{2}\rfloor$ pairs of edges $T_1,\dots,T_{\lfloor\frac{d}{2}\rfloor}$. If there is an unpaired edge, the edge is ignored in this round. For each pair $T_i$, we view the subgraph formed by $T_i$ as a star with center $x$ and apply Lemma~\ref{lemma:StarTsmallStars} to replace the star with the construction. Note that the new added vertex is not marked as an internal one. After processing every pair, $x$ has a combinatorial degree of at most $\lfloor\frac{d}{2}\rfloor + 1$. Let the resulting graph be $G^{(k)}$. Let $\mL^{(\bar{k})}$ be the Laplacian of $G^{k}$. By Lemma~\ref{lemma:StarTsmallStars}, we have $\sc(\mL^{(\bar{k})},V)=\mL$.
    \item Otherwise, we are done $G^{(k)}$ is undefined.
\end{enumerate}
We let $\bar{\mL}$ be $\mL^{(\bar{k})}$ where $\bar{k}$ is the largest $k$ such that $\mL^{(k)}$ is defined. By Lemma~\ref{lemma:StarTsmallStars}, we have $\sc(\bar{\mL},V)=\mL$. 

In $G^{(\bar{k})}$, all internal (original) vertices have a combinatorial degree of at most $2$ because of the graph stretching, and we only apply Lemma~\ref{lemma:StarTsmallStars} with internal vertices. So only the degree of non-internal (newly added) vertices might be increasing. There are two types of newly added vertices. One type comes from the graph stretching in Section~\ref{subsubsec:stretch}, the other type comes from applying Lemma~\ref{lemma:StarTsmallStars}. Now we analyze on these two types of newly added vertices.
\begin{lemma}
    In $G^{(\bar{k})}$, all newly added vertices' combinatorial degree is at most $4$.
\end{lemma}
\begin{proof}
For any vertex $x$ added in the graph stretching stage in Section~\ref{subsubsec:stretch}, its degree is $2$ when the vertex is first added. Since $x$ is not an internal vertex, its degree can only change when an edge connected to $x$ is paired and replaced during the process. Based on the construction of Lemma~\ref{lemma:StarTsmallStars}, each time we replace a star with our construction, $x$ will be connecting to at most $2$ new edges while being disconnected from $1$ edge. Therefore, for such $x$, its combinatorial degree is at most $4$. 

Now suppose vertex $y$ is added when a star with center $z$ and vertices $\{z,v_1,v_2\}$ is replaced by the construction of Lemma~\ref{lemma:StarTsmallStars}. Notice $z$ is an internal vertex, while due to graph stretching, $v_1,v_2$ are no internal vertices. Initially, $y$ will be connected to $\{z,v_1,v_2\}$. Therefore, $y$'s initial degree is 3. Since the combinatorial degree of $z$ might need to be further decreased, $y$ may be involved in another round of such reduction process while its other edges remain. If this is the case, $y$ will be adjacent to two new edges and $y$ is then disconnected from $z$. When $y$ is disconnected from $z$, all edges adjacent to $y$ are not internal vertices, so $y$ won't be affected by further reductions. In summary, the final degree of $y$ is at most $4$ in this case.
\end{proof}

Since all vertices' combinatorial degree is not larger than $4$, we prove that after such reduction, the max combinatorial degree is $O(1)$ level.

Now we need to analyze the number of vertices and edges after the combinatorial degree reduction process. The graph stretching stage is adding $|E|$ edges and vertices, so both the number of vertices and the number of edges after stretching become $O(|E|)$. Since we only deal with the internal vertices, the number of invocations to Lemma~\ref{lemma:StarTsmallStars} is $O(\sum_{i\in V} \deg_i \log \deg_i)=O(|E|\log |E|)$. Each time we replace a star, we add $2$ more edges and $1$ new node. Thus, we will add $O(|E|\log|E|)$ edges and nodes in total in the reduction phase. Therefore, $|V'|$ and $|E'|$ are in $O(|E|\log|E|)$.
\end{proof}

\subsection{Reduction to Unit-Weight Laplacian System}
\label{sec:reductionunitweight}

In this section, we show that the class of Laplacian matrices $\LS_{\mathbb{Z}_p}$ is efficiently and decisional $f$-reducible to the class of unit-weight Laplacians for some $f$ that does not depend on the original $\size$ of the matrix. Specifically, the reduction is by replacing every edge in a Laplacian system in $\LS_{\mathbb{Z}_p}$ by a gadget with $\bound$ unit-weight edges. 

\thmGfUL*

\begin{proof}
We first use the circuit construction from
Lemma~\ref{lem:resistance_construction}
to show that given any Laplacian system $\mL \in \mathbb{Z}_p^{n\times n}$, we can compute a unit-weight Laplacian system $\bar{\mL} \in \mathbb{Z}_p^{k\times k}$ such that 
\begin{enumerate}
    \item $\nnz(\bar{\mL})\le O(\nnz(\mL)\log^2 p/\log \log p)$.
    \item For any linear system $\mL \vx = \vb$, $\left\{\vx\mid \mL \vx = \vb\right\}=\left\{\vy_{[n]}\mid\bar{\mL} \vy = \left[\begin{array}{c}\vb \\ \vzero \end{array}\right]\right\}$.
\end{enumerate}

In addition, our construction also guarantees that
\[
    \maxdeg(\bar{\mL})\le O(\log p) \maxdeg(\mL).
\]
Then we will let $\mathcal{A}_{1\mapsto2}(\mL, \vb)=\left(\bar{\mL}, \left[\begin{array}{c}\vb \\ \vzero \end{array}\right]\right)$, $\mathcal{A}_{2\mapsto1}(\mY)=\{\vy_{[n]}|y\in \mY\}$ and prove their running times.

\paragraph{Construction of $\bar{\mL}$.} We apply Lemma~\ref{lem:replace_edge} repeatedly. We let $\mL^{(0)}$ be $\mL$. We fix $n$ as the number of rows in $\mL^{(0)}$. $\mL^{(k)}$ is defined recursively:
\begin{enumerate}
    \item If $\mL^{(k-1)}_{i, j}$ is not $0$ or $p-1$ for some $1\le i< j\le n$ satisfying $\forall 1\le \ell < k, (i, j)\neq (i(\ell), j(\ell))$, apply Lemma~\ref{lem:replace_edge} for $\mL=\mL^{(k-1)}$, $i_0=i, j_0=j$. Let Lemma~\ref{lem:resistance_construction} constructs the circuit $\mR$ in Lemma~\ref{lem:replace_edge} with size $\bound$ and resistance $-\mL^{(k-1)}_{i, j}$. Let the resulting $\mU$ be $\mL^{(k)}$. We record the indices $i$ and $j$ as $i(k)$ and $j(k)$. 
    \item Otherwise, $\mL^{(k)}$ is undefined. 
\end{enumerate}
Let $\bar{\mL}$ be $\mL^{(\bar{k})}$ where $\bar{k}$ is the largest $k$ such that $\mL^{(k)}$ is defined. $\{(i(1), j(1)), \ldots, \\ (i(\bar{k}), j(\bar{k}))\}$ goes through every entry of $\mL$ which is strictly above the diagonal and is not equal to $0$ or $p-1$ exactly once. Thus, $$\bar{k}\le \nnz(\mL).$$ By Item~\ref{itm:replaceitem1} of Lemma~\ref{lem:replace_edge}, we also have $$\nnz(\mL^{(i)})\le \nnz(\mL^{(i-1)})+\bound$$ for $1\le i\le \bar{k}$. Combining the two inequalities, we get 
$$\nnz(\bar{\mL}) \le \nnz(\mL)+\nnz(\mL)\bound=O\left(\nnz(\mL)\log^2p/\log\log p\right).$$

Since each edge is either preserved or replaced by a circuit with maximum degree $O(\log p)$ constructed by Lemma~\ref{lem:resistance_construction}, the maximum degree $\maxdeg(\bar{\mL})$ is no more than $O(\log p)\maxdeg(\mL)$. This property is vital in the reduction to Low-Degree Laplacian System described in~\ref{section:lowdegree}.

Next we prove that for any linear system $\mL \vx = \vb$, $\left\{\vx\mid \mL \vx = \vb\right\}=\left\{\vy_{[n]}\mid\bar{\mL} \vy = \left[\begin{array}{c}\vb \\ \vzero \end{array}\right]\right\}$ by induction. Fix any vector $\vb$. Suppose 
\[\left\{\vx\mid \mL \vx = \vb\right\}=\left\{\vy_{[n]}\mid\mL^{(k)} \vy = \left[\begin{array}{c}\vb \\ \vzero \end{array}\right]\right\}.
\] By Lemma~\ref{lem:replace_edge}, 
\[
\left\{\vy\mid\mL^{(k)} \vy = \left[\begin{array}{c}\vb \\ \vzero \end{array}\right]\right\}\\
    =\left\{\vz_{[\norm{y}_0]}\mid\mL^{(k+1)} \vz = \left[\begin{array}{c}\vb \\ \vzero \end{array}\right]\right\}
\] where $\norm{\vy}_0$ is the number of coordinates of $\vy$.
Thus,
\begin{align*}
    &\left\{\vx\mid \mL \vx = \vb\right\}\\
    =&\left\{\vy_{[n]}\mid\mL^{(k)} \vy = \left[\begin{array}{c}\vb \\ \vzero \end{array}\right]\right\}\\
    =&\left\{\vz_{[n]}\mid\mL^{(k+1)} \vz = \left[\begin{array}{c}\vb \\ \vzero \end{array}\right]\right\}.
\end{align*}
By induction, we have 
\[\left\{\vx\mid \mL \vx = \vb\right\} = 
\left\{\vy_{[n]}\mid\mL^{(\bar{k})} \vy = \left[\begin{array}{c}\vb \\ \vzero \end{array}\right]\right\}
=\left\{\vy_{[n]}\mid\bar{\mL} \vy = \left[\begin{array}{c}\vb \\ \vzero \end{array}\right]\right\}.
\]

\paragraph{Running Time of $\mathcal{A}_{1\mapsto2}$ and $\mathcal{A}_{2\mapsto1}$.}

$\mathcal{A}_{1\mapsto2}$ simply replaces each non-diagonal non-zero entry by a matrix $\mU$ with $\nnz(\mU)=O(\log^2 p/\log \log p)$. This costs \begin{equation*}
O(\nnz(\bar{\mL}))=O(\nnz(\mL)\log^2 p/\log \log p).
\end{equation*}
We can implement $\mathcal{A}_{2\mapsto1}$ as 
\[
\mathcal{A}_{2\mapsto1}(\vy^*)=(\vy^*_{[n]}).
\] It costs nearly linear time as we only truncate vectors.
\end{proof}
\begin{corollary}
\label{cor:p13}
Let $f(\nnz) = O(\nnz \log^2 p /\log \log p)$. Then $\LS_{\mathbb{Z}_p}$ is decisional $f$-reducible to $\UWLS_{\mathbb{Z}_p}$.
\end{corollary}
\begin{proof}
    We may use the same $\mathcal{A}_{1\mapsto2}$ as in Theorem~\ref{thm:GfUL}. The algorithm runs in the same time and satisfies 
    \[\mathcal{S}(\mathcal{A}_{1\mapsto2}(\mL,\vb)) \le f(\mathcal{S}(\mL,\vb)).\]
    
    If $\mathcal{A}_{1\mapsto2}(\mL,\vb)$ has a solution $\vy$, $(\mL, \vb)$ has a solution $\vy_{[n]}$. If $(\mL, \vb)$ has a solution $\vx$, $\mathcal{A}_{1\mapsto2}(\mL,\vb)$ must have a solution $\vy$ because otherwise \[\left\{\vx\mid \mL \vx = \vb\right\}\neq \emptyset=\left\{\vy_{[n]}\mid\bar{\mL} \vy = \left[\begin{array}{c}\vb \\ \vzero \end{array}\right]\right\}.\] Thus, the outputs to $\mathcal{A}_{1\mapsto2}(\mL,\vb)$ and to $(\mL,\vb)$ are always the same.
\end{proof}

\subsection{Reduction to Low-Degree Laplacian System}
\label{section:lowdegree}
In this section, we show that the class of Laplacian matrices $\LS_{\mathbb{Z}_p}$ is efficiently and decisional $f$-reducible to itself for some $f$ that restrict the diagonal entries to $O(\log p)$, which we also define as the $\maxdegree$ of the matrix.

\thmGfDL*
\begin{proof}
We will prove that for any Laplacian system $\mL \in \mathbb{Z}_p^{n \times n}$, we can compute a low degree Laplacian system $\bar{\mL} \in \mathbb{Z}_p^{k\times k}$ such that
\begin{itemize}
    \item $\nnz(\bar{\mL})\le O(\nnz(\mL) \cdot \log \nnz(\mL) \cdot \log^2 p /\log \log p)$.
    \item $\Delta_{\bar{\mL}}^{\vw}=O(\log p)$.
    \item $\bar{\mL}$ is also a unit-weight matrix.
    \item For any linear system $\mL \vx = \vb$, $\left\{\vx\mid \mL \vx = \vb\right\}=\left\{\vy_{[n]}\mid\bar{\mL} \vy = \left[\begin{array}{c}\vb \\ \vzero \end{array}\right]\right\}$.
\end{itemize}
\paragraph{Construction of $\bar{\mL}$}
Initially, $\mL$ has $\nnz(\mL)$ entries. Therefore, there are $O(\nnz(\mL))$ edges, which means the maximum combinatorial degree is bounded above by $O(\nnz(\mL))$ and the maximum weighted degree is bounded above by $O(p \cdot \nnz(\mL))$. The unit-weight Laplacian reduction in Section~\ref{sec:reductionunitweight} only guarantees the maximum weighted degree is bounded above by $O(\log p \cdot nnz(\mL))$. Therefore, our main idea is to transform $\mL$ to a Laplacian matrix $\hat{\mL}$ such that $\Delta_{\hat{\mL}}=O(1)$. Then we apply our unit-weight reduction and turn it to $\bar{\mL}$.

We construct $\hat{\mL}$ by applying Lemma~\ref{lemma:combdegree}. Therefore, $\nnz(\hat{\mL})=O(\nnz(\mL)\log \nnz(\mL))$ and $\sc(\hat{\mL},[n])=\mL$. Most importantly, $\Delta_{\hat{\mL}}$ is decreased to $O(1)$ level. Note that the edge weights can be $O(p)$ and thus the weighted degree of some nodes could be up to $O(p)$. Then we run the reduction from Lemma~\ref{thm:GfUL} on $\hat{\mL}$, getting our $\bar{\mL}$ eventually. Specifically, the reduction is done by replacing every edge to a gadget with $O(\log^2 p/\log \log p)$ unit-weight edges and $O(\log p)$ max degree. Therefore, every node in  $\hat{\mL}$'s degree is reduced to 

$$O(1) \text{ (combinatorial degree)} \times O(\log p) \text{ (gadget max degree)}=O(\log p).$$

Since the extra vertices in the gadget also have $O(\log p)$ degree, $\Delta_{\bar{\mL}}^{\vw}=O(\log p)$.

By Lemma~\ref{lem:sc_preserve_sol_1},~\ref{lem:sc_preserve_sol_2}, we know that $\sc(\hat{\mL},[n])=\mL$ preserves the solution space to arbitrary linear system $\mL \vx=\vb$. Specifically, for any linear system $\mL \vx = \vb$, $\left\{\vx\mid \mL \vx = \vb\right\}=\left\{\vy_{[n]}\mid\bar{\mL} \vy = \left[\begin{array}{c}\vb \\ \vzero \end{array}\right]\right\}$. This is in the same form of our unit-weight Laplacian reduction, which also preserves the solution space. Therefore, our low degree reduction preserves the solution space after the two-step transform. Since $\nnz(\hat{\mL})=O(\nnz(\mL)\log \nnz(\mL))$, after the unit-weight gadget replacement (each gadget is $O(\log^2 p/\log \log p)$, $\nnz(\bar{\mL})\le O(\nnz(\mL) \cdot \log \nnz(\mL) \cdot \log^2 p /\log \log p)$. Since every edge in the gadget is unit-weight, $\bar{\mL}$ is also a unit-weight Laplacian by definition.

\paragraph{Running Time of $\mathcal{A}_{1\mapsto2}$ and $\mathcal{A}_{2\mapsto1}$.}

In $\mathcal{A}_{1\mapsto2}$, the first step is  do the combinatorial degree decrease, which costs $O(\nnz(\mL)\log \nnz(\mL))$ time. Then in the second step of applying the unit-weight Laplacian reduction, we simply replaces each non-diagonal non-zero entry by a matrix $\mU$ with $\nnz(\mU)=O(\log^2 p/\log \log p)$. This costs \begin{equation*}
O(\nnz(\bar{\mL}))=O(\nnz(\hat{\mL})\log^2 p/\log \log p)=O(\nnz(\mL)\log \nnz(\mL) \cdot \log^2 p/\log \log p).
\end{equation*}
We can implement $\mathcal{A}_{2\mapsto1}$ as 
\[
\mathcal{A}_{2\mapsto1}(\vy^*)=(\vy^*_{[n]}).
\] It costs nearly linear time as we only truncate vectors.
\end{proof}
\begin{corollary}
\label{cor:lowdp}
Let $f(\nnz) = O(\nnz \log \nnz \cdot \log^2 p /\log \log p)$. Then $\LS_{\mathbb{Z}_p}$ is decisional $f$-reducible to $\DLS_{\log p,\mathbb{Z}_p}$.
\end{corollary}
\begin{proof}
    We may use the same $\mathcal{A}_{1\mapsto2}$ as in Theorem~\ref{thm:GfUL}. The algorithm runs in the same time and satisfies 
    \[\mathcal{S}(\mathcal{A}_{1\mapsto2}(\mL,\vb)) \le f(\mathcal{S}(\mL,\vb)).\]
    
    If $\mathcal{A}_{1\mapsto2}(\mL,\vb)$ has a solution $\vy$, $(\mL, \vb)$ has a solution $\vy_{[n]}$. If $(\mL, \vb)$ has a solution $\vx$, $\mathcal{A}_{1\mapsto2}(\mL,\vb)$ must have a solution $\vy$ because otherwise \[\left\{\vx\mid \mL \vx = \vb\right\}\neq \emptyset=\left\{\vy_{[n]}\mid\bar{\mL} \vy = \left[\begin{array}{c}\vb \\ \vzero \end{array}\right]\right\}.\] Thus, the outputs to $\mathcal{A}_{1\mapsto2}(\mL,\vb)$ and to $(\mL,\vb)$ are always the same.
\end{proof}

\section{Hardness of Solving Walk Matrix System Over $\mathbb{Z}_p$}

\label{sec:hardnessW}

\subsection{Reduction From General Linear System}
\thmGfW*

\begin{proof}
\ We first prove that given any linear system $(\mA \in \mathbb{Z}_p^{m \times n}, 
\vb \in \mathbb{Z}_p^{m}) \in \GS_{\mathbb{Z}_p}$, we can compute a walk matrix system 
$(\mW \in \mathbb{Z}_p^{4(m+n) \times 4(m+n)}, \vc \in \mathbb{Z}_p^{4(m+n)})$ such 
that 

\begin{enumerate}
    \item $\nnz(\mW) = O(\nnz(\mA))$ 
    \item $\{\vy_{m+1:m+n} - \vy_{2m + n + 1:2m + 2n} \mid \mW \vy = \vc \} = 
    \{ \vx \mid \mA \vx = \vb \}$
\end{enumerate}

Then we will let $\mathcal{A}_{1 \mapsto 2}(\mA, \vb) = (\mW, \vc)$, $\mathcal{A}_{2 \mapsto 1}(\mW, \vc, \vy) = \vy_{m+1:m+n} - \vy_{2m + n + 1:2m + 2n}$ and show their running times.

\paragraph{Construction of $(\mW, \vc)$}

we construct $(\mW, \vc)$ as 
\begin{align*}
     \mW \defeq 
     \left[ 
            \begin{array}{cccccccc}
                \mI_m & \mA & \vzero &  -\mA & -\mI_m & \vzero & \vzero & \vzero \\
                \mA^T & \mI_n & -\mA^T & \vzero & \vzero & -\mI_n & \vzero & \vzero\\
                \vzero & -\mA & \mI_m & \mA & \vzero & \vzero & -\mI_m & \vzero\\
                -\mA^T & \vzero & \mA^T & \mI_n & \vzero & \vzero & \vzero & -\mI_n \\
               -\mI_m & \vzero & \vzero & \vzero & \mI_m & \vzero & \vzero & \vzero \\
                \vzero & -\mI_n & \vzero & \vzero & \vzero & \mI_n & \vzero & \vzero \\
                \vzero & \vzero & -\mI_m & \vzero & \vzero & \vzero & \mI_m & \vzero \\
                \vzero & \vzero & \vzero & -\mI_n & \vzero & \vzero & \vzero & \mI_n \\
            \end{array}
            \right],\quad
            \vc \defeq  
            \left[
            \begin{array}{c}
            \vb \\
            \vzero \\
            -\vb \\
            \vzero \\
            \vzero \\
            \vzero \\
            \vzero \\
            \vzero
            \end{array}
            \right]
\end{align*}
It's easy to verify $\mW$ is symmetric and with all-ones diagonal, according to Definition~\ref{def:walkmatrix}, $\mW$ is a walk matrix over the finite field $\mathbb{Z}_p$. 
\paragraph{Proof for $\mathcal{S}(\mW, \vc) \le f(\mathcal{S}(\mA, \vb))$}
    By our construction of $\mW$, we have $\nnz(W) = O(\nnz(\mA))$ and  
    \begin{align*}
        \mathcal{S}(\mW, \vc) \le O(\nnz(\mA)) = f(\mathcal{S}(\mA, \vb)). 
    \end{align*}
Next we show $\{\vy_{m+1:m+n} - \vy_{2m + n + 1:2m + 2n} \mid \mW \vy = \vc \} = 
    \{ \vx \mid \mA \vx = \vb \}$. 
Considering a solution $\vy = (\vy_1^T, \vy_2^T, \vy_3^T, \vy_4^T, \vy_5^T, \vy_6^T, \vy_7^T, \vy_8^T)^T$ to the linear system $\mW \vy = \vc$ 
, we have
    \begin{align*}
    \left. 
    \begin{array}{r @{{}={}} l}
        \vy_1 + \mA \vy_2 - \mA \vy_4 - \vy_5 & \vb   \\
        \mA^T \vy_1 +\vy_2 - \mA^T \vy_3 -\vy_6 & \vzero   \\
        -\mA \vy_2 + \vy_3 + \mA \vy_4 -\vy_7 & -\vb   \\
        -\mA^T\vy_1 + \mA^T \vy_3 +\vy_4 -\vy_8 & \vzero   \\
        -\vy_1+\vy_5 & \vzero \\
        -\vy_2+\vy_6 & \vzero \\
        -\vy_3+\vy_7 & \vzero \\
        -\vy_4+\vy_8 & \vzero 
    \end{array}
    \right\}
    \text{ implies } \mA(\vy_2 - \vy_4) = \vb  
    \end{align*}
     Thus 
     \begin{align*}
        \vy_{m+1:m+n} - \vy_{2m + n + 1:2m + 2n} = \vy_2 - \vy_4 \in \{ \vx \mid \mA \vx = \vb   \} 
     \end{align*}
     For every solution $\vx$ to $\mA \vx = \vb  $, we have
     \begin{align*}
        \left[
        \begin{array}{cccccccc}
                \mI_m & \mA & \vzero &  -\mA & -\mI_m & \vzero & \vzero & \vzero \\
                \mA^T & \mI_n & -\mA^T & \vzero & \vzero & -\mI_n & \vzero & \vzero\\
                \vzero & -\mA & \mI_m & \mA & \vzero & \vzero & -\mI_m & \vzero\\
                -\mA^T & \vzero & \mA^T & \mI_n & \vzero & \vzero & \vzero & -\mI_n \\
               -\mI_m & \vzero & \vzero & \vzero & \mI_m & \vzero & \vzero & \vzero \\
                \vzero & -\mI_n & \vzero & \vzero & \vzero & \mI_n & \vzero & \vzero \\
                \vzero & \vzero & -\mI_m & \vzero & \vzero & \vzero & \mI_m & \vzero \\
                \vzero & \vzero & \vzero & -\mI_n & \vzero & \vzero & \vzero & \mI_n \\
            \end{array}
        \right]
        \left[
        \begin{array}{c}
            \vzero \\
            \vx \\
            \vzero \\
            \vzero \\
            \vzero \\
            \vx \\
            \vzero \\
            \vzero 
        \end{array}
        \right]
        = 
        \left[
        \begin{array}{c}
            \mA \vx \\
            \vzero \\
            -\mA \vx \\
            \vzero \\
            \vzero \\
            \vzero \\
            \vzero \\
        \end{array}
        \right]
        = 
        \left[
        \begin{array}{c}
            \vb \\
            \vzero \\
            -\vb \\
            \vzero \\
            \vzero \\
            \vzero \\
            \vzero \\
            \vzero \\
        \end{array}
        \right]
        = \vc  
 \end{align*}
which implies
\begin{align*}
    \vy = \left[
        \begin{array}{c}
            \vzero \\
            \vx \\
            \vzero \\
            \vzero \\
            \vzero \\
            \vx \\
            \vzero \\
            \vzero \\
        \end{array}
        \right]
\end{align*}
is a solution to the linear system $\mW \vy = \vc  $. Since $\vy_{m+1:m+n} - \vy_{2m + n + 1:2m + 2n} = \vx - \vzero = \vx$,
\begin{align*}
    \vx \in\{\vy_{m+1:m+n} - \vy_{2m + n + 1:2m + 2n} \mid \mW \vy = \vc \}.
\end{align*}
\paragraph{Running Time of $\mathcal{A}_{2 \mapsto 1}$, $\mathcal{A}_{1 \mapsto 2}$}

By our construction of $(\mW, \vc)$, we have $\mathcal{A}_{1 \mapsto 2}$ run in $O(\nnz(\mA))$ time. Since $\vx = \vy_{m+1: m+n} - \vy_{2m + n + 1:2m + 2n}$,
$\mathcal{A}_{2 \mapsto 1}$ runs in $O(n) = O(\nnz(\mA))$ time.

\end{proof}

\begin{corollary}
\label{cor:GFW}
      Let $f(\nnz) = O(\nnz)$, then $\GS_{\mathbb{Z}_p}$ is decisional $f$-reducible to $\WS_{\mathbb{Z}_p}$.
\end{corollary} 
\begin{proof}
We may use the same $\mathcal{A}_{1 \mapsto 2}$ as in Theorem~\ref{theorem_GfW}. The algorithm runs in the same time and satisfies
\begin{align*}
    \mathcal{S}(\mathcal{A}_{1 \mapsto 2}(\mA, \vb)) \le f(\mathcal{S}(\mA, \vb))
\end{align*}
If $\mathcal{A}_{1 \mapsto 2}(\mA, \vb)$ has a solution $\vy$, $(\mA, \vb)$ has a solution $\vy_{m+1:m+n} - \vy_{2m+n+1:2m+2n}$. If $(\mA, \vb)$ has a solution $\vx$, $\mathcal{A}_{1 \mapsto 2}(\mA, \vb)$ must have a solution $\vy$ because otherwise
\begin{align*}
    \{ \vx \mid \mA \vx = \vb \} \neq \emptyset = \{ \vy_{m+1:m+n} - \vy_{2m+n+1:2m+2n} \mid \mW \vy = \vc \}
\end{align*}
Thus, the outputs to $\mathcal{A}_{1 \mapsto 2}(\mA, \vb)$ and to $(\mA, \vb)$ are always the same.
\end{proof}

\subsection{Reduction From Laplacian System}
\label{subsec:walkreducefroml}
In this section, we give a reduction from the class of Laplacian $\LS_{\mathbb{Z}_p}$ to the class of the walk matrix $\WS_{\mathbb{Z}_p[\sqrt{t}]}$. $\mathbb{Z}_p[\sqrt{t}]$ denotes the extension field from $\mathbb{Z}_p$ by adjoining $\sqrt{t}$ where $t$ is a primitive root modulo $p$ (or any quadratic non-residue modulo $p$). By definition, $\mathbb{Z}_p[\sqrt{t}] \cong \mathbb{F}_{p^2}$ is also a finite field. To simplify the analysis, throughout this section we assume $p>3$.

\thmLfW*

\begin{proof}
For any given Laplacian system $\mL \in \mathbb{Z}_p^{n \times n}$ with non-zero diagonals, we can compute a walk matrix system $\mW \in \mathbb{Z}_p[\sqrt{t}]^{k\times k}$ such that 

\begin{enumerate}
    \item $\nnz(\mW) = O(\nnz(\mL))$ 
    \item We define a function $\varphi: \mathbb{F}_p(\sqrt{t}) \mapsto \mathbb{F}_p(\sqrt{t})$ given by $\varphi(a + b\sqrt{t}) = a - b \sqrt{t}$. For any Laplacian linear system $\mL \vx = \vb$, $\{\frac{1}{2}(\mD^{-\frac{1}{2}}\vy)+\frac{1}{2}\varphi(\mD^{-\frac{1}{2}}\vy) \mid \mW \vy = \vc \} = 
    \{ \vx \mid \mL \vx = \vb \}$. 
\end{enumerate}

\paragraph{Construction of ($\mW, \vc$).} First, we need to ensure the diagonals are non-zero. We use the method described in Section~\ref{subsub:nonzerodiag} to achieve so. This process keeps $\nnz =O(\nnz(\mL))$. Then we write $\mL$ as $\mD-\mA$. Notice that the difference between Laplacian and our target (which is a normalized Laplacian) is we would like to have $\mI$ instead of $\mD$. If the diagonal entries of $\mL$ are all non-zero quadratic residues modulo $p$, we can simply normalize $\mL$ by letting $\mW = \mD^{-1/2} \mL \mD^{-1/2}$. However, not all in $\mathbb{Z}_p$ are quadratic residues modulo $p$, so we must extend the field to make sure all in $\mathbb{Z}_p$ have a square root in the extended field. Specifically, we can let $t$ be the primitive root of $p$, which is the minimum $t$ in $\mathbb{Z}_p$ satisfying $t^k$ iterates through all in $\mathbb{Z}_p - 0$ when $k$ goes through $\mathbb{N}$. We extend the finite field $\mathbb{Z}_p$ by adjoining $\sqrt{t}$. The following lemma shows that this extension is sufficient to allow any $x \in [0,\dots,p-1]$ to have a square root in the new field.

\begin{lemma}[Quadratic residue guarantee in extended field]
\label{lem:extendfield}
    For any $x \in \mathbb{Z}_p (p>2)$, there exists $y \in \mathbb{Z}_p[\sqrt{t}]$ such that $y^2 \equiv x$ in $\mathbb{Z}_p[\sqrt{t}]$.
\end{lemma}
\begin{proof}
    By the nature of the primitive root over $\mathbb{Z}_p$, we know any $x \in \mathbb{Z}_p$ can be express as $t^k (k\geq 0)$. By the construction,  $\sqrt{t}=t^{\frac{1}{2}}$ is in the extended field. Therefore, any $x \in \mathbb{Z}_p$ can be express as $(\sqrt{t})^{2k}=(\sqrt{t}^k)^2$. By definition, $\sqrt{t}^k$ is in $\mathbb{Z}_p[\sqrt{t}]$, thus we prove our lemma.
\end{proof}

By the extended field, we guarantee $\mD^{\frac{1}{2}}$'s existence. Then we have our construction of $\mW$ and $\vc$ as follows:
\begin{align*}
\mW \defeq \mD^{-\frac{1}{2}}\mL\mD^{-\frac{1}{2}}, \vc\defeq \mD^{\frac{1}{2}}\vb
\end{align*}

Since $\mD^{-\frac{1}{2}}\mL\mD^{-\frac{1}{2}}=\mD^{-\frac{1}{2}}\mD\mD^{-\frac{1}{2}}-\mD^{-\frac{1}{2}}\mA\mD^{-\frac{1}{2}}=\mI-\mD^{-\frac{1}{2}}\mA\mD^{-\frac{1}{2}}$, $\mW=\mD^{-\frac{1}{2}}\mL\mD^{-\frac{1}{2}}$ is a walk matrix.

For $\mW\vy=\vc$, we have $\mD^{-\frac{1}{2}}\mL\mD^{-\frac{1}{2}}\vy=\mD^{\frac{1}{2}}\vb$, which leads to $\mL\mD^{-\frac{1}{2}}\vy=\vb$. Therefore, for any solution $\vx$ of $\mL\vx=\vb$ in $\mathbf{Z}_p$, $\mD^{\frac{1}{2}}\vx$ is a solution to $\mW \vy=\vc$ in $\mathbb{Z}_p \subset \mathbb{Z}_p[\sqrt{t}]$. On the other side, we can see that for any solution $\vy$ of $\mW\vy=\vc$ in $\mathbb{Z}_p[\sqrt{t}]$, $\mD^{-\frac{1}{2}}\vy$ is a solution to $\mL\vx=\vb$ in $\mathbb{Z}_p[\sqrt{t}]$. 

Now we need to further prove that any solution $\vx$ to $\mL \vx=\vb$ in $\mathbb{Z}_p[\sqrt{t}]$ corresponds to a solution to $\mL\vx'=\vb$ in $\mathbb{Z}_p$. We can represent any element in $\mathbb{Z}_p[\sqrt{t}]$ as $a+b\sqrt{t} (a,b \in \mathbb{Z}_p)$. We define the conjugation map $\varphi: \mathbb{F}_p(\sqrt{t}) \mapsto \mathbb{F}_p(\sqrt{t})$ as the map given by $\varphi(a + b\sqrt{t}) = a - b \sqrt{t}$. By definition, $\varphi$ is a field isomorphism. Now let $\phi(\mL)$ be the matrix obtained by conjugating all entries of $\mL$ using $\varphi$. Since $\varphi$ is a field isomorphism, if $\mL\vx=\vb$, $\varphi(\mL)\varphi(\vx)=\varphi(\vb)$. If $\vx=\va+\vb\sqrt{t}$ is a solution to the linear system $\mL \vx = \vb$ in $\mathbb{Z}_p[\sqrt{t}]$,  $\varphi(\vx)=\va-\vb\sqrt{t}$ is a solution to $\varphi(\mL)\varphi(\vx)=\varphi(\vb)$. Since all elements of $\mL$ and $\vb$ in $\mathbb{Z}_p$, $\phi(\mL) = \mL$ and $\phi(\vb) = \vb$, so $\varphi(\vx)$ is also a solution to $\mL\vx=\vb$ in  $\mathbb{Z}_p[\sqrt{t}]$. Then we can construct a solution $\vx' = \frac{1}{2}(\vx+\varphi(\vx))$ in $\mathbb{Z}_p$ to the linear system $\mL\vx=\vb$. Note that  $\frac{1}{2}(\vx+\varphi(\vx))$ simply extracts the part in $\mathbb{Z}_p[\sqrt{t}]$ for $\vx$ as long as $p \neq 2$.

\paragraph{Running Time of $\mathcal{A}_{2 \mapsto 1}$, $\mathcal{A}_{1 \mapsto 2}$}

By our construction of $(\mW, \vc)$, we have $\mathcal{A}_{1 \mapsto 2}$ run in $O(\nnz(\mL))$ time. Since $\vx = \frac{1}{2}(\mD^{-\frac{1}{2}}\vy)+\frac{1}{2}\varphi(\mD^{-\frac{1}{2}}\vy) $,
$\mathcal{A}_{2 \mapsto 1}$ runs in $O(n) = O(\nnz(\mL))$ time.

\end{proof} 

\begin{corollary}
\label{cor:LFW}
      Let $f(\nnz) = O(\nnz)$, then $\LS_{\mathbb{Z}_p}$ is decisional $f$-reducible to $\WS_{\mathbb{Z}_p[\sqrt{t}]}$.
\end{corollary} 
\begin{proof}
We may use the same $\mathcal{A}_{1 \mapsto 2}$ as in Theorem~\ref{theorem_LfW}. The algorithm runs in the same time and satisfies
\begin{align*}
    \mathcal{S}(\mathcal{A}_{1 \mapsto 2}(\mL, \vb)) \le f(\mathcal{S}(\mL, \vb))
\end{align*}
If $\mathcal{A}_{1 \mapsto 2}(\mL, \vb)$ has a solution $\vy$, $(\mA, \vb)$ has a solution $\frac{1}{2}(\mD^{-\frac{1}{2}}\vy)+\frac{1}{2}\varphi(\mD^{-\frac{1}{2}}\vy)$. If $(\mL, \vb)$ has a solution $\vx$, $\mathcal{A}_{1 \mapsto 2}(\mA, \vb)$ must have a solution $\vy$ because otherwise
\begin{align*}
    \{ \vx \mid \mL \vx = \vb \} \neq \emptyset = \{\frac{1}{2}(\mD^{-\frac{1}{2}}\vy)+\frac{1}{2}\varphi(\mD^{-\frac{1}{2}}\vy) \mid \mW \vy = \vc \}
\end{align*}
Thus, the outputs to $\mathcal{A}_{1 \mapsto 2}(\mL, \vb)$ and to $(\mL, \vb)$ are always the same.
\end{proof}

\section{Hardness of Solving Symbolic Matrices}
\label{sec:symbolic}
\subsection{Motivation}
Previous discussions focus on structured linear systems with scalar coefficients. Symbolic matrices can also inherit graph structures. Two examples are the Edmonds matrices and the Tutte matrices, which have deep connections with graph matchings. The determinant of these matrices encode information about perfect matchings of the graphs. Testing if the determinant is zero help us decide if the graph has perfect matchings. We are first going to describe relevant work using a less formal language. We will revisit the results after introducing the formal definitions related to symbolic matrices.

\paragraph{Edmonds Matrices and Bipartite Graph Matching.} The first class of symbolic matrices we are going to discuss is the Edmonds matrices. 
\begin{definition}[Edmonds Matrix of Bipartite Graph]
\label{def:edmonds}
Given a balanced bipartite graph $G = ((V_L, V_R), E)$ with $|V_L| = |V_R| = n$ (i.e.~$G$ has $2n$ vertices), the corresponding Edmonds matrix $\mA$ is an $n \times n$ symbolic matrix with variables $x_e$ where the index $e \in E$ ranges over the edges of $G$. The Edmonds matrix of the bipartite graph, $\mG$, is simply given by:
\begin{equation*}
\mG_{ij} := \begin{cases}
x_e & \text{if $e$ is the edge connecting vertex $V_{L,i}$ on the left and vertex $V_{R,j}$ on the right. }\\
0 & \text{otherwise.}
\end{cases}
\end{equation*}
\end{definition}
It follows that $G$ admits a perfect matching if and only if $\det(\mG) \neq 0$ \cite{motwani1995randomized}.

Therefore, if we have an $n \times n$ symbolic matrix $\mA$ where $\mA_{ij} = 0$ or $\mA_{ij} = x_{ij}$ for variables $x_{ij}$ that occur exactly once in $\mA$, $\mA$ will correspond to a balanced bipartite graph $G = ((V_L, V_R), E)$ where $|V_L| = |V_R| = n$ and there is an edge between $V_{L,i}$ and $V_{R,j}$ if and only if $\mA_{ij} = x_{ij} \neq 0$. Notice that the correspondence between matrix $\mA$ and the bipartite graph $G$ is slightly different from the notion of ``underlying graph" \cite{AY10} used earlier: each non-zero entry of $\mA$ will correspond to an edge in its corresponding bipartite graph $G$. Furthermore, if $\mA_{ij} \neq 0$, the edge corresponding to $\mA_{ij}$ is $(V_{L,i}, V_{R,j})$, while if $\mA_{ji} \neq 0$, the edge corresponding to $\mA_{ji}$ is $(V_{L,j}, V_{R,i})$. Therefore, two elements $\mA_{ij}$ and $\mA_{ji}$ symmetric about the diagonal do not correspond to the same edge unless $i = j$, while in the definition of underlying graphs, $\mA_{ij}$ and $\mA_{ji}$ gets mapped to the same undirected edge $(i,j).$

Deciding whether $\det(\mA) = 0$ can be done efficiently using known bipartite matching algorithms. The almost-linear time network flow algorithm by Chen et al. \cite{CKL22} can be used to solve bipartite matching in almost-linear time with high probability. 

\begin{lemma}
\label{lem:edmonds-symbolic}
Let $\mA$ be a symbolic matrix whose entries are polynomials without non-zero constant terms in $x_1,\ldots,x_t$ such that the variables occurring in different entries of $\mA$ are disjoint. Let $G = ((V_L, V_R), E)$ be the corresponding graph such that there is an edge between $V_{L,i}$ and $V_{R,j}$ if and only if $\mA_{ij} \neq 0$. Then $\det(\mA) \neq 0$ if and only if $G$ admits a perfect bipartite matching.
\end{lemma}

\begin{proof}
For each non-entry of $\mA$,  we may substitute $\mA_{ij} \neq 0$ with variable $z_{ij}$. Let $\hat{\mA}$ be the matrix after substitution. It follows that $\hat{\mA}$ is an Edmonds matrix, so $\det (\hat \mA) = 0$ if and only if $G$ admits a perfect matching. 

Furthermore, $\det (\hat \mA) = 0$ if and only if $\det(\mA) = 0$. If $\det(\hat \mA) = 0$, then substituting $z_{ij}$ with $\mA_{ij}$ immediately gives us $\det(\mA) = 0$. Conversely, if $\det (\hat \mA) = \sum_{\sigma \in S_n} \prod \hat \mA_{i\sigma(i)} \neq 0$, we can pick a $\sigma \in S_n$ such that $\prod \hat \mA_{i\sigma(i)} = \prod z_{i \sigma(i)} \neq 0$. Substituting $\mA_{i \sigma(i)}$ back to $z_{ij}$ gives us $\prod \mA_{i \sigma(i)} \neq 0$. Since variables occurring in different entries are disjoint in $\mA$, if $\tau \neq \sigma$, $\prod \mA_{i\tau(i)}$ cannot be a scalar multiple of $\prod \mA_{i \sigma(i)} \neq 0$, so the term will have non-zero contribution to $\det (\mA)$. 
\end{proof}

In the worst case, if $\mG$ is the Edmonds matrix of perfect bipartite graph $K_{n,n}$, the polynomial $\det(\mG)$ would contain $n!$ terms. In fact, even computing the exact number of terms in the polynomial $\det(\mG)$ (in its canonical form) is $\#P$-complete. This is because the terms of $\det(\mG)$ correspond to permutations $\sigma \in S_n$ such that $\mG_{i \sigma(i)} \neq 0$ for all $i$, so the terms essentially correspond to the number of perfect matchings. Valiant's theorem shows computing the number of perfect matchings for bipartite graphs is $\#P$-complete \cite{V79}. 

\paragraph{Tutte Matrices and General Graph Matching.} The Tutte matrix \cite{T47} is a generalization of the Edmonds matrix to general undirected graphs. First, let us describe the correspondence between undirected graphs $G = (V,E)$ and their corresponding Tutte matrices $\mA$. We should see soon that this correspondence is exactly the correspondence between a matrix $\mA$ and its underlying graph $G_{\mA}$ occuring in the context of hardness of Laplacian systems.

\begin{definition}[Tutte Matrix]
\label{def:tutte}
Given a  graph $G = (V,E)$ where $|V| = n$ and $|E|$, its Tutte matrix $\mA$ is an $n \times n$ skew-symmetric matrix in variables $x_e$ where the index $e \in E$ ranges over $E$:
\begin{equation*}
\mA_{ij} := \begin{cases}
x_e & \text{if $e = (v_i,v_j) \in E$ and $i < j$,}\\
-x_e & \text{if $e = (v_j,v_i) \in E$ and $i > j$,} \\
0 & \text{otherwise.}
\end{cases}
\end{equation*}
\end{definition}

It follows that $\det(\mA) \neq 0$ if and only if $G$ has a perfect matching \cite{T47}. 

Now we may think about the reverse process: given an $n \times n$ skew-symmetric symbolic matrix $\mA$, construct the graph corresponding to $\mA$ such that $\mA$ is the Tutte matrix of the graph (or show this is impossible). Notice that by the construction given above, the entries above the diagonal in $\mA$ must be independent variables or zero. If this is the case, then we know that the corresponding graph $G = (V,E)$ should have an undirected edge between $v_i$ and $v_j$ ($i < j$) if and only if $\mA_{ij} = -\mA_{ji} \neq 0$. Therefore, the graph corresponding to $\mA$ in this case is the underlying graph $G_{\mA}$, which we have been dealing with when discussing the hardness of special classes of Laplacian systems over $\mathbb{Z}_p$. Furthermore, checking if a general undirected graph admits a perfect matching can be done in polynomial time by Edmonds' blossom algorithm \cite{E65}.

\paragraph{Graph-Structured Symbolic Matrices With Intermediate Hardness.} Tutte matrices is an example where deciding if $\det(\mA) = 0$ can be directly reduced to combinatorial properties of its underlying graph $G_{\mA}$. What remains unclear is whether there exists graph-structured symbolic matrices with intermediate hardness from a fine-grained complexity perspective. Here, ``intermediate" can mean it is harder than solving Edmonds matrices but easier than Tutte matrices, or harder than solving Tutte matrices but easier than general symbolic matrices with polynomial entries. Deterministically deciding if general symbolic matrices with polynomial entries have determinant zero may require exponential time.

\subsection{Formalization of Symbolic Matrix Classes}
In this section, we are going to investigate the hardness of special classes of symbolic matrices. Intuitively, symbolic matrices are matrices whose entries are in the polynomial ring $\mathbb{F}[x_1,\ldots,x_t]$ where $\mathbb{F}$ is some field. Two complexity measures we care about are the maximum (total) degree of a term in a matrix entry and the maximum number of occurrences of any variable in the matrix. First, we are going to formally define the metrics mentioned above.

\subsubsection{Polynomial Degree}

\begin{definition}[Maximum polynomial degree of a symbolic matrix] Given a matrix $\mA$ whose entries live in $R = \mathbb{F}[x_1,\ldots,x_m]$, the maximum polynomial degree of $\mA$ is the maximum total degree of entries of $\mA$ when viewed as polynomials in $\mathbb{F}[x_1,\ldots,x_m]$. In other words,
\begin{equation*}
\pdeg_{\mA} \defeq \max_{i,j} \deg(\mA_{ij}).
\end{equation*}

\end{definition}
The notion of ``degree" is a bit overused in mathematics, which causes naming conflicts if we want to study the relationship between symbolic matrices and their corresponding graphs. For a graph-structured symbolic matrix,
we might want to reason about both the maximum degree of the vertices in its corresponding graph and the total degrees of entries in the matrix as polynomials in $x_1,\ldots,x_n$.
We use ``polynomial degree", $\pdeg$, to emphasize the degree we are discussing refers to the degree of a polynomial instead of the degree of a vertex in a graph.

\subsubsection{Variable Multiplicity}
To formally define the number of occurrences of a variable in a matrix, we need to properly define how to count the number of times a variable appears in a polynomial. This would require us to choose a specific canonical form of a polynomial. Without a chosen canonical form of a polynomial, the notion of occurrences count of a variable is not well-defined. For our purposes, the canonical form of a polynomial is simply the polynomial written as a sum of distinct (non-zero) monomials. To make the definition go through, we work on the case of monomials first.

\begin{definition}[Monomial expressed by exponent vector] Let $R = \mathbb{F}[x_1,\ldots,x_t]$ be a polynomial ring. Let $I \in \mathbb{Z}^t$ be the exponent vector. The monomial $X^I$ is defined to be 
\begin{equation*}
X^I := \prod_{i=1}^m x_i^{I_i}.
\end{equation*}
\end{definition}

\begin{definition}[Variable multiplicity in a monomial] Let $R = \mathbb{F}[x_1,\ldots,x_t]$ be a polynomial ring. Let $c \in \mathbb{F}^*$ be a non-zero scalar in $\mathbb{F}$. The occurrence count of $x_i$ in the monomial $cX^I$ where $I \in \mathbb{Z}^m$ is a exponent vector is simply given by $I_i$:
\begin{equation*}
\m(x_i, cX^I) \defeq I_i.
\end{equation*}
\end{definition}

\begin{definition}[Canonical form of a polynomial] The canonical form of a polynomial $p \in \mathbb{F}[x_1,\ldots,x_t]$ is simply the form such that $p$ is written as a sum of distinct (non-zero) monomials, that is 
\begin{equation*}
p \defeq \sum_{i} c_i X^{I^i} 
\end{equation*}
where $I^i \in \mathbb{Z}^t$ are exponent vectors and $c_i \in \mathbb{F} - \{0\}$ are non-zero coefficients. Furthermore, the exponent vectors $I^i$ are distinct.

If $p$ is the zero polynomial, we may treat the above sum as an empty sum, in which case the canonical form is simply given by
\begin{equation*}
    p \defeq 0.
\end{equation*}

With a canonical form, we may formally define the variable multiplicity in a polynomial.

\begin{definition}[Variable multiplicity in a polynomial] Let $R = \mathbb{F}[x_1,\ldots,x_t]$ be a polynomial ring. Let $\displaystyle p = \sum_{i} c_{i} X^{I^i}$ be a polynomial in $R$ in its canonical form. The multiplicity of $x_j$ in $p$ is given by
\begin{equation*}
\m(x_j, p) \defeq \sum_{i} \m(x_j, c_i X^{I^i}) = \sum_i I^i_{j}.
\end{equation*}
\end{definition}

\end{definition}
We may now define the variable multiplicity in a symbolic matrix to be the sum of its multiplicities over the entries in the matrix:
\begin{definition}[Variable multiplicity in a symbolic matrix]Let $R = \mathbb{F}[x_1,\ldots,x_t]$ be a polynomial ring, and let $\mA \in R^{h \times w}$ be a matrix whose entries are in $R$. The variable multiplicity of $x_j$ in $\mA$ is defined to be 
\begin{equation*}
\m(x_j,\mA) \defeq \sum_{s = 1,\ldots,h;t = 1,\ldots,w} \m(x_j, \mA_{s,t}).
\end{equation*}
The maximum variable multiplicity of $\mA$ is simply the maximum variable multiplicity of $x_j$ in $\mA$ over variables $x_j$:
\begin{equation*}
\maxm(\mA) \defeq \max_{j = 1,\ldots,m} \m(x_j, \mA).
\end{equation*}
Additionally, we define the $\nnz(\mA)$ here as the sum of variable multiplicity of all variable $x_j$ in $\mA$. Notice that in this context, $\nnz(\mA)$ is not necessarily equal to the non-zero entries of $\mA$:
\begin{equation*}
\nnz(\mA) \defeq \sum_{j = 1,\ldots,m} \m(x_j, \mA).
\end{equation*}
\end{definition}

\subsubsection{Symbolic Matrix Class}

Finally, we may define the class of symbolic matrices subject to polynomial degree and variable occurence constraints:
\begin{definition}[Symbolic matrix classes parameterized by polynomial degree and variable multiplicity] Given a field $\mathbb{F}$ and a list of variables $\{x_1,\ldots,x_t\}$, let $R = \mathbb{F}[x_1,\ldots,x_t]$ be the ring of polynomials in $x_1,\ldots,x_t$ over $\mathbb{F}$, and let $I = (x_1,\ldots,x_t) \subseteq R$ be the ring generated by $x_1,\ldots,x_t$ (i.e.~$I$ consists of polynomials in $R$ without non-zero constant term). The class of symbolic matrices $\mathcal{SM}[R,  d,  m]$ are the class of matrices satisfy all following constraints: 
\begin{enumerate}
    \item The entries of the matrix are in the ideal $I$.
    \item The maximum total degree of entries in the matrix is bounded by $d$.
    \item The maximum multiplicity of any variable $x_i$ in the matrix is bounded above by $m$.
\end{enumerate}
\end{definition}

With a simple reduction, we can convert a symbolic matrix $\mA \in R^{n \times n}$ with non-zero constant term to a symbolic matrix $\hat \mA \in \mathbb{F}[x_0,x_1,\ldots,x_t]$ without non-zero constant term while preserving whether $\det(\mA) = 0$. The trick is simply multiply elements of $\mA$ with variable $x_0$. Since $\det \hat \mA = t^n \det(\mA)$, we know that $\det(\hat{\mA})= 0$ if and only if $\det(\mA) = 0$.

\subsubsection{Solving Symbolic Matrices}
After formally defining the notion of symbolic matrices, we are interested in how to define the meaning of solving symbolic matrices. If the entries of $\mA$ and $\vb$ are both in $R = \mathbb{F}[x_1,\ldots,x_t]$, one possible definition of ``solving" a symbolic linear system $\mA \vx = \vb$ is to find a solution $\vx$ where entries of $\vx$ are in $\mathbb{F}(x_1,\ldots,x_t)$, the field of rational functions in $x_1,\ldots,x_t$, satisfying that $\mA \vx = \vb$ under the operations of $\mathbb{F}(x_1,\ldots,x_t)$. Notice that since we are working with the field of rational functions in $x_1,\ldots,x_t$, we do not necessarily need to worry about what will happen if after plugging in values for $x_1,\ldots,x_t$, we end up with zero denominators. We only work with formal polynomials and rational functions, so we don't need to worry about plugging in actual values. 

One potential issue with this approach is the cost of representing an entry of $\vx$. For example, if the number of variable is $t = n^2$ and $\mA \in R^{n \times n}$ is given by $\mA_{i,j} = x_{(i-1)n+j}$, then the canonical form of $\det(\mA) \in R$ consists of $n!$ terms since each permutation $\sigma \in S_n$ gives rise to a distinct term $\sgn(\sigma) x_{1\sigma(1)}x_{2\sigma(2)}\cdots x_{n\sigma(n)}$ where $\sgn(\sigma) = 1$ if and only if $\sigma$ is an even permutation and $\sgn(\sigma) = -1$ otherwise. It would be extremely unwise to manipulate the denominators of $\vx$ in case $\det(\mA)$ is complicated. Fortunately, in most cases, we are more interested in the coefficient matrix $\mA$ than $\vb$, so we give a relaxed definition of ``solving" symbolic matrices below, which essentially asks if $\det(\mA) = 0$ as a polynomial:

\begin{definition}[Solving Symbolic Matrices, \smp]
Let $\mathbb{F}$ be a field and let $R = \mathbb{F}[x_1,\ldots,x_t]$ be a polynomial ring. Let $\mA \in R^{n \times n}$ be a symbolic matrix with entries in $R$. Solving a symbolic matrix $\mA$ is defined to be checking if $\det(\mA) = 0$ where $\det(\mA) \in R$ is a polynomial in $x_1,\ldots,x_t$.
\end{definition}

We will justify this definition through its applications in graph algorithms and the relationship between non-zero determinants and existence of inverses and non-trivial solutions to the homogeneous linear system $\mA \vx = \vzero$. The graph-theoretic applications motivates us to study the hardness of solving symbolic matrices. 
If $\mA$ is the Edmonds matrix of a corresponding bipartite graph, then $\det(\mA) \neq 0$ if and only if the corresponding bipartite graph admits a perfect matching.  If $\mA$ is the Tutte matrix of a corresponding graph, then $\det(\mA) \neq 0$ if and only if the corresponding graph admits a perfect matching. Furthermore, if  $\mA \in R^{n \times n}$ is a square matrix with entries in the polynomial ring $R = \mathbb{F}[x_1,\ldots,x_t]$, we may view $\mA$ as a matrix with entries in the field of rational functions $\mathbb{F}(x_1,\ldots,x_t)$. Working in a field rather than a polynomial ring makes reasoning about inverses easier. It follows that $\mA^{-1}$ exists if and only if $\det(\mA) \neq 0$ in $\mathbb{F}(x_1,\ldots,x_t)$. Since operations in $\mathbb{F}(x_1,\ldots,x_t)$ inherit the operations of polynomials in $F[x_1,\ldots,x_t]$, we know that there exists an $\mA^{-1}$ whose entries are in $\mathbb{F}(x_1,\ldots,x_t)$ if and only if $\det(\mA) \neq 0$ in $R = \mathbb{F}[x_1,\ldots,x_t]$. Finally, we are going to show that $\mA \vx = \vzero$ has a non-trivial solution $\vx$ such that all entries are in $R = \mathbb{F}[x_1,\ldots,x_t]$ if and only if $\det(\mA) = 0$ as a polynomial in $R$. Notice we are working with polynomial ring instead of a field here, so we need to embed $\mA$ as a matrix with entries in the field of rational functions in $x_1,\ldots,x_t$.

\begin{lemma}
Let $\mathbb{F}$ be a field and let $R = \mathbb{F}[x_1,\ldots,x_t]$ be the ring of polynomials in $x_1,\ldots,x_t$. Let $\mA \in R^{n \times n}$ be a symbolic matrix. $\mA \vx = \vzero$ has a non-trivial (i.e.~not identically zero) solution $\vx \in R^n$ if and only if $\det(\mA) = 0$.
\end{lemma}

\begin{proof}
We can treat $\mA$ as a matrix whose entries live in the field of rational functions in $x_1,\ldots,x_t$, $\mathbb{F}(x_1,\ldots,x_t)$. In this case, $\mA \vx = \vzero$ has a non-trivial solution $\vx$ with entries in $\mathbb{F}(x_1,\ldots,x_t)$ if and only if $\det(\mA) = 0$ in $\mathbb{F}(x_1,\ldots,x_t)$. Since $R$ is contained in $\mathbb{F}(x_1,\ldots,x_t)$ and operations of elements in $R$ in $\mathbb{F}(x_1,\ldots,x_t)$ inherits the operations in $R$, we conclude that $\det(\mA) = 0$ in $\mathbb{F}(x_1,\ldots,x_t)$ if and only if $\det(\mA) = 0$ in $R$. Furthermore, if we are given a solution $\vx$ in $(\mathbb{F}(x_1,\ldots,x_t))^n$, we may rescale the solution $\vx$ to get $\vx'$ by multiplying a common multiple $D$ of the denominators of the entries in $\vx$. 
It follows that we continue to have $\mA \vx' = \mA (D\vx) = 0$, while entries in $\vx'$ are all in $R$.

As a result, we conclude that $\mA \vx = \vzero$ has a non trivial solution $\vx \in R^n$ if and only if $\det(\mA) = 0$ in $R$.

\end{proof}
\subsection{Existing Results}
We can now rephrase the results about Edmonds matrices and Tutte matrices using the notation defined above.

\begin{lemma}
Solving symbolic matrices of class $\mathcal{SM}[R,1,1]$ is equivalent to determining if a balanced bipartite graph admits a perfect matching.
\end{lemma}

\begin{proof}
Suppose we are given $\mA \in R^{n \times n}$ of class $\mathcal{SM}[R,1,1]$. Since each variable appears at most once in $\mA$, we conclude that the variables appearing in different entries of $\mA$ are disjoint. By Lemma~\ref{lem:edmonds-symbolic}, we conclude that deciding if $\det(\mA) = 0$ can be reduced to deciding if the corresponding bipartite graph has a perfect matching. Conversely, given a bipartite graph, its Edmonds matrix is in class $\mathcal{SM}[\mathbb{F}[x_1,\ldots,x_{|E|}],1,1]$ where $|E|$ is the number of edges in the graph. Checking if the bipartite graph has a perfect matching can be reduced to checking if its Edmonds matrix has determinant 0. 
\end{proof}

\begin{lemma}
Solving symbolic matrices in class $\mathcal{SM}[R, 1, 2]$ is at least as hard as solving general graph matchings.
\end{lemma}

\begin{proof}
Given a graph $G = (V,E)$, we may construct its Tutte matrix $\mA \in \mathcal{SM}[\mathbb{F}[x_1,\ldots,x_{|E|}], 1,2]$. The entries of the Tutte matrix are linear, and each variable appears exactly twice: once above the diagonal and once under the diagonal.
\end{proof}

\subsection{Hardness Reduction}
Here we provide a reduction from the class $\GS_{\mathbb{R}}$ to a symbolic matrix class in solving the determinant decisional problem.

\begin{definition}[Zero Determinant Decision Problem over $\mathbb{R}$, \ddp]
\label{def:ddp}
    Given a matrix $\mA \in \mathbb{R}^{n \times n}$, we define the \ddp{} problem as checking if $\det(\mA)=0$.
\end{definition}

\begin{definition}[Symbolic Decisional $f$-reducibility] 
    Suppose we have a matrix class $\mathcal{M}^1$, a symbolic matrix class $\mathcal{SM}^2$, and an algorithm $\mathcal{A}_{1 \mapsto 2}$ such that given a \ddp{} instance $\mM^1$ where $\mM^1 \in \mathcal{M}^1$, the call $\mathcal{A}_{1 \mapsto 2}(\mM^1)$ returns an \smp{} instance $\mM^2$ that has the same logical output as $\mM^1$ and satisfies $\mM^2 \in \mathcal{SM}^2$. The logical output here refers to the boolean verification of $\ddp$ and $\smp$.
    
    Consider a function of $f: \mathbb{Z}_{+} \mapsto \mathbb{Z}_{+}$ such that every output coordinate is an non-decreasing function of every input coordinate. Suppose we always have
\begin{align*}
    \nnz(\mM^2) \le f(\nnz(\mM^1))
\end{align*}
and the running times of $\mathcal{A}_{1 \mapsto 2}$ is bounded by $\Otil(nnz(\mM^1))$. 

Then we say $\mathcal{M}^1$ is decisional $f$-reducible to $\mathcal{SM}^2$.
\end{definition}

\thmsymb*

\begin{proof}
Let $R=\mathbb{R}[x_1,\ldots,x_{n^2}]$. For any given matrix system $\mA \in \mathcal{\GS_{\mathbb{R}}}^{n \times n}$ we can compute a symbolic matrix $\mB \in \mathcal{SM}[R, 1, 3]$ such that 

\begin{enumerate}
    \item $\nnz(\mB) = O(\nnz(\mA))$ 
    \item $\det(\mA)$ is zero if and only if $\det(\mB)$ (as a polynomial) is zero.
\end{enumerate}
\paragraph{Construction of $\mB$} By basics of algebra, $\det(\mA)$ is zero if and only if there exists a solution $\vy \neq \vzero$ to $\mA \vy = \vzero$.

Let $k_i$ be the number of non-zero entries on row $i$. For each row $i (1 \leq i \leq n)$, one can write it as an equation
\begin{equation*}
    \mA_{i,p_1}\vy_{p_1}+\mA_{i,p_2}\vy_{p_2}+\dots+\mA_{i,p_{k_i}}\vy_{p_{k_i}}=0
\end{equation*}

For simplicity, we assume no row has all zero entries, i.e. $k_i>0$. Furthermore, we can rewrite this equation into a series of equations:

\begin{align*}
    \mA_{i,p_1}\vy_{p_1}+\mA_{i,p_2}\vy_{p_2}-\vy_{n+(\sum_{j<i}k_j)-i+2}=0 \\
    \vy_{n+(\sum_{j<i}k_j)-i+2}+\mA_{i,p_3}\vy_{p_3}-\vy_{n+(\sum_{j<i}k_j)-i+3}=0 \\
    \dots \\
    \vy_{n+(\sum_{j<i}k_j)-i+k_i-1}+\mA_{i,p_{k_i}}\vy_{p_{k_i}}-\vy_{n+(\sum_{j<i}k_j)-i+k_i}=0 \\
    \vy_{n+(\sum_{j<i}k_j)-i+k_i}=0
\end{align*}
In this way, we can rewrite the $i$-th equation $\mA_{i,p_1}\vy_{p_1} + \cdots + \mA_{i,p_{k_i}} \vy_{p_{k_i}}$ into $k_i$ equations with $k_i-1$ additional variables $\vy_{n+(\sum_{j<i}k_j)-i+2}$,$\dots$,$\vy_{n+(\sum_{j<i}k_j)-i+k_i}$. Each new equation contains at most $3$ terms on the left hand side and the right hand side is $0$.

After rewriting all rows, we will have $\sum_{i=1}^{n} k_i=\nnz(\mA)$ new equations with $\vy$ such that the number of variables is $\nnz(\mA)$:
$$|\vy|=n \text{ (the original $y_1,\dots,y_n$)}+ \sum_{i=1}^{n} (k_i-1)=n+\nnz(\mA)-n=\nnz(\mA).$$ The new equations form a linear system $\hat{\mA} \vy=\vzero$ such that $\hat{\mA} \in \mathbb{R}^{\nnz(\mA) \times \nnz(\mA)}$. Since each equation only contains $3$ items, $\nnz(\hat{\mA})=O(\nnz(\mA))$.

Similarly, by definition, $\det(\hat{\mA})$ is zero if and only if $\hat{\mA}\vy=\vzero$ has a non-zero solution $\vy$.

\begin{lemma}
    $\det(\mA)$ is zero if and only if $\det(\hat{\mA})$ is zero.
\end{lemma}
\begin{proof}
    If $\det(\mA)$ is zero, there exists a solution $\vy=\{\vy_1,\dots,\vy_n\} \neq \vzero$ for $\mA \vy=\vzero$. For each $y_i (i > n)$, we can compute the value by the corresponding equations respectively. We first compute $\vy_{n+(\sum_{j<i}k_j)-i+2}$ by the relation $\vy_{n+(\sum_{j<i}k_j)-i+2} = \mA_{i,p_1}\vy_{p_1}+\mA_{i,p_2}\vy_{p_2}$. $\vy_{n+(\sum_{j<i}k_j)-i+k_i}$ is $0$ by definition. Then for any $\vy_{n+(\sum_{j<i}k_j)-i+j} (3 \leq j <k_i)$, we have 
    $$\vy_{n+(\sum_{j<i}k_j)-i+j}=\vy_{n+(\sum_{j<i}k_j)-i+j-1}+\mA_{i,p_j}\vy_{p_j}$$
    
    All new $\vy_i$ can be attached to $\vy$ and continue being a solution $\vy \neq \vzero$ to $\hat{\mA}\vy=\vzero$. Since there exists such a solution, $\det(\hat{\mA})$ is zero.
    
    On the other direction, if $\det(\hat{\mA})$ is zero, there exists a $\vy \neq \vzero$ such that $\hat{\mA}\vy=\vzero$. We truncate all $\vy_i$ for $i>n$. By construction, the truncated vector is a solution to $\mA \vy =\vzero$. If after truncation, $\vy$ becomes $\vzero$, by the process we compute $\vy_i (i>n)$ described above, we will have $\vy_i = 0$ holds for all $i$, which gives a contradiction.
\end{proof}

The last step will be transform $\hat{\mA}$ into a symbolic matrix. We need the following lemma:

\begin{lemma}
    For any matrix $\mA \in \mathcal{\GS_{\mathbb{R}}}^{n \times n}$, let $\mC=diag(b_1,b_2,\dots,b_n)$ be a diagonal matrix of $n$ symbolic variables $b_i$. Let $\mB=\mA\mC$. We have $\det(\mA)$ is zero if and only if $\det(\mB)$ is zero.
\end{lemma}

\begin{proof}
When $\det(\mA)$ is zero, by $\det(\mB)=\det (\mA\mC)=\det(\mA) \cdot \det(\mC)$ is zero. Reversely, when $\det(\mB)$ is zero, since $\det(\mC)=\prod_{i=1}^{n}b_i$ is non-zero, $\det(\mA)$ must be zero. 
\end{proof}

Therefore, Let $\mB=\hat{\mA}\mC$. Because $\det(\mA) = 0$ if and only if $\det(\hat{\mA}) = 0$ and $\det(\hat{\mA}) = 0$ if and only if $\det(\mB) = 0$, $\det(\mA)$ is zero if and only if $\det(\mB)$ is zero. $\mB$ is defined by $O(\nnz(\mA))$ symbolic variables, which the polynomial degree is $1$ and max multiplicity is $\leq 3$.

\paragraph{Running Time of $\mathcal{A}_{1 \mapsto 2}$}

By our construction, $\mathcal{A}_{1 \mapsto 2}$ is decomposing the matrix to $\nnz(\mA)$ rows with $O(\nnz(\mA))$ entries, which takes $O(\nnz(\mA))$ time. Computing $\hat{\mA}\vb$ takes $O(\nnz(\hat{\mA}))=O(\nnz(\mA))$.
\end{proof}
\bibliography{ref}

\newcommand{\etalchar}[1]{$^{#1}$}
\begin{thebibliography}{BMSVH06}

\bibitem[AY10]{AY10}
Noga Alon and Raphael Yuster.
\newblock Solving linear systems through nested dissection.
\newblock In {\em 51th Annual {IEEE} Symposium on Foundations of Computer
  Science, {FOCS} 2010, October 23-26, 2010, Las Vegas, Nevada, {USA}}, pages
  225--234. {IEEE} Computer Society, 2010.

\bibitem[BCPT05]{BCPT05}
Erik~G. Boman, Doron Chen, Ojas Parekh, and Sivan Toledo.
\newblock On factor width and symmetric {H}-matrices.
\newblock {\em Linear Algebra and its Applications}, 405:239--248, 2005.

\bibitem[BMSVH06]{BMSV06}
D.~A. Bini, B.~Meini, S.~Steff\'{e}, and B.~Van~Houdt.
\newblock Structured markov chains solver: Software tools.
\newblock In {\em Proceeding from the 2006 Workshop on Tools for Solving
  Structured Markov Chains}, SMCtools '06, page 14–es, New York, NY, USA,
  2006. Association for Computing Machinery.

\bibitem[CFM{\etalchar{+}}14]{Coh+14}
Michael~B. Cohen, Brittany~Terese Fasy, Gary~L. Miller, Amir Nayyeri, Richard
  Peng, and Noel Walkington.
\newblock Solving 1-{L}aplacians in nearly linear time: Collapsing and
  expanding a topological ball.
\newblock In Chandra Chekuri, editor, {\em Proceedings of the Twenty-Fifth
  Annual {ACM-SIAM} Symposium on Discrete Algorithms, {SODA} 2014, Portland,
  Oregon, USA, January 5-7, 2014}, pages 204--216. {SIAM}, 2014.

\bibitem[CGR05]{DGR05}
Gianna M.~Del Corso, Antonio Gull{\'\i}, and Francesco Romani.
\newblock Fast pagerank computation via a sparse linear system.
\newblock {\em Internet Mathematics}, 2(3):251--273, 2005.

\bibitem[Chu96]{C96}
Fan~RK Chung.
\newblock Lectures on spectral graph theory.
\newblock {\em CBMS Lectures, Fresno}, 6(92):17--21, 1996.

\bibitem[CK22]{CK22}
S{\'\i}lvia Casacuberta and Rasmus Kyng.
\newblock {Faster Sparse Matrix Inversion and Rank Computation in Finite
  Fields}.
\newblock In {\em 13th Innovations in Theoretical Computer Science Conference
  (ITCS 2022)}, volume 215, pages 33:1--33:24, 2022.

\bibitem[CKK{\etalchar{+}}18]{Coh+18}
Michael~B. Cohen, Jonathan~A. Kelner, Rasmus Kyng, John Peebles, Richard Peng,
  Anup~B. Rao, and Aaron Sidford.
\newblock Solving directed {L}aplacian systems in nearly-linear time through
  sparse {LU} factorizations.
\newblock In Mikkel Thorup, editor, {\em 59th {IEEE} Annual Symposium on
  Foundations of Computer Science, {FOCS} 2018, Paris, France, October 7-9,
  2018}, pages 898--909. {IEEE} Computer Society, 2018.

\bibitem[CKL{\etalchar{+}}22]{CKL22}
Li~Chen, Rasmus Kyng, Yang~P Liu, Richard Peng, Maximilian~Probst Gutenberg,
  and Sushant Sachdeva.
\newblock Maximum flow and minimum-cost flow in almost-linear time.
\newblock {\em 2022 IEEE 61rd Annual Symposium on Foundations of Computer
  Science (FOCS)}, 2022.

\bibitem[CKM{\etalchar{+}}11]{CKMST11}
Paul Christiano, Jonathan~A. Kelner, Aleksander Madry, Daniel~A. Spielman, and
  Shang-Hua Teng.
\newblock Electrical flows, {L}aplacian systems, and faster approximation of
  maximum flow in undirected graphs.
\newblock In {\em Proceedings of the Forty-Third Annual ACM Symposium on Theory
  of Computing}, STOC '11, page 273–282, New York, NY, USA, 2011. Association
  for Computing Machinery.

\bibitem[CKP{\etalchar{+}}17]{CKP17}
Michael~B Cohen, Jonathan Kelner, John Peebles, Richard Peng, Anup~B Rao, Aaron
  Sidford, and Adrian Vladu.
\newblock Almost-linear-time algorithms for markov chains and new spectral
  primitives for directed graphs.
\newblock In {\em Proceedings of the 49th Annual ACM SIGACT Symposium on Theory
  of Computing}, pages 410--419, 2017.

\bibitem[CLL11]{CLL11}
Ho~Yee Cheung, Lap~Chi Lau, and Kai~Man Leung.
\newblock Graph connectivities, network coding, and expander graphs.
\newblock In {\em 2011 IEEE 52nd Annual Symposium on Foundations of Computer
  Science}, pages 190--199. IEEE, 2011.

\bibitem[Cop93]{C93}
Don Coppersmith.
\newblock Solving linear equations over gf(2): block lanczos algorithm.
\newblock {\em Linear Algebra and its Applications}, 192:33--60, 1993.

\bibitem[CS11]{CS11}
William Cook and Daniel Steffy.
\newblock Solving very sparse rational systems of equations.
\newblock {\em ACM Trans. Math. Softw.}, 37:39, 02 2011.

\bibitem[Dix82]{Dix82}
John~D Dixon.
\newblock Exact solution of linear equations using {P}-adic expansions.
\newblock {\em Numerische Mathematik}, 40(1):137--141, 1982.

\bibitem[DKGZ22]{DGKZ22}
Ming Ding, Rasmus Kyng, Maximilian~Probst Gutenberg, and Peng Zhang.
\newblock Hardness results for {L}aplacians of simplicial complexes via
  sparse-linear equation complete gadgets.
\newblock {\em CoRR}, abs/2202.05011, 2022.

\bibitem[DKP{\etalchar{+}}17]{DKP+17}
David Durfee, Rasmus Kyng, John Peebles, Anup~B. Rao, and Sushant Sachdeva.
\newblock Sampling random spanning trees faster than matrix multiplication.
\newblock STOC 2017, page 730–742, New York, NY, USA, 2017. Association for
  Computing Machinery.

\bibitem[DL78]{DL78}
Richard~A. Demillo and Richard~J. Lipton.
\newblock A probabilistic remark on algebraic program testing.
\newblock {\em Information Processing Letters}, 7(4):193--195, 1978.

\bibitem[dlVP96]{P1896}
Ch.J. de~la Vallée~Poussin.
\newblock Recherches analytiques la théorie des nombres premiers.
\newblock {\em Ann. Soc. scient. Bruxelles}, 20:183--256, 1896.

\bibitem[DS07]{DS07}
Samuel~I. Daitch and Daniel~A. Spielman.
\newblock Support-graph preconditioners for 2-dimensional trusses.
\newblock {\em CoRR}, abs/cs/0703119, 2007.

\bibitem[DS08]{DS08}
Samuel~I. Daitch and Daniel~A. Spielman.
\newblock Faster approximate {L}ossy generalized flow via interior point
  algorithms.
\newblock In {\em Proceedings of the Fortieth Annual ACM Symposium on Theory of
  Computing}, STOC '08, page 451–460, New York, NY, USA, 2008. Association
  for Computing Machinery.

\bibitem[Edm65]{E65}
Jack Edmonds.
\newblock Paths, trees, and flowers.
\newblock {\em Canadian Journal of Mathematics}, 17:449–467, 1965.

\bibitem[EGG{\etalchar{+}}06]{EGGSV06}
Wayne Eberly, Mark Giesbrecht, Pascal Giorgi, Arne Storjohann, and Gilles
  Villard.
\newblock Solving sparse rational linear systems.
\newblock In Barry~M. Trager, editor, {\em Symbolic and Algebraic Computation,
  International Symposium, {ISSAC} 2006, Genoa, Italy, July 9-12, 2006,
  Proceedings}, pages 63--70. {ACM}, 2006.

\bibitem[EGG{\etalchar{+}}07]{EGGSV07}
Wayne Eberly, Mark Giesbrecht, Pascal Giorgi, Arne Storjohann, and Gilles
  Villard.
\newblock Faster inversion and other black box matrix computations using
  efficient block projections.
\newblock In Dongming Wang, editor, {\em Symbolic and Algebraic Computation,
  International Symposium, {ISSAC} 2007, Waterloo, Ontario, Canada, July 28 -
  August 1, 2007, Proceedings}, pages 143--150. {ACM}, 2007.

\bibitem[Fus06]{Fus06}
Andrea Fusiello.
\newblock Elements of geometric computer vision.
\newblock {\em Available fro m: http://homepages. inf. ed. ac.
  uk/rbf/CVonline/LOCAL\_COPIES/FUSIELLO4/tutorial. html}, 2006.

\bibitem[Geo73]{Geo73}
Alan George.
\newblock Nested dissection of a regular finite element mesh.
\newblock {\em SIAM J. Numer. Anal.}, 10(2):345–363, apr 1973.

\bibitem[GGOW16]{GGOW16}
Ankit Garg, Leonid Gurvits, Rafael Oliveira, and Avi Wigderson.
\newblock A deterministic polynomial time algorithm for non-commutative
  rational identity testing.
\newblock In {\em 2016 IEEE 57th Annual Symposium on Foundations of Computer
  Science (FOCS)}, pages 109--117, 2016.

\bibitem[Had96]{H1896}
J.~Hadamard.
\newblock Sur la distribution des z\'eros de la fonction $\zeta (s)$ et ses
  cons\'equences arithm\'etiques.
\newblock {\em Bulletin de la Soci\'et\'e Math\'ematique de France},
  24:199--220, 1896.

\bibitem[JJT{\etalchar{+}}07]{JJT07}
Hrvoje Jasak, Aleksandar Jemcov, Zeljko Tukovic, et~al.
\newblock Openfoam: A {C}++ library for complex physics simulations.
\newblock In {\em International workshop on coupled methods in numerical
  dynamics}, volume 1000, pages 1--20. IUC Dubrovnik Croatia, 2007.

\bibitem[JS21]{JS21}
Arun Jambulapati and Aaron Sidford.
\newblock Ultrasparse ultrasparsifiers and faster laplacian system solvers.
\newblock In {\em Proceedings of the Thirty-Second Annual ACM-SIAM Symposium on
  Discrete Algorithms}, SODA '21, page 540–559, USA, 2021. Society for
  Industrial and Applied Mathematics.

\bibitem[KLP{\etalchar{+}}16]{KLPSS16}
Rasmus Kyng, Yin~Tat Lee, Richard Peng, Sushant Sachdeva, and Daniel~A.
  Spielman.
\newblock Sparsified cholesky and multigrid solvers for connection laplacians.
\newblock In {\em Proceedings of the Forty-Eighth Annual ACM Symposium on
  Theory of Computing}, STOC '16, page 842–850, New York, NY, USA, 2016.
  Association for Computing Machinery.

\bibitem[KLS20]{KLS20}
Tarun Kathuria, Yang~P. Liu, and Aaron Sidford.
\newblock Unit capacity maxflow in almost $o(m^{4/3})$ time.
\newblock In {\em 2020 IEEE 61st Annual Symposium on Foundations of Computer
  Science (FOCS)}, pages 119--130, 2020.

\bibitem[KM09]{KM09}
Jonathan~A. Kelner and Aleksander Madry.
\newblock Faster generation of random spanning trees.
\newblock In {\em 2009 50th Annual IEEE Symposium on Foundations of Computer
  Science}, pages 13--21, 2009.

\bibitem[KMP11]{KMP11}
Ioannis Koutis, Gary~L. Miller, and Richard Peng.
\newblock A nearly-m log n time solver for {SDD} linear systems.
\newblock In {\em 2011 IEEE 52nd Annual Symposium on Foundations of Computer
  Science}, pages 590--598, 2011.

\bibitem[KMP14]{KMP10}
Ioannis Koutis, Gary~L. Miller, and Richard Peng.
\newblock Approaching optimality for solving {SDD} linear systems.
\newblock {\em SIAM Journal on Computing}, 43(1):337--354, 2014.

\bibitem[KOSZ13]{KOSZ13}
Jonathan~A Kelner, Lorenzo Orecchia, Aaron Sidford, and Zeyuan~Allen Zhu.
\newblock A simple, combinatorial algorithm for solving {SDD} systems in
  nearly-linear time.
\newblock In {\em Proceedings of the forty-fifth annual ACM symposium on Theory
  of computing}, pages 911--920, 2013.

\bibitem[KPSZ18]{KyngPSZ18}
Rasmus Kyng, Richard Peng, Robert Schwieterman, and Peng Zhang.
\newblock Incomplete nested dissection.
\newblock In Ilias Diakonikolas, David Kempe, and Monika Henzinger, editors,
  {\em Proceedings of the 50th Annual {ACM} {SIGACT} Symposium on Theory of
  Computing, {STOC} 2018, Los Angeles, CA, USA, June 25-29, 2018}, pages
  404--417. {ACM}, 2018.
\newblock Available at:~\url{https://arxiv.org/abs/1805.09442}.

\bibitem[KRSS15]{KRSS15}
Rasmus Kyng, Anup Rao, Sushant Sachdeva, and Daniel~A. Spielman.
\newblock Algorithms for {L}ipschitz learning on graphs.
\newblock In Peter Grünwald, Elad Hazan, and Satyen Kale, editors, {\em
  Proceedings of The 28th Conference on Learning Theory}, volume~40 of {\em
  Proceedings of Machine Learning Research}, pages 1190--1223, Paris, France,
  03--06 Jul 2015. PMLR.

\bibitem[KZ20]{KZ17}
Rasmus Kyng and Peng Zhang.
\newblock Hardness results for structured linear systems.
\newblock {\em SIAM Journal on Computing}, 49(4):FOCS17--280, 2020.

\bibitem[Lan50]{L50}
Cornelius Lanczos.
\newblock An iteration method for the solution of the eigenvalue problem of
  linear differential and integral operators.
\newblock {\em Journal of research of the National Bureau of Standards},
  45:255--282, 1950.

\bibitem[LRT79]{LRR79}
Richard~J. Lipton, Donald~J. Rose, and Robert~Endre Tarjan.
\newblock Generalized nested dissection.
\newblock {\em SIAM Journal on Numerical Analysis}, 16(2):346--358, 1979.

\bibitem[LS19]{LS19}
Yin~Tat Lee and Aaron Sidford.
\newblock Solving linear programs with sqrt(rank) linear system solves, 10
  2019.

\bibitem[LZL03]{LZL03}
Jeonghwa Lee, Jun Zhang, and Cai-Cheng Lu.
\newblock Incomplete lu preconditioning for large scale dense complex linear
  systems from electromagnetic wave scattering problems.
\newblock {\em Journal of Computational Physics}, 185(1):158--175, 2003.

\bibitem[Lá79]{L79}
Lovász László.
\newblock On determinants, matchings and random algorithms.
\newblock volume~79, pages 565--574, 01 1979.

\bibitem[Mon95]{M95}
Peter~L. Montgomery.
\newblock A block {L}anczos algorithm for finding dependencies over gf(2).
\newblock In {\em EUROCRYPT}, 1995.

\bibitem[MR95]{motwani1995randomized}
Rajeev Motwani and Prabhakar Raghavan.
\newblock {\em Randomized algorithms}.
\newblock Cambridge university press, 1995.

\bibitem[Nie22]{N22}
Zipei Nie.
\newblock Matrix anti-concentration inequalities with applications.
\newblock In Stefano Leonardi and Anupam Gupta, editors, {\em {STOC} '22: 54th
  Annual {ACM} {SIGACT} Symposium on Theory of Computing, Rome, Italy, June 20
  - 24, 2022}, pages 568--581. {ACM}, 2022.
\newblock Available at:~\url{https://arxiv.org/abs/2111.05553}.

\bibitem[Odl84]{O84}
Andrew~M. Odlyzko.
\newblock Discrete logarithms in finite fields and their cryptographic
  significance.
\newblock In {\em EUROCRYPT}, 1984.

\bibitem[Ost71]{OSTROWSKI1971}
Alexander Ostrowski.
\newblock A new proof of {H}aynsworth's quotient formula for schur complements.
\newblock {\em Linear Algebra and its Applications}, 4(4):389--392, 1971.

\bibitem[PV21]{PV21}
Richard Peng and Santosh~S. Vempala.
\newblock Solving sparse linear systems faster than matrix multiplication.
\newblock In D{\'{a}}niel Marx, editor, {\em Proceedings of the 2021 {ACM-SIAM}
  Symposium on Discrete Algorithms, {SODA} 2021, Virtual Conference, January 10
  - 13, 2021}, pages 504--521. {SIAM}, 2021.
\newblock Available at:~\url{https://arxiv.org/abs/2007.10254}.

\bibitem[San04]{S04}
P.~Sankowski.
\newblock Dynamic transitive closure via dynamic matrix inverse: extended
  abstract.
\newblock In {\em 45th Annual IEEE Symposium on Foundations of Computer
  Science}, pages 509--517, 2004.

\bibitem[San05]{Sankowski05}
Piotr Sankowski.
\newblock Shortest paths in matrix multiplication time.
\newblock In {\em European Symposium on Algorithms}, pages 770--778. Springer,
  2005.

\bibitem[Sch80]{S80}
J.~T. Schwartz.
\newblock Fast probabilistic algorithms for verification of polynomial
  identities.
\newblock {\em J. ACM}, 27(4):701–717, oct 1980.

\bibitem[Sch18]{Schlid18}
Aaron Schild.
\newblock An almost-linear time algorithm for uniform random spanning tree
  generation.
\newblock In {\em Proceedings of the 50th Annual ACM SIGACT Symposium on Theory
  of Computing}, STOC 2018, page 214–227, New York, NY, USA, 2018.
  Association for Computing Machinery.

\bibitem[ST04]{ST04}
Daniel~A. Spielman and Shang-Hua Teng.
\newblock Nearly-linear time algorithms for graph partitioning, graph
  sparsification, and solving linear systems.
\newblock In {\em Proceedings of the Thirty-Sixth Annual ACM Symposium on
  Theory of Computing}, STOC '04, page 81–90, New York, NY, USA, 2004.
  Association for Computing Machinery.

\bibitem[Sto05]{Storjohann05}
Arne Storjohann.
\newblock The shifted number system for fast linear algebra on integer
  matrices.
\newblock {\em Journal of Complexity}, 21(4):609--650, 2005.
\newblock Festschrift for the 70th Birthday of Arnold Schonhage.

\bibitem[Ten10]{Teng10}
Shang-Hua Teng.
\newblock The {L}aplacian paradigm: Emerging algorithms for massive graphs.
\newblock In {\em Proceedings of the 7th Annual Conference on Theory and
  Applications of Models of Computation}, TAMC'10, page 2–14, Berlin,
  Heidelberg, 2010. Springer-Verlag.

\bibitem[Tut47]{T47}
W.~T. Tutte.
\newblock {The Factorization of Linear Graphs}.
\newblock {\em Journal of the London Mathematical Society}, s1-22(2):107--111,
  04 1947.

\bibitem[Val79]{V79}
L.G. Valiant.
\newblock The complexity of computing the permanent.
\newblock {\em Theoretical Computer Science}, 8(2):189--201, 1979.

\bibitem[vdBNS19]{JNS19}
Jan van~den Brand, Danupon Nanongkai, and Thatchaphol Saranurak.
\newblock Dynamic matrix inverse: Improved algorithms and matching conditional
  lower bounds.
\newblock In {\em 2019 IEEE 60th Annual Symposium on Foundations of Computer
  Science (FOCS)}, pages 456--480. IEEE, 2019.

\bibitem[Vis12]{Vish12}
Nisheeth Vishnoi.
\newblock Lx=b. {L}aplacian solvers and their algorithmic applications.
\newblock {\em Foundations and Trends in Theoretical Computer Science}, 8, 01
  2012.

\bibitem[VLO73]{V73}
M~Van~Lier and RHJM Otten.
\newblock Planarization by transformation.
\newblock {\em IEEE Transactions on Circuit Theory}, 20(2):169--171, 1973.

\bibitem[ZGL03]{ZGL03}
Xiaojin Zhu, Zoubin Ghahramani, and John Lafferty.
\newblock Semi-supervised learning using gaussian fields and harmonic
  functions.
\newblock In {\em Proceedings of the Twentieth International Conference on
  International Conference on Machine Learning}, ICML'03, page 912–919. AAAI
  Press, 2003.

\bibitem[Zip79]{Z79}
Richard Zippel.
\newblock Probabilistic algorithms for sparse polynomials.
\newblock In Edward~W. Ng, editor, {\em Symbolic and Algebraic Computation},
  pages 216--226, Berlin, Heidelberg, 1979. Springer Berlin Heidelberg.

\end{thebibliography}
\bibliographystyle{alpha}

\end{document}